\documentclass[namedreferences]{solarphysics}

\usepackage[optionalrh]{spr-sola-addons} 
\usepackage{graphicx}        
\usepackage{color}           
\usepackage{amsmath}
\usepackage{appendix}
\usepackage{adjustbox}

\defcitealias{Liewer2019}{I}

\renewcommand{\vec}[1]{{\mathbfit #1}}

\begin{document}
\begin{article}
\begin{opening}

\title{Quantifying Chromosphere Response to Flare Energy Release Using AIA Observations in 1600~\AA\ and 304~\AA\ Passbands}


\author[addressref=aff1]{\inits{}\fnm{Jiong}~\lnm{Qiu}}
\author[addressref=aff1]{\inits{}\fnm{Rhiannon}~\lnm{Fleming}}

\address[id=aff1]{Department of Physics, Montana State University, Bozeman, MT, 59717, USA}

\runningauthor{Qiu \& Fleming}

\begin{abstract}

Imaging observations of the solar lower atmosphere by the Atmosphere Imaging Assembly (AIA) have been mostly used as the context, and their quantitative information has been much less explored. The chromosphere responds rapidly to energy release by magnetic reconnection during flares. Furthermore, a flare is a collection of multiple energy release events that can be identified in spatially resolved chromosphere observations. In this paper, we conduct a statistical and semi-quantitative study of the relative photometry in the UV 1600~\AA\ and EUV 304~\AA\ passbands for 18 flares observed by AIA. In each flare, we have identified thousands of flare ribbon pixels in the UV 1600~\AA\ images, and measured their brightness (counts per second) and the rise and decay timescales, which are indicative of heating properties in flare loops. The analysis shows that bright flare pixels, characterized by peak brightness larger than ten times the quiescent brightness, exhibit sharp light curves with the half rise time below 2 min, followed by a two-phase decay with a rapid decay on timescales comparable to the rise time and then a more gradual decay. Flare ribbon pixels identified in both UV 1600 ~\AA\ and EUV 304~\AA\ images exhibit similar time profiles during the rise, and their peak brightness appear to be related by a power law. Our analysis shows that AIA observed flare brightness in UV 1600~\AA\ relative to the quiescent brightness is a meaningful measurement of the flare chromosphere photometry, and AIA observations for over a decade thus provide a unique and extensive database for systematic and semi-quantitative study of flaring chromosphere, either in the context of the Sun as a star, or in spatially resolved manner that helps to probe the nature of flare energy release on elementary scales.

\end{abstract}


\end{opening}

\section{Introduction}
\label{sec:intro}

\begin{figure}    
   \includegraphics[width=0.98\textwidth,clip=]{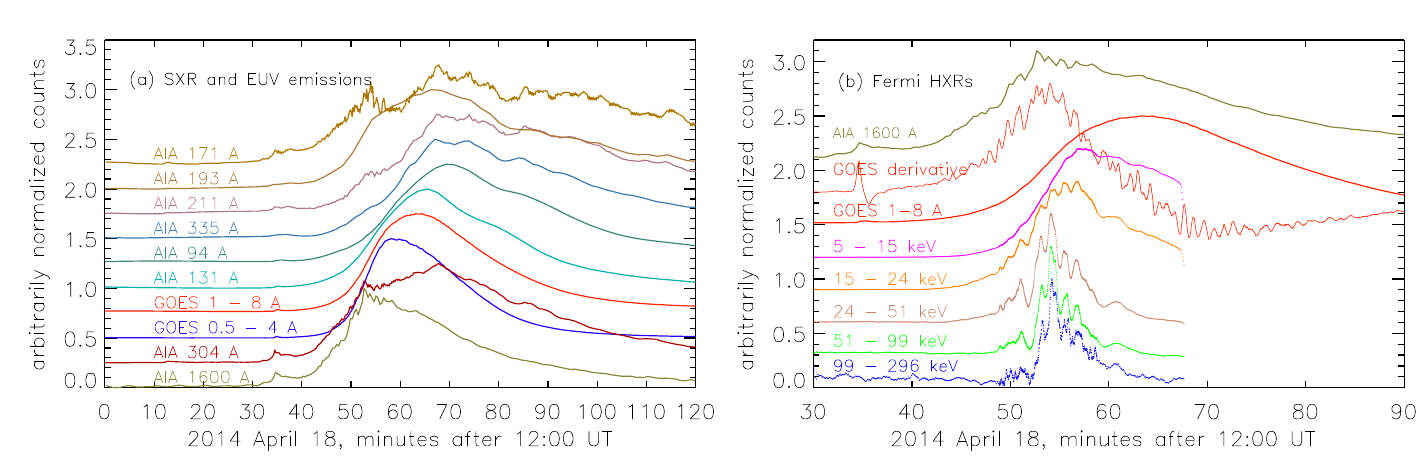}
   \includegraphics[width=0.98\textwidth,clip=]{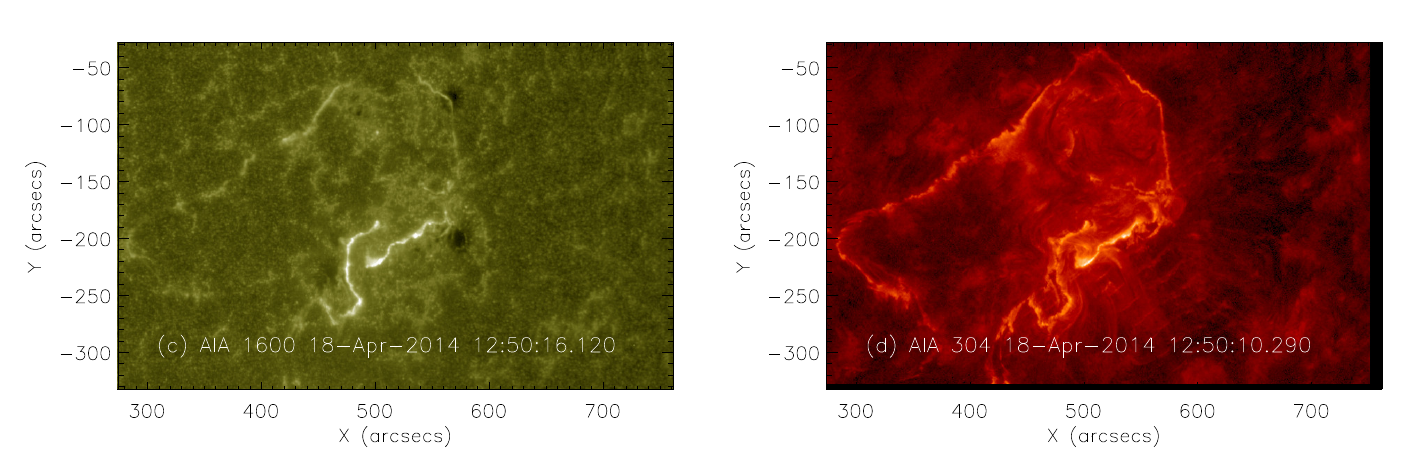}

	\caption{Overview of the SOL2014-04-18 M7.3 flare observed by GOES, AIA, and Fermi. (a) Total light curves of the flare in GOES 0.5 - 4~\AA\ and 1 - 8~\AA\, and in AIA UV 1600~\AA\ passband and 7 EUV passbands. (b) Flare light curves in photon energies 5 - 300 keV by Fermi, in comparison with the GOES 1 - 8~\AA light curve, its time derivative, and the UV light curve from AIA 1600~\AA\ passband.
    (c-d) Snapshot of the active region during the flare in AIA 1600~\AA\
    and 304~\AA\ passbands, respectively.
    } 
    \label{fig:overview}
   \end{figure}

The solar and stellar lower atmosphere is the mass reservoir and magnetic boundary of the corona. During a flare, the chromosphere reacts more rapidly and prominently to energy release by magnetic reconnection occurring in the corona. Figure~\ref{fig:overview} shows an example of flare emission in ultraviolet (UV) 1600~\AA\ and extreme ultraviolet (EUV) 304~\AA\ by the Atmosphere Imaging Assembly \citep[AIA;][]{Lemen2012}, and hard X-rays (HXRs) by the Fermi Gamma-ray Burst Monitor \citep[GMB;][]{Meegan2009}, followed by soft X-rays (SXRs) in two channels of the Geostationary Operational Environmental Satellites (GOES), and then by EUV emissions from the other six channels by AIA. It demonstrates a clear sequence of the emissions. In particular, UV 1600~\AA\ and EUV 304~\AA\ emissions often rise and peak together with HXRs, and every now and then may rise even before HXRs \citep{Cheng1990, Warren2001, Alexander2006, Qiu2021, Naus2022}. The chromosphere emission therefore can be used to infer the temporal profile of impulsive flare energy release. \citet{Kowalski2013} have defined an ``impulsiveness" index from the total light curve of the chromosphere emission in Near Ultraviolet (NUV) wavelength to characterize the rate of energy release in stellar flares. \citet{Tamburri2024} applied the ``impulsiveness" analysis to the total chromosphere emission in EUV 304~\AA\ in over 1000 solar flares observed by the Extreme ultraviolet Variability Experiment \citep[EVE;][]{Woods2011} onboard the Solar Dynamics Observatory \citep[SDO;][]{Pesnell2012}, and verified that the ``impulsiveness" index is correlated with the flare peak reconnection rate, which is inferred from imaging observations of the morphological evolution of flare ribbons in the chromosphere \citep{Fletcher2001, Qiu2002, Asai2004, Saba2006, Temmer2007, Kazachenko2017, Hinterreiter2018, Zhu2020}.

Spatially resolved solar flare observations have for long demonstrated that a flare is a collection of individual reconnection events, forming a multitude of flare loops and releasing energy in them on timescales shorter than the duration of the total flare emission \citep{Aschwanden2001, Warren2006, Qiu2012}. It is, however, less clear what are the spatio-temporal scales of these ``elementary" energy release events, the building blocks of the flare. The flaring chromosphere has been observed with ever increasing resolving capabilities, down to 1-2" by AIA \citep{Lemen2012}, below 1" by the Interface Region Imaging Spectrometer \citep[IRIS;][]{DePontieu2014}, and $\le $ 0.1" by the Goode Solar Telescope \citep[GST;][]{Cao2010} and the Daniel K. Inouye Solar Telescope \citep[DKIST;][]{Rimmele2020}. In this regard, the chromosphere observations can be used to help underpin the spatio-temporal scales of the elementary bursts \citep{Rast2021}.

Furthermore, temporally and spatially resolved photometry and spectroscopy in the chromosphere are directly related to heating rates and heating mechanisms in flare loops and provide critical observational constraints for modeling flaring atmosphere -- both the chromosphere and the corona \citep{Fletcher2024, Kowalski2024}. Decades of numerical simulation effort has been focused on studying plasma evolution in the corona and chromosphere, using one-dimensional hydrodynamic or radiative hydrodynamic models \citep{Allred2015, Reale2014} to reproduce or interpret observed spectroscopic signatures at given locations. Some recent studies also model the flare total emission in the corona as the superposition of emissions from multiple flare loops with their heating rates either prescribed, or inferred from or constrained by the chromosphere emissions at the feet (or flare ribbons) of the loops \citep{Hori1997, Warren2006, Liu2013, Kerr2020, Graham2020, Qiu2021}. Ultimately, flares are powered magnetically, and the three-dimensional MHD approach such as MURaM \citep{Rempel2017} incorporated with appropriate treatment of the chromosphere will be able to model flare energy release and subsequent plasma heating. For such global approaches, flare observations by AIA in both the chromosphere and the corona, accompanied with simultaneous magnetic field observations by the Helioseismic and Magnetic Imager \citep[HMI;][]{Scherrer2012}, have unique advantages to help constrain and test models \citep{Cheung2019, Rempel2023}. To date, AIA observations in the six EUV channels have been widely used to diagnose plasma properties in the solar corona, for example, to derive the Differential Emission Measure (DEM) of flare plasmas 
\citep{Weber2004, Hannah2012, Plowman2013, Cheung2015, Massa2023}.  On the other hand, images of the chromosphere obtained in the UV passbands and with the EUV 304~\AA\ passband have been mostly used merely as the context. Given the importance and potential of chromosphere observations in diagnosing energy release in flares, in this study, we attempt to extract the quantitative information, including the temporal evolution and relative photometry, of spatially resolved flare ribbon emission in the UV 1600~\AA\ and EUV 304~\AA\ passbands by AIA.

In terms of photometry, the AIA UV observations are obtained with broadband filters; as such, it is difficult to distinguish contributions to the gross emission in these channels by various lines and continuum emission. In the following text, we use the term ``brightness" to refer to the CCD counts rate (Data Number per second) at each pixel of the images. In non-flaring active regions, it is generally recognized that UV 1600~\AA\ and 1700~\AA\ brightness is dominated by quasi-blackbody radiation from the temperature minimum region with the brightness temperature $T_B$ between 3500 - 4500 K \citep{Brekke1994}. An experiment was conducted by \citet{Qiu2013} to find effective $T_B$ in active regions, returning results consistent with published by \citep{Brekke1994}.  During flares, situations are complicated, as many chromosphere and transition region lines are excited, and the continuum is also enhanced. In the UV 1600~\AA\ passband, the optically-thin C{\sc iv} doublet is significantly enhanced, together with other lines and continuum. 

Spectral observations of solar flares in the ultraviolet 
wavelengths have been sparse, and the very few flare spectra in the 1000 - 2000~\AA\ range were obtained decades ago.  
\citet{Simoes2019} conducted a unique study exploring the legacy high-resolution UV spectral observations by Skylab NRL SO82B
spectrograph \citep{Cook1979, Doyle1992}. They re-calibrated the spectra of an active region plage observed on 1973 September 11 and of  
an X1 flare SOL1973-09-07 during its decay phase, and then convolved the specific intensity spectra (in units 
of erg s$^{-1}$ \AA$^{-1}$ cm$^{-2}$ Sr$^{-1}$) with the AIA instrument response functions. In this way,
\citet{Simoes2019} could determine the spectral content in AIA images in UV 1600~\AA\ and 1700~\AA\ passbands. 
The EUV 304~\AA\ images are obtained with a narrowband filter; however, the spectral content
is even more complex with its response function including contributions from a few optically-thick lines formed in the chromosphere and corona \citep{Odwyer2010, Milligan2015, Antolin2024}. 
Lacking spectroscopic information, in this paper, we cannot discuss these details; instead, we compare the observed brightness (in units of counts per second) in UV 1600~\AA\ and EUV 304~\AA\ passbands across various solar features, to obtain insight into the chromosphere response during the flare, with reference to the unique study by \citet{Simoes2019}. For this purpose, we examine over a dozen solar active regions observed by AIA from 2010 through 2014. An active region in 2021 is also included to demonstrate the change of the instrument performance in a decade. In the following text, we will describe the data for 18 active regions (S\ref{sec:data}), the statistical and semi-quantitative study of the brightness and timescales of flare ribbon pixels in UV 1600~\AA\ in 18 active regions/flares (S\ref{sec:1600}), and the cross-comparison between the UV 1600~\AA\ and EUV 304~\AA\ observations in a few flares (S\ref{sec:304}). Conclusions and discussions are given in S\ref{sec:discussion}. 

\section{Overview of Observations}
\label{sec:data}
In this paper, we analyze 18 solar active regions that produce flares. These regions and their information are summarized in Table~\ref{table:summary}. They were observed by SDO from 2010 through 2014. The sample combines events from three studies by \citet{Vievering2023}, \citet{Qiu2021}, and \citet{Qiu2022}, including both eruptive and confined flares. In addition, we have also included an eruptive X1.0 flare and its host active region observed by SDO on 2021 October 28 to provide some perspectives on the AIA CCD degradation over the decade. 

\begin{table}
\begin{adjustbox}{angle=90}
\caption{List of Events$^a$ and Properties}
\label{table:summary}
\begin{tabular}{cccccccccccccc} 
\hline
  & time$^b$ and magnitude & position ($\mu$)$^c$  & \multicolumn{2}{c}{brightness$^d$} & \multicolumn{4}{c}{median rise and decay timescales (min)$^e$} & \# of pxls \\
  & & & $I_q$ & $\langle R_m\rangle $ & $\langle \tau_r^h \rangle$ & $\langle \tau_d^h\rangle$ & $\langle\tau_r^t\rangle$ & $\langle \tau_d^t\rangle$ &   \\
\hline
1 & 2010-08-01 08:20 C3.2 & N10E38 (0.77) & 57.5  & 6. & 4.5 & 7.8 & 1.5/1.8/3.8 & 2.2/2.5/7.8 & 7,398 \\
2 & 2010-08-07 18:14 M1.0 & N07E40 (0.76) & 58.5 & 17. & 1.2 & 1.8 & 1.5/1.8/2.5 & 6.2/8.8/13.8 & 16,760 \\
3 & 2011-06-21 02:55 C7.8 & N10W06 (0.98) & 72.5 & 5. & 3.2 & 4.8 & 1.2/1.2/2.2  & 1.2/1.5/2.8 & 15,249 \\
4 & 2011-08-02 06:15 M1.5 & N10W06 (0.98) & 69.5 & 8. & 2.5 & 3.8 & 1.8/1.8/3.2 & 3.5/3.8/6.2 & 12,516 \\
5 & 2011-11-09 13:20 M1.2 & N24E34 (0.75) & 52.5 & 12.  & 1.2 & 1.8 & 1.5/1.8/2.5 & 2.5/3.2/5.2 & 24,992\\
6 & 2011-11-26 07:04 C1.2 & N11W46 (0.69) & 50.5  & 6. & 3.2 & 5.2 & 0.8/1.2/2.5 & 1.2/1.8/4.2 & 2,585 \\
7 & 2012-03-09 03:41 M6.4 & N27W00 (0.89) & 45.5 & 49. & 1.2 & 1.8 & 1.8/2.2/2.8 & 4.8/5.5/7.5 & 21,376\\
8 & 2013-02-06 00:10 C8.7 & E30N23 (0.79) & 38.5 & 12. & 1.2 & 2.2 & 1.5/1.5/2.5  & 3.2/4.2/6.2 & 10,410 \\
9 & 2013-08-06 02:00 B4.4 & E21N23 (0.86)  & 37.5 & 4. & 4.2 & 5.5 & 0.8/0.8/2.2 & 0.8/1.2/2.8 & 491\\
10 & 2013-08-12 10:40 M1.5 & S23E18 (0.87)  & 38.5  & 22. & 1.2 & 1.8 & 2.5/3.2/4.2 & 3.8/4.8/6.8 & 2,541 \\
11 & 2013-08-30 02:26 C8.4 & N08E47 (0.67) & 37.5 & 10. & 1.8 & 3.5 & 1.5/1.8/2.8 & 3.2/4.8/9.2 & 10,879 \\
12 & 2014-04-18 12:52 M7.3 & S11W33 (0.83) & 43.5 & 14. & 1.5 & 2.8 & 2.2/2.8/3.5 & 6.2/7.5/12.5 & 14,420\\
13 & 2014-05-10 06:59 C8.7 & N06E31 (0.85)  & 43.5  & 17. & 1.2 & 1.5  & 1.5/1.8/2.5 & 3.8/4.8/6.8 & 7,563\\
14 & 2014-08-25 15:01 M2.0 & N09W36 (0.81)  & 39.5  & 14. & 1.8 & 3.5  & 2.5/3.2/4.2 & 9.5/10.5/13.8 & 4,762\\
15 & 2014-09-28 02:47 M5.0 & S19W22 (0.87)  & 43.5  & 18.  & 1.5 & 2.8 & 2.5/2.8/3.8 & 11.5/12.2/15.2 & 9,250 \\
16 & 2014-12-04 18:17 M6.1 & S20W30 (0.81)  & 37.5  & 24. & 1.5 & 2.2 & 2.8/3.5/4.2 & 6.8/7.8/10.5 & 13,391\\
17 & 2014-12-18 21:57 M6.9 & S14E08 (0.96) & 43.5  & 15. & 1.2 & 2.2 & 1.8/2.5/3.8 & 5.2/6.5/9.5 & 11,548 \\
18 & 2021-10-28 15:29 X1.0 & S28W02 (0.88)  & 22.5  & 21. & 0.8 & 1.8 & 1.5/1.8/2.2 & 6.2/7.8/10.8 & 10,339 \\

\hline
\end{tabular}
\end{adjustbox}
\\
$a$: Events 1-9 were studied in \citet{Vievering2023}; events 10-13, 15, and 16 in \citet{Qiu2021}, and event 14 and 17 in \citet{Qiu2022}, and references in those papers. Event 18 has been reported in numerous studies \citep[e.g.][and references therein]{Chikunova2023, Guo2023}.  \\
$b$: The time reported here refers to the peak time of the total UV brightness in the AIA 1600~\AA\ passband. \\
$c$: The position refers to the center of the cut-out AIA images analyzed in this study.\\
$d$: The quiescent brightness $I_q$ in units of DN s$^{-1}$ is the mode of histograms of the active region pixel brightness (see Figure~\ref{fig:histogram1600}). The median of the peak brightness $I'_m$ of all flaring pixels is normalized to the quiescent brightness in the active region $\langle R_m\rangle \equiv \langle I'_m\rangle /I_q$. Here $I'$ refers to the residual pixel brightness $I' \equiv I - I_p$, with $I_p$ being the average brightness of the same pixel in the first five minutes. \\
$e$: $\langle \tau_{r}^h \rangle$ and $\langle \tau_{d}^h\rangle $ indicate the median half-time in the rise and decay phase of flaring ribbon pixels. $\langle \tau_{r}^t \rangle $ and $\langle \tau_{d}^t \rangle $ are the median threshold-time of the ribbon pixel light curves measured with $I' = (5, 4, 3)I_q$ respectively; see text in S\ref{subsec:1600time}. \\

\end{table}

For each event, we have obtained cut-out AIA images observed in the UV 1600~\AA\ and EUV 304~\AA\ passbands of size up to $\sim 400$~Mm, centered around the active region, and 
for several hours, starting from about half an hour before the flare onset. These images are processed with the SDO software pipeline that performs the dark current subtraction, flat-fielding, plate-scale correction, and image registration. Afterwards, we apply the correction of differential rotation to align images to a reference time, and normalize the images with respect to the exposure time. With these level 1.5 data, we define the {\em brightness} $I(\vec{r}, t)$ of the features in the images in units of data number (DN) per second per pixel, and conduct quantitative analysis of the flare ribbon brightening in temporally and spatially resolved manner.
The time cadence of the AIA 1600~\AA\ and 304~\AA\ images is 24~s and 12~s, respectively. The pixel scale is 0.6" with the point spread function (PSF) of typical size $\sim$~1-2" \citep{Poduval2013, Hofmeister2024}. In this paper, we do not correct the PSF of the images, and the temporal and spatial resolutions are constrained by these instrument specifications.

In the following analysis, we first derive the distribution of the pixel brightness in each active region during the evolution from pre-flare to post-flare. From the brightness distribution, we identify various features, define the average non-flaring quiescent brightness $I_q$ (in units of DN s$^{-1}$) and flaring brightness with respect to $I_q$, and study the variations of the brightness of the flaring pixels.

\section{Semi-quantitative Photometry in UV 1600~\AA\ Images}
\label{sec:1600}

\subsection{Distribution of Pixel Brightness}
\label{subsec:1600brightness}

\begin{figure}    
   \includegraphics[width=0.98\textwidth,clip=]{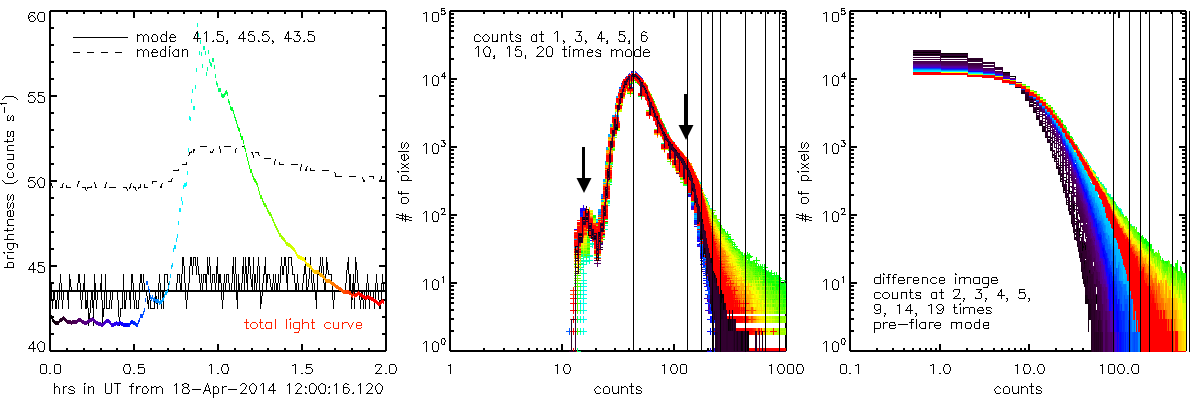}
    \includegraphics[width=0.98\textwidth,clip=]{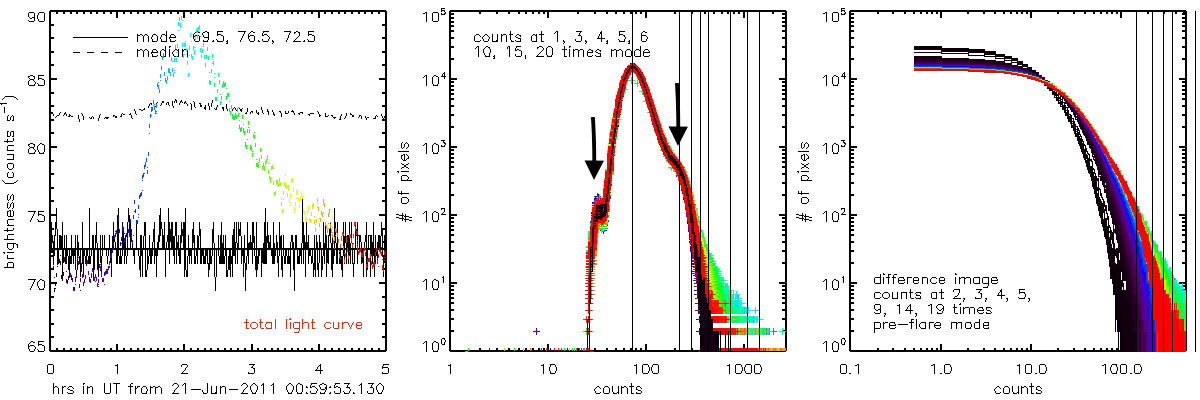}
    \includegraphics[width=0.98\textwidth,clip=]{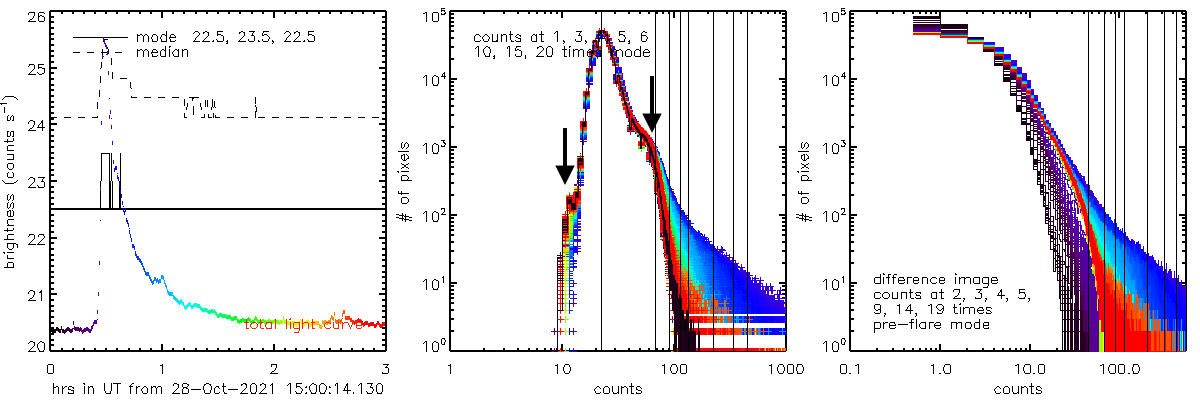}

	\caption{Statistics of UV 1600~\AA\ brightness in three active regions producing the SOL2014-04-18 M7.3 flare (top), the SOL2011-06-21 C7.8 flare (middle), and the SOL2021-10-28 X1.0 flare (bottom), respectively. Left: the time-dependent mode (solid) and median (dashed) of the pixel brightness of the active region, compared with the arbitrarily normalized total light curve (color) of the region. The color code indicates the time evolution from pre-flare (black and purple) through the flare (green to orange) and post-flare (red) phase. Middle: time-dependent histograms of the pixel brightness $I$ (DN s$^{-1}$) of the active region. Vertical solid lines denote pixel brightness at $1, 3, 4, 5, 6, 10, 15,$ and 20 times the quiescent brightness $I_q$, defined as the pre-flare mode. The two black arrows indicate the populations of sunspot and plage brightness to the left and right of the mode, respectively. Right: time-dependent histograms of the pixel brightness of the base-difference images $I' = I - I_{p}$, or the residual pixel brightness, $I_p$ being the mean pixel brightness in the first five minutes. Vertical solid lines denote residual pixel brightness at $2, 3, 4, 5, 9, 14,$ and 19 times the quiescent brightness $I_q$. The times of the histograms in the middle and right panels are indicated by the color code same as in the left panels, and the time cadence of the histograms (images) is 24~s. The same three-panel plots are generated for each of the 18 active regions, provided in the supplementary matereial. } 
    \label{fig:histogram1600}
   \end{figure}

 Figure~\ref{fig:histogram1600} shows the histogram of the pixel brightness (in units of DN s$^{-1}$) in each of the few hundred time frames of the UV 1600~\AA\ images for three active regions that produced an M-, C-, and X-class flare, respectively. In each row, the left panel shows the time evolution of the mode (solid), median (dashed), and arbitrarily normalized total (color) pixel brightness of the active region, the middle panel shows the time-dependent histogram of the pixel brightness $I$, and the right panel shows the time-dependent histogram of the residual brightness $I' \equiv I - I_p$, or pixel brightness from the difference image, produced by subtracting the pre-flare image $I_p$, which is an average over the first five minutes. Various features including the quiescent region, plage, sunspot, and flare ribbons, can be identified directly from the histograms. Identification of non-flaring features will be described in detail in the following paragraphs, and an automated algorithm
to identify flare ribbon pixels will be described in Section~\ref{subsec:1600flare}.

Seen from the middle column in Figure~\ref{fig:histogram1600}, the pre-flare brightness distribution (black to purple) ranges from a few to a few hundred DN s$^{-1}$, and peaks at a few tens DN s$^{-1}$ -- this is the mode of the histogram. We define the mode of the histogram as the quiescent brightness $I_q$. As the active region evolves into the flare (blue to orange), the mode of the histograms does not change, as long as the size of the images is sufficiently large. During the decay of the flare (orange to red), the distribution changes back toward the pre-flare distribution. The mode of the active region over time is plotted in solid black in the left panel, and the three numbers in the legend mark the minimum, maximum, and pre-flare mode. In the same active region, throughout the flare, the mode does not vary significantly\footnote{The difference between the maximum and minimum values of the mode is roughly comparable with statistical fluctuations of the data counts with a Poisson distribution, $\delta C \approx \sqrt{C}$, $C$ being the counts without exposure normalization.}. 
In comparison, the median of the active region may vary during the flare. Therefore, we use the pre-flare mode $I_q$ of the active region as a reference to normalize the pixel brightness throughout the flare.

In the pre-flare and post-flare histograms, we can also identify sunspots and plages, marked by black arrows in the middle column in Figure~\ref{fig:histogram1600}. The minor peak to the left of the mode (pixels less bright than the quiescent region) represents sunspot umbra and penumbra, part of which may become brightened during the flare. We also observe a hump at $I \approx 3I_q$, which represents plage brightening. Similarly, \citet{Qiu2010} found the plage population at $I \sim 3.5 I_q$ in UV 1600~\AA\ broadband images by the Transition Region And Coronal Explorer \citep[TRACE;][]{Handy1999}.  

It is noted that, although $I_q$ does not change over time in the same active region, it varies significantly in different active regions 
-- as seen in the figure, $I_q = 72.5, 43.5, 22.5$ DN s$^{-1}$ in the three active regions observed in 2011, 2014, and then 2021.
On the other hand, the relative brightness of the plage is rather stable at $\sim$3I$_q$, independent of the positions of active regions and the
times of observations. To understand the nature of the change in $I_q$, we use the legacy spectral observation of an active region plage, provided
in \citet{Simoes2019}, to synthesize the plage brightness observed by AIA in UV 1600~\AA\ passband. 
The results demonstrate that the variation of the quiescent brightness $I_q$ (found to be one third of the plage brightness) across different active
regions is largely due to the degradation of the CCD quantum efficiency over a decade, and partly due to the center-to-limb variation 
of the chromosphere brightness. The details of the analysis and results are given in Section~\ref{subsec:quiescent}.

During the flare, the distribution at $I \ge 4 I_q$ starts to flatten with increased number of bright pixels of up to $10^2 I_q$ at the peak of the flare (cyan to green). At the threshold $4I_q$, the ribbon-identification algorithm (see Section~\ref{subsec:1600flare}) still picks up a significant number of non-flaring pixels at the tail of the plage. Therefore, we also examine the residual brightness from the base difference image $I' \equiv I - I_{p}$, where $I_{p}$ is the mean brightness of the same pixel during the first five minutes.
The right panels show the histograms of the residual brightness. During the flare the distribution at $I' \ge 3 I_q$ starts to flatten, and better distinguishes flaring pixels from the non-flaring plage. The evolution of the pixel brightness distribution in the active region therefore guides us how to identify flaring pixels.

\subsection{Identification of Flare Ribbon Brightening}
\label{subsec:1600flare}

We identify flaring pixels whose residual brightness reaches $I' \ge N I_q$ and stays bright continuously for an extended time period $\tau_c$. The first instant the pixel brightness satisfies these criteria is taken as the onset time of the flare at the pixel. Based on the histograms, we experiment on the range of $N = 3, 4, 5$, which effectively separates flaring pixels from the plage while minimizing the loss of mildly brightened pixels. The requirement for enhanced brightness over an extended time period $\tau_c$ helps exclude pixels that are sporadically or randomly brightened for only one or two time frames, such as due to hot pixels, cosmic rays, saturation, or ejecta, which are not features of flare ribbons. In practice, we use the empirical value $\tau_c = 4$~min, or ten AIA 1600~\AA\ frames, and the reason for this choice will be discussed in the following section. 

Applying such thresholding to the SOL2014-04-18 M7.3 flare, we have identified more than 10,000 flaring pixels in the field of view of 800$\times$500 pixels. The total area of the flare ribbons occupies $\le $5\% of the total area of the active region included in the image. In general, the size of the image is chosen so that the total area of the flare ribbons makes a similarly small fraction (less than 10\%) of the image, and therefore does not skew the computation of the mode of the pixel brightness distribution during the flare.

\begin{figure}    
    \includegraphics[width=0.46\textwidth,clip=]{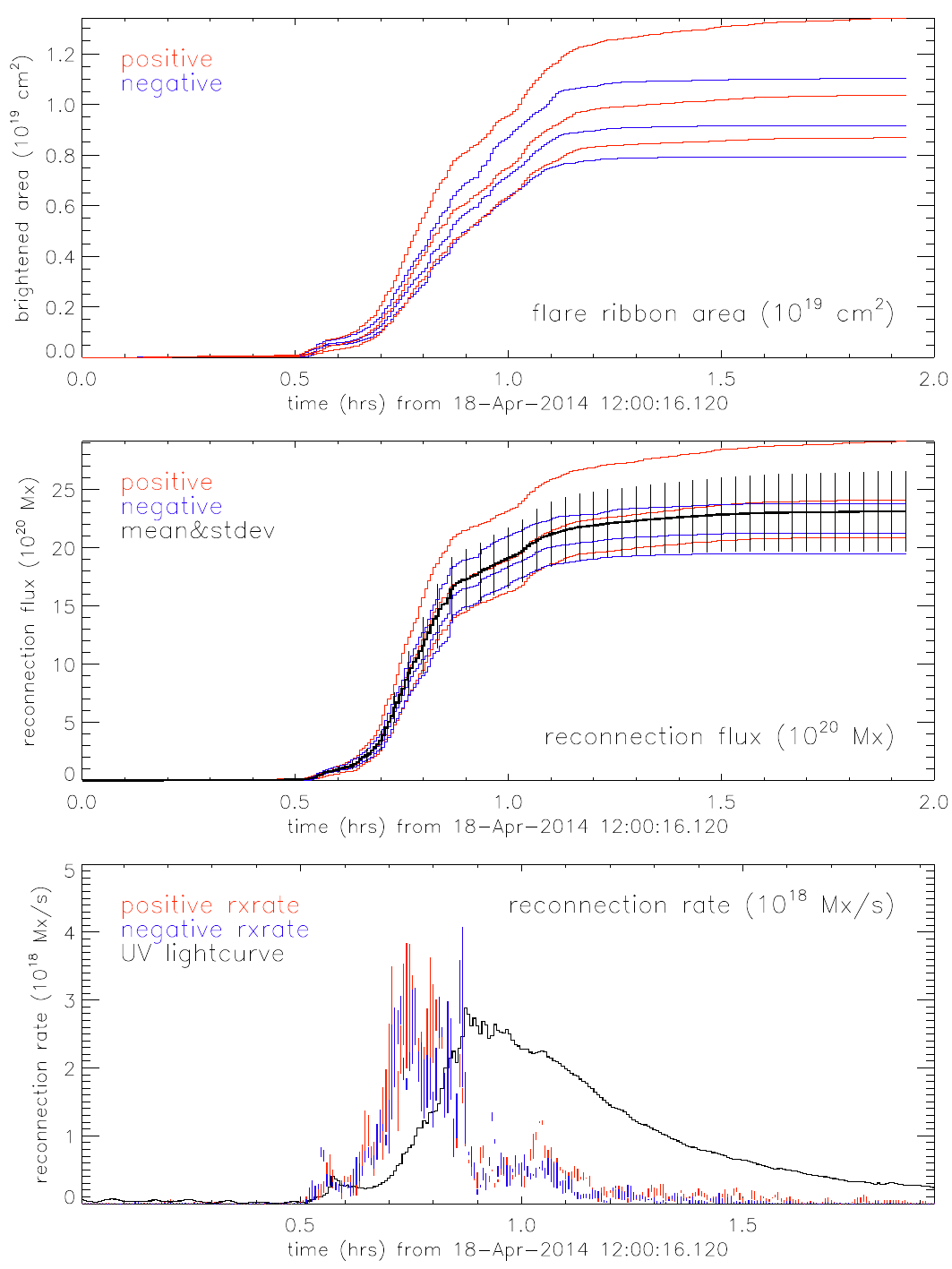}
    \includegraphics[width=0.50\textwidth,clip=]{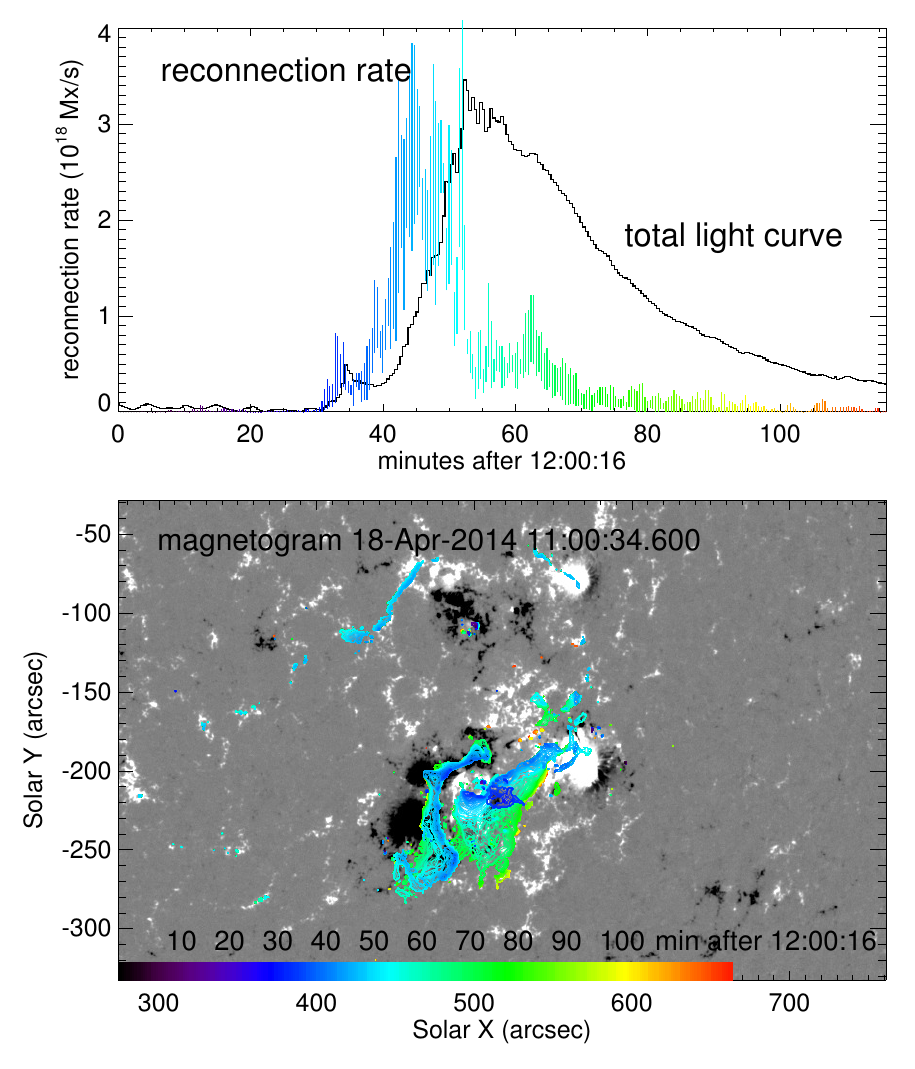}

	\caption{Left: the total area of flare ribbons (top), the cumulative magnetic flux $\psi_{\pm}$ in flaring ribbons (middle), and the reconnection flux rate $\dot{\psi}_{\pm}$ (bottom) in positive (red) and negative (blue) magnetic fields with $I' \ge 3, 4, 5 I_q$. In the bottom panel, $\dot{\psi}_{\pm}$ ranges from the minimum to the maximum of the three measurements, and the total UV light curve in comparison with $\dot{\psi}$ is arbitrarily scaled. Top right: reconnection rate $\dot{\psi}$ from all six measurements, ranging from the minimum to the maximum of the six $\dot{\psi}$ at each time. Bottom right: evolution of the flare ribbon brightening in UV 1600~\AA\ (color), identified with $I' \ge 3I_q$ for $\tau_c = 240$~s, superimposed on a pre-flare line-of-sight magnetogram by HMI. For clarity of display, the magnetogram is saturated at $\pm 300$~G. The color code in the right panels denote the onset time of the ribbon brightening.
    } 
    \label{fig:rxflx1600}
   \end{figure}

With $N = 3, 4, 5$ applied to the base difference image $I'$, the number of flaring pixels varies. The top left panel in Figure~\ref{fig:rxflx1600} shows the time evolution of the total area (in units of cm$^2$) of flaring pixels in positive (red) and negative (blue) magnetic fields for $N = 3, 4, 5$.  The middle panel shows the cumulative magnetic flux $\psi$ integrated in flaring pixels, again, in positive and negative magnetic fields separately. $\psi$ measures the total amount of magnetic flux participating in magnetic reconnection in the corona, which forms flare loops and deposits energy in them, leading to enhanced brightening at the feet (ribbons) of the loops. The bottom left panel shows the global reconnection rate $\dot{\psi} \equiv d\psi/dt$ in units of Mx s$^{-1}$ measured in positive and negative magnetic fields separately.

Figure~\ref{fig:rxflx1600} shows the range of $\psi_{\pm}$ measured with different $N$ applied to $I'$. In principle, $\psi_+ = \psi_-$ -- equal amount of the positive flux and negative flux are reconnected; in reality, due to a variety of uncertainties in the measurements \citep[see discussions in][]{Naus2022}, the two fluxes are not equal. Seen from Figure~\ref{fig:rxflx1600}, $\psi$ (and consequently $\dot{\psi}$) measured in different ways may vary by a factor of two \citep[also see][]{Zhu2020}; however, the time evolution of the measurements are usually consistent. 
We usually quote the average of these measurements as the reconnection flux or reconnection rate in a flare, and the variation will be taken as a conservative estimate of the uncertainty in $\psi$ or $\dot{\psi}$. It is noted that $\psi$ measured this way is larger than in previous studies that used a larger threshold $I \ge 6I_q$ to identify flaring pixels \citep{Qiu2010, Kazachenko2017}; a lower threshold used here picks up more pixels of weak brightenings.

We also note that, despite different $I_q$ values in different active regions observed over a decade, the relative threshold brightness for flaring pixels does not vary significantly, $I' \ge 3I_q$ (see Figure~\ref{fig:histogram1600}). Therefore, we adopt the same set of relative thresholds $I' \ge (3,4,5) I_q$ to identify flare ribbon pixels, using them to infer reconnection rates and reconnection energy release rates. Again, we note that using residual brightness $I'$ better distinguishes flaring pixels from non-flaring plage pixels.

\subsection{Time Evolution of Flare Ribbon Brightness}
\label{subsec:1600time}

Flare ribbons are brightened due to impulsive energy deposition at the feet of flare loops formed by reconnection in the corona. It has been considered that energy flux is transported to the lower atmosphere by either non-thermal particles \citep{Fisher1985a, Fisher1985b, Fisher1985c, Kowalski2017}, or thermal conduction \citep{Gan1991, Longcope2014}, or Alfven waves \citep{Fletcher2008, Kerr2016, Reep2016}. With the typical length-scale of flare loops in the order of 10$^1$~Mm, the timescale of the transport by all these mechanisms is comparable, of order $10^{0-1}$ seconds. Therefore, the chromosphere brightening observed by AIA at the cadence 10-20~s almost ``instantaneously" maps the locations of the energy release in the corona, forming impulsively brightened flare ribbons. Subsequently, the ribbon brightness starts to decay, reflecting the ``cooling" of the overlying flare loops. It is difficult to reconstruct the absolute photometry from the broadband AIA UV images, which has mixed contributions from multiple emission lines and the enhanced continuum \citep{Simoes2019}. Nevertheless, it is possible, and it is an useful practice, to infer the spatio-temporal sequence of reconnection energy release from the time evolution of pixel brightness. 
In particular, the continuous observations of a large number of flares by AIA for more than a decade provide an opportunity to systematically examine the chromosphere brightness during the flare.

\begin{figure}    
   \includegraphics[width=0.98\textwidth,clip=]{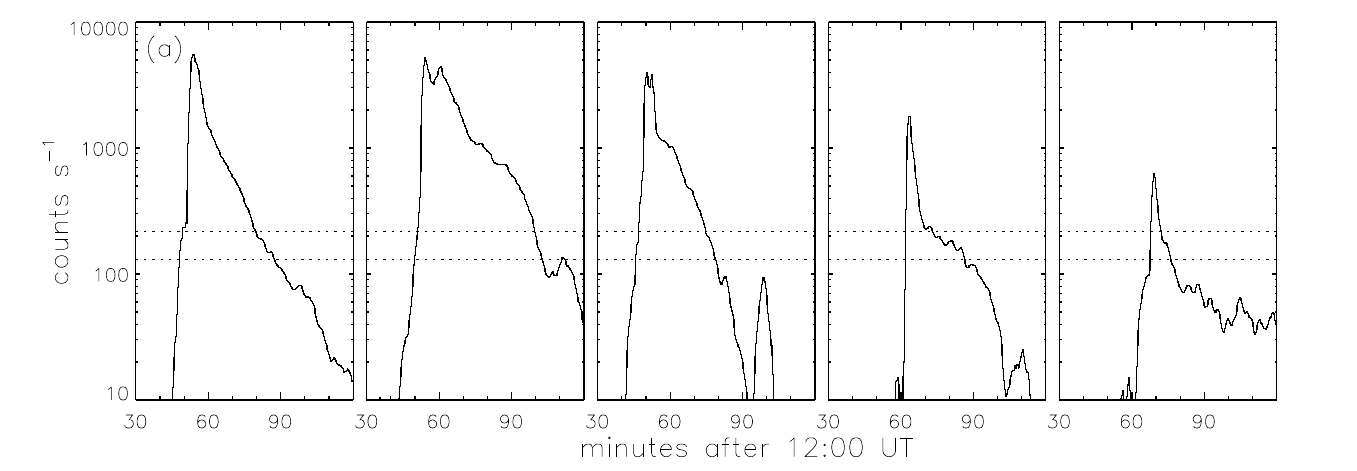}
    \includegraphics[width=0.98\textwidth,clip=]{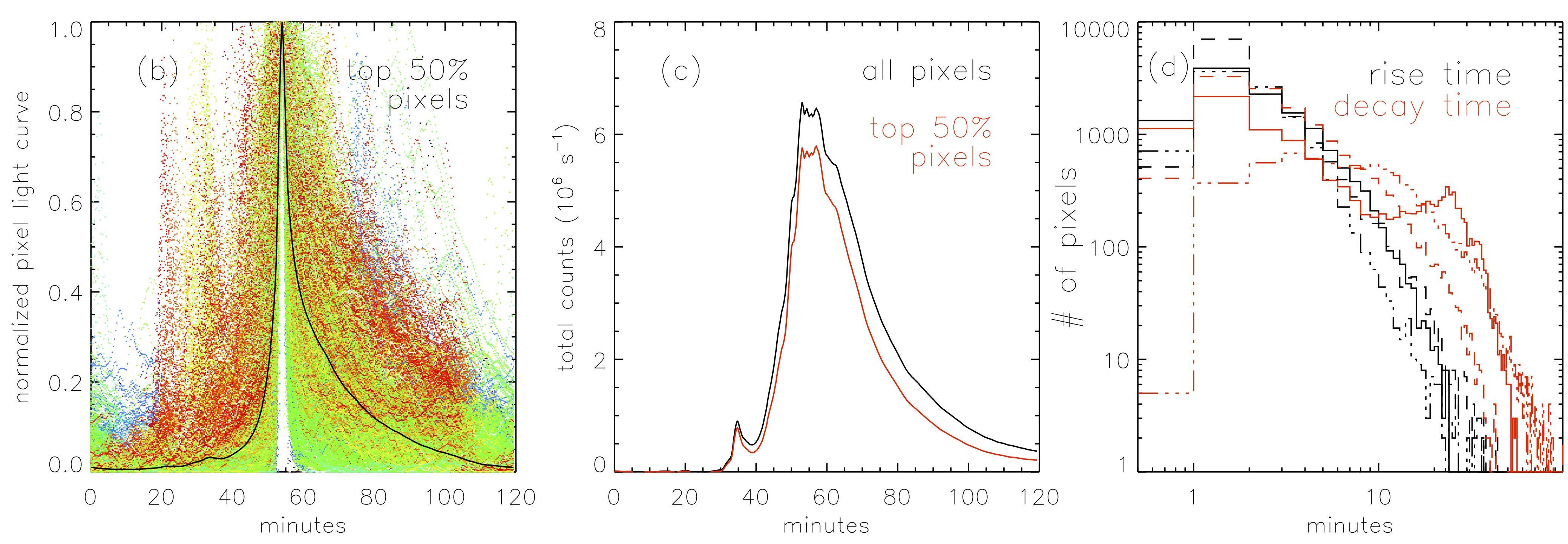}

	\caption{Statistics of pixel light curves. (a) Sample pixels light curves of the residual brightness $I'$, in logarithmic scale. 
The two horizontal dotted lines mark the residual brightness at $I' = 3I_q$ and $I' = 5 I_q$. 
(b) Epoch plots of the pixel light curves of the top 50\% brightest flaring pixels. Each light curve $I'$ is normalized to its own peak, and is shifted in the time domain so that the peak times of all light curves are aligned. Color indicates the time of the peak brightness: earlier brightened pixels are marked in cold colors, and later brightened pixels in warm colors. The solid black curve is the average of all the displayed light curves. 
(c) Total light curves of all flaring pixels (black) and of the top 50\% brightest flaring pixels (red). 
(d) Histograms of the rise time $\tau_{r}$ (black) and decay time $\tau_{d}$ (red) of all pixel light curves. The rise and decay times are estimated in three different ways. Histograms in solid lines show the rise and decay times (or threshold time) determined as the time interval between the peak brightness $I'_m$ and $5I_q$ on either side of the peak. Histograms in dashed lines show the time interval between the peak brightness $I'_m$ and one-half of $I'_m$ on either side of the peak, or the half-times. Histograms in dash-dotted lines show the Gaussian rise time and the e-slope decay time (see text). The same plots as in (b-d) for each of the 18 flares are provided in the supplementary material, where the histograms of the half rise time, half decay time, and of the threshold rise and decay times at $I' = 4I_q$ are shown.
    } 
    \label{fig:pxlcv1600}
   \end{figure}

In the following analysis, we examine the time evolution of flaring pixel brightness. We use $I' \ge 3I_q$ continuously for $\ge \tau_c = 4$~min to select flaring pixels, and derive the rise time $\tau_{r}$ and decay time $\tau_{d}$ of each of the flaring pixels. Figure~\ref{fig:pxlcv1600} shows the statistics of flaring pixels for the SOL2014-04-18 M7.3 flare. Altogether $\sim$14,000 flaring pixels are identified at this threshold throughout the flare, making less than 4\% of the total number of pixels in a cutout image of size 800 by 500 pixels. As illustrated in the bottom right panel in Figure~\ref{fig:rxflx1600} (also see Figure~\ref{fig:overview}c), these pixels outline the two ribbons along the major magnetic polarity inversion line (PIL) at the center of the active region, and also mark out parts of a rectangle-shaped remote ribbon in the diffused positive magnetic fields to the north of the active region. Note that in this study, we use the line-of-sight (LOS) magnetogram from HMI, 
rather than the magnetogram of the radial component $B_r$ from the Spaceweather HMI Active Region Patch \citep[SHARP;][]{Bobra2014}, because of the requirement for a large field of view (FOV) of the images. The maps of $B_r$ from the SHARP database are often smaller than the AIA cut-out images in this analysis, and sometimes do not cover features such as the remote ribbon in Figure~\ref{fig:overview}c.

The time evolution of each flaring pixel illustrates the history of flare energy release, and heating and cooling of flaring plasmas in discrete reconnection energy release events -- here the spatial scale of these events is restricted by the spatial resolution of the AIA imaging observation. 
Figure~\ref{fig:pxlcv1600}a shows a few randomly selected pixel light curves of the residual brightness $I'$, which typically exhibit an impulsive rise followed by a gradual decay. The two horizontal dotted lines indicates $I' = 3I_q$ and $I' = 5I_q$, the thresholds to identify flaring pixels. It is seen that during the rise, the pixel brightness rapidly evolves between these two thresholds, and the pixel brightness decays much slower. 

The global behavior of the flaring pixels can be illustrated by the epoch plot in Figure~\ref{fig:pxlcv1600}b, made by time-shifting the $I'(t)$ profiles for all flaring pixels so that their peak times are aligned. For clarity, we only display the top 50\% brightest pixels, which contributes to more than 85\% of the total flare brightness (Figure~\ref{fig:pxlcv1600}c). In the epoch plot, each of the time-shifted $I'(t)$ is also normalized to its peak brightness $I'_m$, and the color of the plot indicates the time of $I'_m$, with earlier brightened pixels in cold color, and later brightened pixels in warm color. 

The average of these epoch profiles is plotted in the solid black curve, which illustrates the average behavior of pixel light curves characterized by an impulsive rise followed by a gradual decay. In close inspection, the decay of the pixel brightness appears to exhibit two phases: the brightness initially decays rapidly on a timescale comparable to the rise time, and then switches to a much slower decay on a timescale of more than 10 minutes. The exact nature of the decay timescale is not fully understood; it may be correlated with the coronal evolution, that the chromosphere is continuously heated by thermal conduction from the corona which cools over timescales of tens of minutes \citep[see discussion in][]{Qiu2016, Reep2019}. 

To demonstrate the statistical properties, we estimate the rise and decay times of the pixel brightness. There is not a single best way to measure these timescales, and we experiment on three different measurements. First, we may directly determine the time interval between $I' = NI_q  (N = 3, 4, 5)$ and the peak brightness $I'_{m}$ on the two sides of $I'_{m}$. The histograms of the rise and decay times determined with $N = 5$  are displayed in solid lines, and the median rise and decay times are 2.5~min and 6.5~min respectively. Varying $N$ from 5 to 3 does not change significantly the rise time, but can significantly raise the decay time in a significant number of pixels. 

Next, we may define the half-time during the rise or decay, which is the duration between the peak brightness and one half of the peak on either side of the peak. The distributions of the half times are shown in dashed lines. There is little change in the rise time distribution (dashed black), and the median half time is 1.8~min. In this flare, the decay half time (dashed red) is shorter than found with the threshold method, with the median at 3.5~min. 

Finally, we also fit the relevant part of the light curve to an analytical function to estimate characteristic timescales. We fit the rise phase of $I'(t)$ to a half-gaussian $I'(t) \sim I'_m e^{-(t - t_m)^2/(2\tau_{g}^2)}, ~~ (t \le t_m)$ and the gaussian width $\tau_g$ is approximated as the characteristic rise time. Figure~\ref{fig:pxlcv1600}d shows that the distribution of $\tau_{r}$ derived this way is quite similar to the other two methods, and the median $\tau_g$ is $2.3$~minutes. For the decay time, we fit the decay profile to an exponential function $I' \sim I'_{m} e^{-(t-t_m)/\tau_{d}} ~~( t \ge t_m)$. Figure~\ref{fig:pxlcv1600}d shows that the distribution of the e-slope decay time is generally longer, and has the median at $10.3$~min. 

In summary, because of the relative simple profile in the rise phase, all different methods would return similar rise times which peak at 1-2~min, with the median at about 2~min in this flare. The decay profile is more complex, yielding different decay times depending on how they are measured, which may characterize the two phases of the decay, a rapid decay (e.g. 3.5~min half time) followed by a gradual decay (e.g., 10~min e-slope time).

The extended timescale (including both the rise and decay) of the pixel brightening at flare ribbons has guided our selection of $\tau_c$ to be in the order of a few minutes -- at present, we take $\tau_c = 4$~min, or 10 time frames of the AIA UV 1600~\AA\ images. It is likely that with this duration constraint, some genuine flaring pixels that evolve very fast, such as pixels at the feet of reconnection-opened field lines, may be missed. We consider that the small number of those pixels do not make a significant contribution to the estimate of the total reconnection flux, nor to the estimate of the total heating energy released in flare plasmas, as the majority of flare energy is within closed loops.

\subsection{Analysis of Multiple Flares}
\label{subsec:multiple}

\begin{figure}    
   \includegraphics[width=0.98\textwidth,clip=]{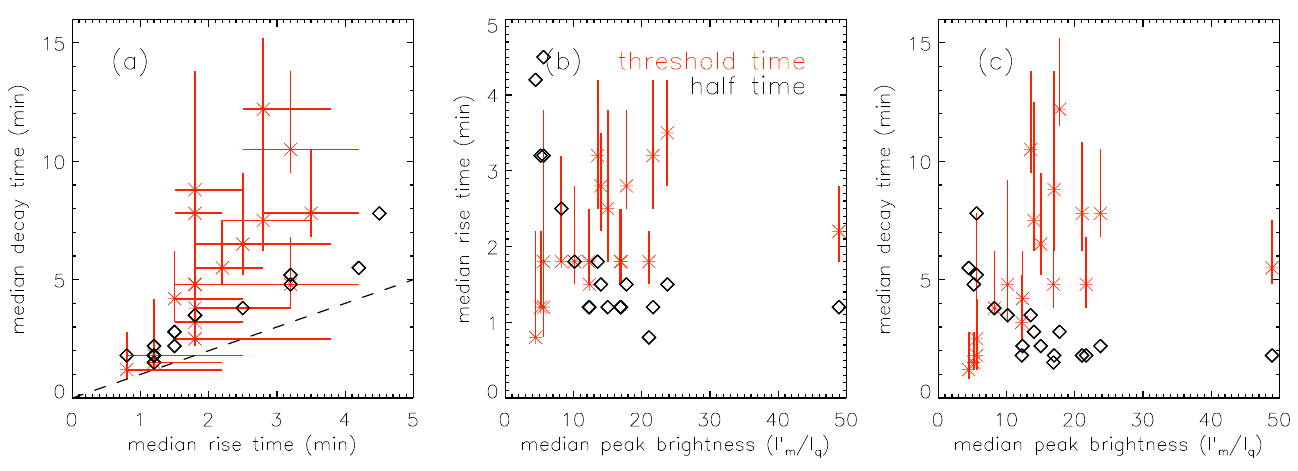}
   
	\caption{ Median rise and decay times of flaring pixel brightness. (a) The median decay time versus the median rise time for 18 flares. The dashed line is the unity line. (b) The median rise time $\langle \tau_r \rangle$ against the median (relative) peak brightness $\langle I'_m/I_q \rangle $ for the 18 flares. (c) The median decay time $\langle \tau_d \rangle$ against the median relative peak brightness $\langle I'_m/I_q \rangle $ for the 18 flares. In all three panels, black diamonds show the half-time measurements, and red asterisks show the threshold-time measurements at $4I_q$, with the horizontal and vertical bars indicating the range of the threshold measurements from $5I_q$ to $3I_q$. 
    } 
    \label{fig:timescales}
   \end{figure}

The same analysis is applied to 18 flares listed in Table~\ref{table:summary}. The magnitude of these flares ranges from B4.4 to X1.0. For each flare, the quiescent brightness $I_q$, defined as the mode of the pixel brightness distribution of the active region, ranges from 20 to 70 DN s$^{-1}$ over a decade. The histograms of these active regions are provided in the supplementary material. Despite the variation of $I_q$ value in different active regions, the plage population is evident and consistently found at $I \approx 3I_q$ in all these histograms. Therefore, flare ribbon pixels can be identified by applying the same {\em relative} threshold to the residual brightness $I' \equiv I - I_p \ge 3I_q$ for $\tau_c = 4$~min. The number of flaring pixels determined at this threshold is listed in the table. For each flare, we measure the rise and decay times for all flaring pixels. We generate the epoch plots of $I'$ and the histograms of the rise and decay timescales, like shown in Figure~\ref{fig:pxlcv1600}b-d, which are also provided in the supplementary material.

Table~\ref{table:summary} lists the median rise time $\langle \tau_r \rangle $ and median decay time $\langle \tau_d \rangle $ of all flaring pixels for each flare. Learning from Figure~\ref{fig:pxlcv1600}d, we measure the rise and decay timescales in two ways, the half-time measurement ($\tau_r^h$ and $\tau_d^h$), and the threshold measurement ($\tau_r^t$ and $\tau_d^t$). The threshold measurements use three thresholds to derive the interval between the peak time and the time the brightness is reduced to $I' = (3, 4, 5)I_q$ on either side of the peak, so three pairs of $\tau_r^t$ and $\tau_d^t$ are reported. It is noted that the rise time derived from the half-gaussian fit is generally comparable with the other two methods, whereas the decay of the pixel brightness often exhibits two phases, and does not fit very well to an exponential profile (see Figure~\ref{fig:pxlcv1600}a). Therefore, we do not conduct the function fitting any more.

Figure~\ref{fig:timescales} show the median rise and decay times for each of the 18 flares. Shown in panel (a), in all events, however they are measured, $\langle \tau_r \rangle \le \langle \tau_d \rangle$ -- pixel brightness decays slower than its rise. Comparing different flares, $\langle \tau_d \rangle $ and $\langle \tau_r \rangle$ appear to grow together -- the longer the rise, the longer the decay. In most flares, the half-time is shorter than the threshold-time. The half-time is a relative measurement with respect to the peak brightness, and tends to characterize the rapid rise or decay of a single impulsive burst at a ribbon pixel. The threshold measurement, on the other hand, tends to characterize the gradual part of the rise or decay, and is therefore longer than the half-time. However, we note that in a few flares of weak brightening, the half-time becomes longer than the threshold time. Figure~\ref{fig:timescales}b and c show the timescales with respect to the median peak pixel brightness $\langle I'_m \rangle$ normalized to $I_q$ in each flare. Clearly the timescale measurements are dependent on the peak brightness, or how sharp the brightness light curve is. The median half-time $\langle \tau^h \rangle$ appears to be nearly inversely proportional to the peak brightness, whereas the median threshold-time $\langle \tau^t \rangle$ grows with the peak brightness. In particular, $\langle \tau^h\rangle < \langle \tau^t \rangle$ for events with $\langle I'_m/I_q \rangle > 10$. Therefore, we call these flares with $\langle I'_m/I_q \rangle > 10$ {\em bright} flares. Seen from Table~\ref{table:summary}, flares with greater magnitude (high C or above) are usually {\em bright} flares. On the other hand, events \#4 and \#10 are both M1.5 flares, yet \#4 is not a bright flare by this definition. 

\begin{figure}    
   \includegraphics[width=0.32\textwidth,clip=]{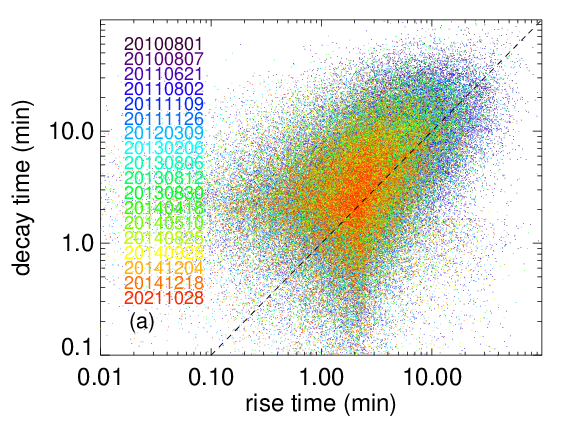}
    \includegraphics[width=0.32\textwidth,clip=]{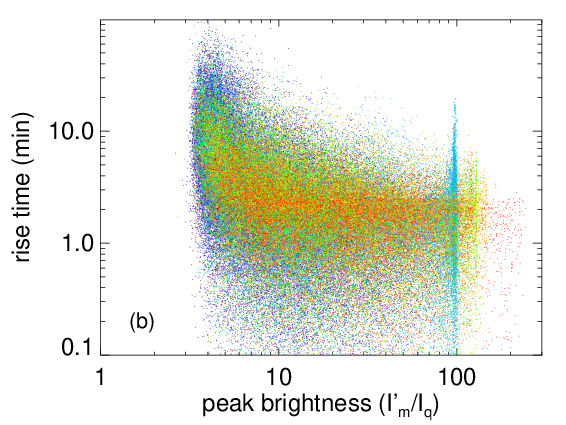}
     \includegraphics[width=0.32\textwidth,clip=]{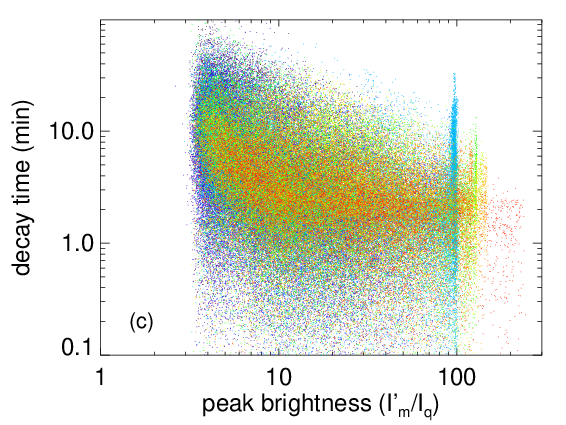}
	\caption{Same as Figure~\ref{fig:timescales}, but for each of the ribbon pixels for each flare, and only half times are presented. The dashed line in (a) denotes the $x=y$ unity line.
    } 
    \label{fig:moretimescales}
   \end{figure}


The temporal properties of individual flaring pixels for each flare can be further illustrated in Figure~\ref{fig:moretimescales}, which shows the relationships among the half rise time, half decay time, and peak residual brightness $I'_m$ normalized to $I_q$ for each active region. At individual pixel level, the $\tau_d \sim \tau_r$, $\tau_r \sim I'_m/I_q$, and $\tau_d \sim I'_m/I_q$ patterns shown in Figure~\ref{fig:moretimescales} are similar to the patterns with median values in Figure~\ref{fig:timescales}, and does not really distinguish one flare from another. Again, $\tau_d$ and $\tau_r$ tend to grow together, $\tau_r$ and $\tau_d$ decrease with $I'_m/I_q$ for weak brightening, yet for bright pixels with $I'_m/I_q \ge 10$, $\tau_r$ and $\tau_d$ distribution flattens at $\le 2$~min. 

The timing analysis is conducted on the rise and decay of the pixel profile with respect to the maximum brightness $I'_m$. We note that flare ribbon pixels do not always exhibit a single peak. Depending on the flare, about 30\% of flaring pixels may exhibit multiple peaks separated by $\ge 3$ minutes \citep[e.g.][]{Qiu2012, Zhu2025}. It is very likely that each of these AIA pixels consists of multiple energy release (and deposition) events either co-spatial (i.e. in the same flare loop), or not co-spatial yet unresolved at the AIA pixel scale (i.e., each AIA-scale flare loop consists of multiple unresolved threads). Measurements of the rise and decay timescales, especially the threshold-time measurements, can be complicated by the presence of multiple bursts or fluctuations during the rise or decay of the strongest burst. In the example of the SOL2014-04-18 M7.3 flare, the distribution of the threshold decay time exhibits an increased population (of about one third of the total number of pixels) with long decay times $> 10$~min (Figure~\ref{fig:pxlcv1600}d). By visual inspection, we recognize that a fraction of this population demonstrates another peak or more during the decay of the first (and often the strongest) peak leading to a longer decay time. 

Identification of multiple peaks, such as quasi-periodic pulsations (QPPs), in the pixel light curve is itself an interesting and important topic; however, it is beyond the scope of this study, and therefore will not be pursued in this paper. It is though worth noting that, as shown in Figure~\ref{fig:pxlcv1600}b, the mean light curve of the normalized epoch light curves is singly peaked and does not exhibit any structure in the rise and decay, suggesting that, although QPPs are present in this event \citep{Brannon2015, Brosius2016}, they are not global, or not in phase. This observation holds for other flares as well (see epoch plots and histograms for 18 flares in the supplementary material).


In summary, the survey of the 18 flares shows that the pixel brightness usually exhibits an impulsive rise, and the decay often consists of two phases with a rapid decay followed by a gradual decay. However measured, the rise time is usually shorter than the decay time. Histograms of the rise time typically peak at 1-2~min, and the median half-time of the rise is within 2~min for bright flares characterized by $\langle I'_m/I_q\rangle > 10$. 

\subsection{Discussion on Non-flaring Pixel Brightness} 
\label{subsec:quiescent}

\begin{figure}    
   \includegraphics[width=0.76\textwidth,clip=]{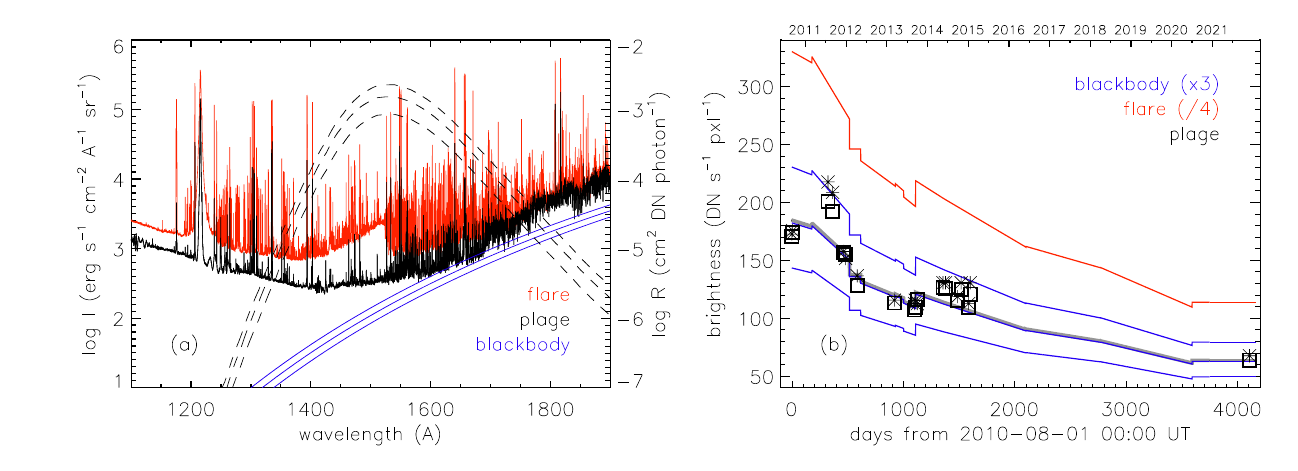}
   \hspace{-0.5cm}
   \includegraphics[width=0.24\textwidth,clip=]{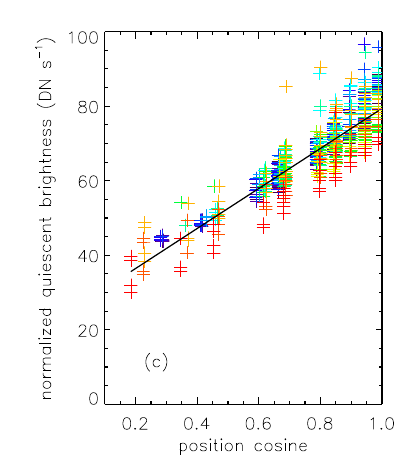}
   
	\caption{(a) The observed specific intensity spectrum of a plage (black; observed at $\mu_0 = 0.74$) and a flare (red) in the UV 1200-2000~\AA\ range, processed and provided by \citet{Simoes2019}. The black dashed curves show the response function of the AIA 1600~\AA\ passband obtained at three times, on 2011 June 21 (top), 2014 April 18 (middle), and 2021 October 28 (bottom), respectively, indicating the reduced CCD quantum efficiency over a decade. The blue curves show the blackbody radiation at the temperature $T_B = 4300, 4350, 4400$~K. (b) The synthetic AIA observed brightness of the plage (black curve), flare (red), and blackbody radiation (blue) by convolving the spectra in (a) with the time-dependent AIA response function from 2010 August through 2021 December. Superimposed in black asterisks are actually observed plage brightness, taken as 3 times the quiescent brightness $I_q$ derived from the pixel brightness histograms, for the 18 active regions analyzed in this study. The black square symbols show the observed plage brightness with the center-to-limb variation corrected with respect to $\mu_0$ (see text). (c) Quiescent brightness $I_q$ observed at various locations (denoted by the position cosine $\mu$) on the solar disk on different days between 2010 and 2021, normalized to the maximum AIA response. The center to limb variation of the quiescent brightness can be fitted to a linear function $I_q \approx (25.7\pm0.8)\mu + (53.7\pm5.2) $ (DN s$^{-1}$), indicated by the solid black line.
    } 
    \label{fig:plage}
   \end{figure}

The brightness of flare ribbons is a proxy of energy deposition. Therefore, the time profile and the photometry of the pixel brightness can be used to infer the heating rate of the flare loop, or to help constrain numerical models. The survey of the 18 flares observed over a decade has established that the plage and flare pixels can be identified with a stable threshold brightness normalized to the quiescent brightness $I_q$, although $I_q$ varies significantly from 2010 to 2021. 

The large variation of $I_q$ is primarily due to the varying AIA instrument response over a decade. The center-to-limb variation also plays a role in the observed emission of the quiescent Sun in this wavelength range \citep{Brekke1994}. To examine the effect of the AIA response over a decade, we repeat the study by \citet{Simoes2019}, who have processed and provided a specific intensity spectrum (in units of erg s$^{-1}$ \AA$^{-1}$ cm$^{-2}$ Sr$^{-1}$) in the 1200-2000~\AA\ range of a plage region observed on 1973 September 11 \citep[the Skylab NRL SO82B Observation;][]{Cook1979, Doyle1992}. 
Following \citet{Simoes2019}, we convolve the plage spectrum with the time dependent response function of the AIA 1600~\AA\ filter to obtain the synthetic plage brightness (in units of DN s$^{-1}$ per pixel) as would be observed by AIA in this passband. Figure~\ref{fig:plage}a shows the plage spectrum, and also a flare spectrum from the same database, both provided by \citet{Simoes2019}. The dashed curves show the response function of the AIA 1600~\AA\ broadband (in units of cm$^2$ DN photons$^{-1}$) at three times, on 2011-06-21 (top), 2014-04-18 (middle), and 2021-10-28 (bottom), respectively, showing the CCD degradation over time. The plage and quiescent emission in UV 1600~\AA\ spectral range is dominated by the continuum formed in the temperature minimum region \citep{Vernazza1981, Brekke1994, Simoes2019}. As a reference, we compute the blackbody radiation spectrum at effective temperatures $T_B = 4300, 4350, 4400$~K, respectively, plotted in blue curves. Figure~\ref{fig:plage}b shows the synthetic brightness of the plage (black curve), blackbody radiation (blue), and flare (red) that would be observed in the AIA 1600~\AA\ passband from 2010 through 2021. Due to the reduced CCD quantum efficiency, the synthetic AIA observed brightness of the (same) plage shown in Figure~\ref{fig:plage}a varies by a factor of three over a decade. Superimposed in black asterisk symbols are the actually observed plage brightness, taken as three times the quiescent brightness $I_q$ determined from the pixel brightness histograms (Figure~\ref{fig:histogram1600} middle columns, also listed in Table~\ref{table:summary}), for the 18 active regions analyzed in this paper. It is seen that the observed plage brightness nearly falls on the synthetic plage brightness, confirming that the variations of $I_q$ is primarily the effect of CCD degradation. It is also noted that the blackbody radiation at the temperature $T_B \approx 4350$~K matches one third of the synthetic plage brightness, suggesting that the quiescent radiation can be characterized by the blackbody continuum at an effective temperature between 4300 and 4400~K, in agreement with reported from spectral observations in the past decades \citep[e.g.][]{Brekke1994}.

\begin{figure}    
      \includegraphics[width=0.47\textwidth,clip=]{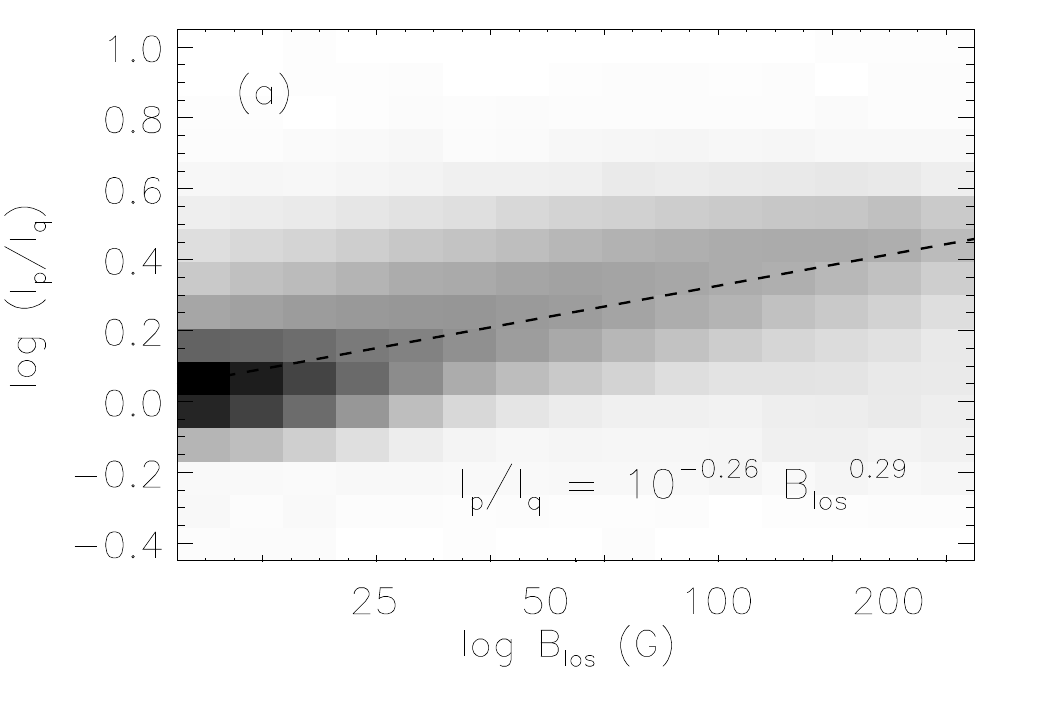}
   \includegraphics[width=0.47\textwidth,clip=]{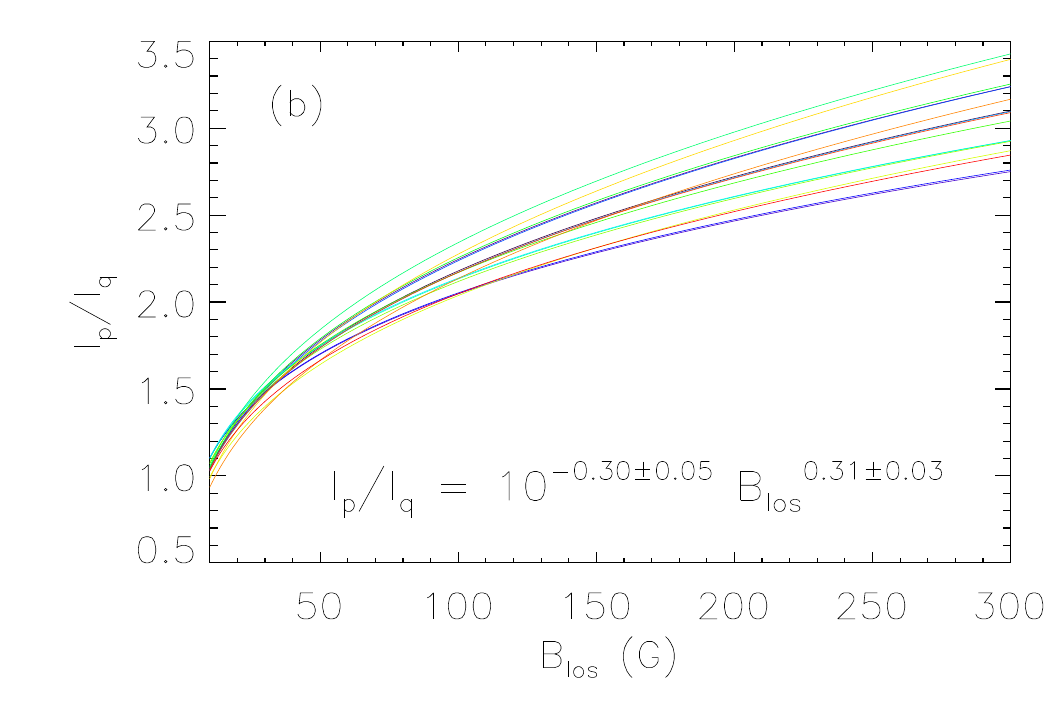}
	\caption{(a) 2D density histogram of normalized non-flaring pixel brightness $I/I_q$ and the line-of-sight magnetic field $|B_{los}|$ (in units of Gauss) for the active region hosting the SOL2014-04-18 M7.3 flare. $I/I_q$ and $|B_{los}|$ can be fitted to a power-law. (b) The power-law scaling of $I/I_q \approx 10^{\beta} |B_{los}|^{\alpha}$ for the 18 active regions analyzed in this study, with $\alpha = 0.31 \pm 0.03$ (ranging from 0.27 to 0.36), and $\beta = -0.30 \pm 0.05 $ (ranging from -0.39 to -0.23) derived from the least-square fit of the data pair in each active region.
    } 
    \label{fig:magfield}
   \end{figure}

Furthermore, the 18 active regions analyzed in this study are at different disk locations with the position cosine ($\mu$) varying between $\sim$0.67 and 0.98 (see Table~\ref{table:summary}), and the plage in panel (a) was obtained in an active region with $\mu_0 = 0.74$ \citep{Simoes2019}. The quiescent and plage emission from the lower solar atmosphere is subject to the center-to-limb variation \citep{Brekke1994}. We evaluate this effect in AIA observations. Figure~\ref{fig:plage}c shows $I_q$ values from several tens square boxes of size 100X100~Mm$^2$ centered at different locations ($\mu$) on the solar disk. The $I_q - \mu$ plots are produced on different days, marked by different colors, from 2010 through 2021, with $I_q$ normalized to the maximum AIA response over the decade. The plots show that, with the CCD response corrected, the center-to-limb variation of $I_q$ (in DN s$^{-1}$) versus $\mu$ can be described by a linear function. We apply the center-to-limb variation determined in Figure~\ref{fig:plage}c, and correct the AIA observed plage brightness in 18 active regions to $\mu_0 = 0.74$, shown by square symbols in Figure~\ref{fig:plage}b. The correction, of up to 10\%, produces a better agreement between the actual observations and the synthetic observation of active region plages, suggesting that the calibrated AIA observations of the quiescent Sun and plage in the 1600~\AA\ broadband have been stable. Therefore the AIA observed flare brightness relative to $I_q$ is a meaningful measurement of the flare chromosphere photometry.

For reference, Figure~\ref{fig:plage}b also shows the synthetic flare brightness by convolving the AIA response function with the observed spectrum of an X1.0 flare during its decay phase \citep{Cook1979}, also processed and provided by \citet{Simoes2019}. For this event, the synthetic flare brightness is about 7 times the plage brightness, or 21 times the quiescent brightness, comparable to the brightest flares listed in Table~\ref{table:summary}. 

The non-flaring chromosphere emission in the active region is known to be associated with the network magnetic field. Figure~\ref{fig:magfield} explores the empirical relation between the observed non-flaring pixel brightness normalized to $I_q$ and the line-of-sight magnetic field $B_{los}$ in the range $10 \le |B_{los}| \le 300$, which characterize the quiescent and plage features. Again, we use the line-of-sight magnetic field measurements, since they are obtained in a larger field of view than the SHARP map and have less calibration uncertainties associated with the transverse magnetic field measurements. Figure~\ref{fig:magfield}a shows an example of the active region hosting the SOL2014-04-18 M7.3 flare. The observed non-flaring brightness $I_p/I_q$ is scaled with the magnetic field roughly by a power-law. Figure~\ref{fig:magfield}b plots the power-law scaling between the brightness and line-of-sight magnetic field for all the 18 active regions analyzed in this study, and from the least-square fit, the scaling law is found to be $I_p/I_q \approx 10^{-0.30\pm 0.05} |B_{los}|^{0.31\pm0.03}$. We regard that the difference in the power-law parameters between different active regions is largely due to the limitation of using $B_{los}$. A more rigorous study should be pursued using physical quantities, like the specific intensity (and spectrum) and well calibrated vector magnetic field measurements. It is also noted that a scaling law between flaring pixel brightness and the magnetic field at the pixel level cannot be established \citep[see discussions in][]{Qiu2021}, confirming that the chromosphere heating during the flare is fundamentally different from heating of plage and quiescent regions.



   






\section{Semi-quantitative Photometry of EUV 304~\AA\ Images}
\label{sec:304}

\begin{figure}    
   \includegraphics[width=0.98\textwidth,clip=]{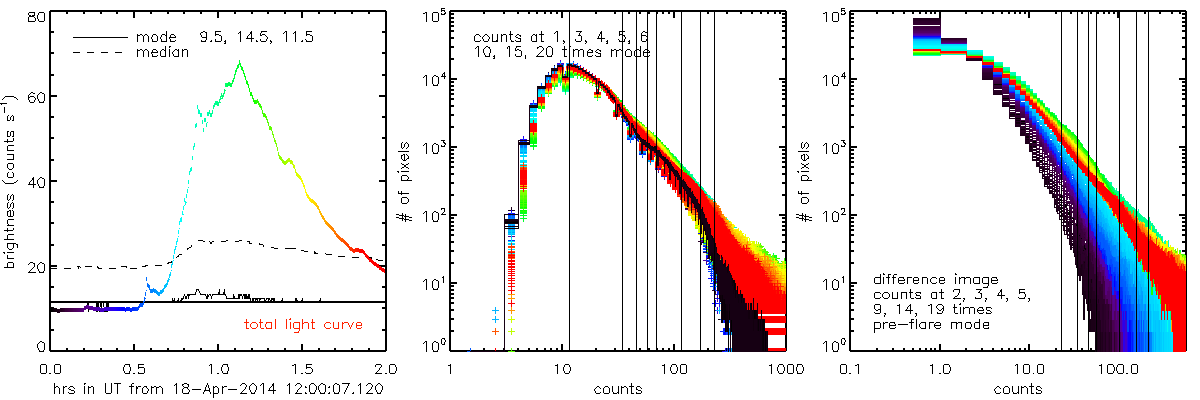}

	\caption{Statistics of pixel brightness in EUV 304~\AA\ in the active region producing the SOL2014-04-18 M7.3 flare, same as in Figure~\ref{fig:histogram1600}.} 
    \label{fig:histogram304}
   \end{figure}

The EUV 304~\AA\ broadband images by AIA also have mixed contributions from multiple lines, including the optically thick He${\sc II}$  line, and the continuum \citep{Odwyer2010}. Whereas flare ribbon signatures, or enhanced emission from the upper chromosphere, can be well detected in both bandpasses, the UV 1600~\AA\ has a large contribution from the lower chromosphere down to the temperature minimum region, and the EUV 304~\AA\ images capture emissions from the corona at the temperature around 1-2~MK \citep{Antolin2024}. Such difference is clearly visible in the snapshots of the active region in Figure~\ref{fig:overview}c and \ref{fig:overview}d. 

\subsection{Distribution of Pixel Brightness}
\label{subsec:304brightness}

\begin{figure}    
   \includegraphics[width=0.46\textwidth,clip=]{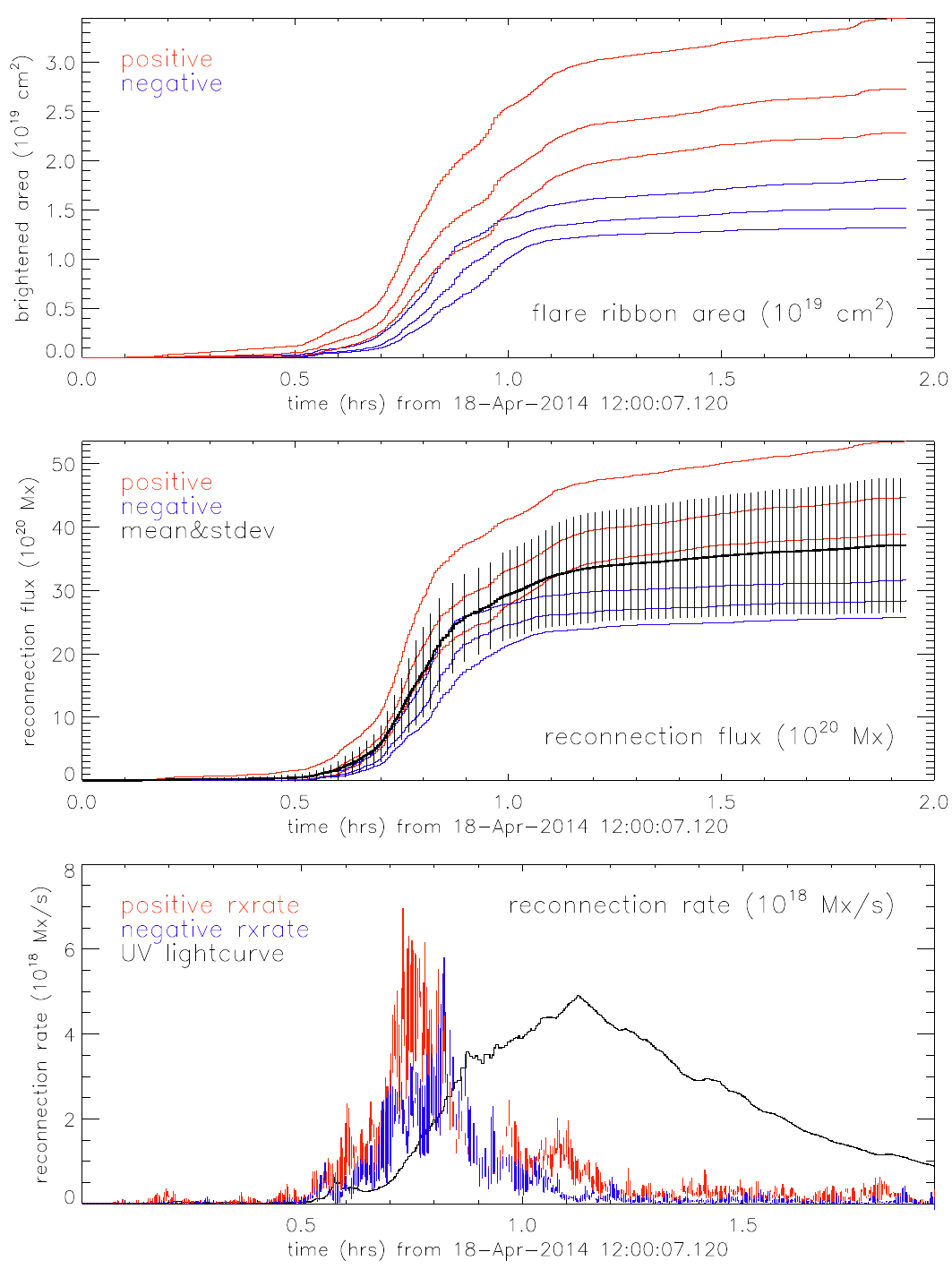}
   \includegraphics[width=0.50\textwidth,clip=]{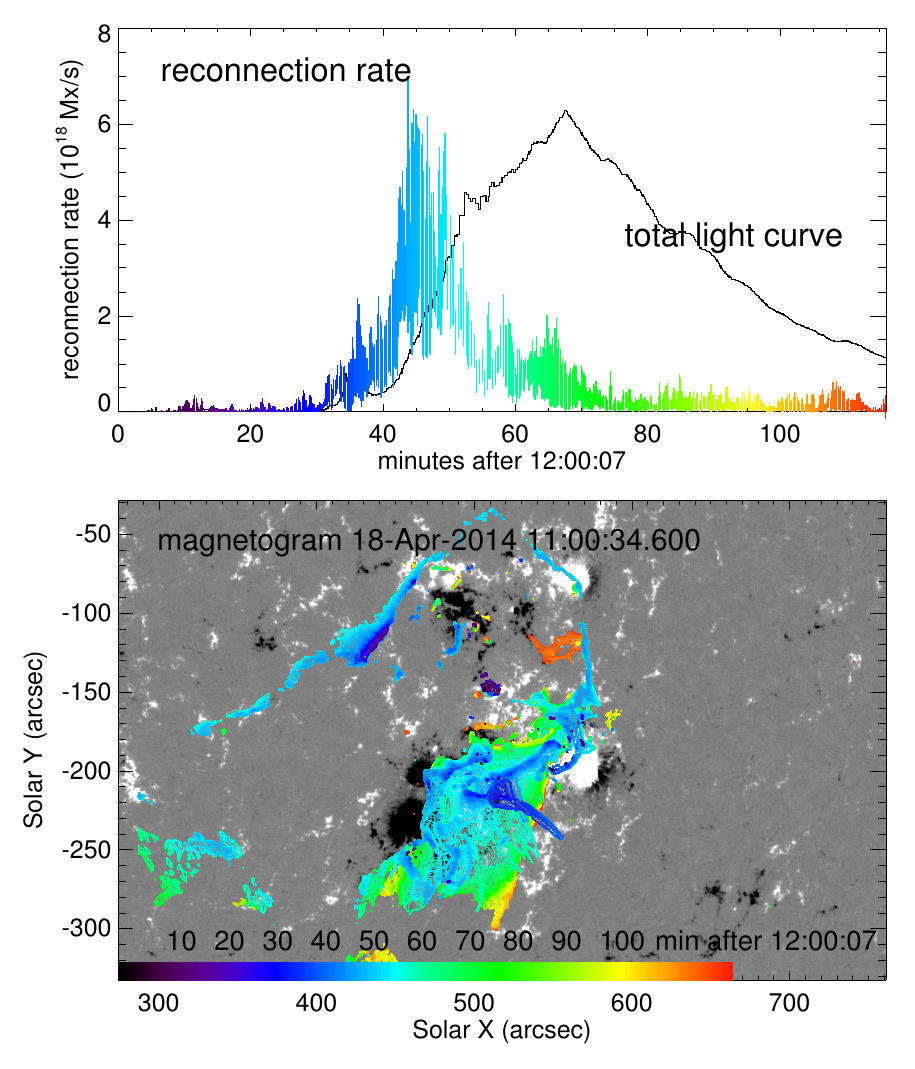}
	\caption{Same as Figure~\ref{fig:rxflx1600}, but using EUV 304~\AA\ images, with $I' \ge 9I_q,  \tau_c = 4~$min, for the SOL2014-04-18 M7.3 flare. }
    \label{fig:rxflx304}
   \end{figure}

Figure~\ref{fig:histogram304} shows the histograms of the EUV 304~\AA\ pixel brightness in the active region producing the SOL2014-04-18 M7.3 flare. The non-flare pixel brightness ranges from a few to a few hundred DN s$^{-1}$, with a relatively stable mode at 11.5 DN s$^{-1}$ for this active region. Again, we quote this number as the quiescent brightness $I_q$. Compared with $I_q$ in the UV 1600~\AA\ passband, the ``quiescent brightness" exhibits a bit more variations in the EUV 304~\AA\ passband. Furthermore, the EUV 304~\AA\ histograms do not exhibit the plage population evident in the UV 1600~\AA\ histograms. 

During the flare, the dynamic range of the image photometry is larger compared with the UV 1600~\AA\ images, with the flare brightness reaching 10$^4$ DN s$^{-1}$ -- this is partly due to the adjustable exposure time in this bandpass; during the flare, the exposure time changes from 3~s to a fraction of a second, hence very bright flaring pixels are not saturated. The distribution at $I \ge 10I_q$ starts to depart from the non-flare histograms; and again, using residual brightness better distinguishes flaring pixels from non-flaring pixels. Therefore, we use $I' \ge 9I_q$ as the lowest threshold to identify flaring pixels. Similar to the experiments with the 1600~\AA\ images, we also use a range of thresholds $ I'\ge (9, 14,19)I_q $ with $\tau_c = 4$~min to probe the sensitivity of the measurement to the relative thresholds.


\subsection{Identification of Flaring Pixels}
\label{subsec:304flare}

\begin{figure}    
   \includegraphics[width=0.98\textwidth,clip=]{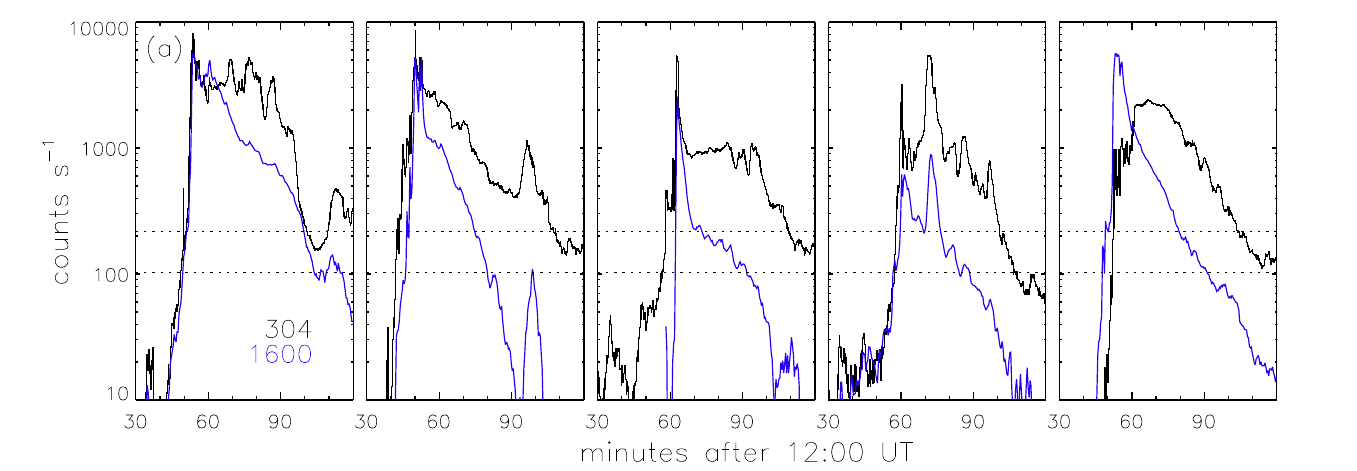}
    \includegraphics[width=0.98\textwidth,clip=]{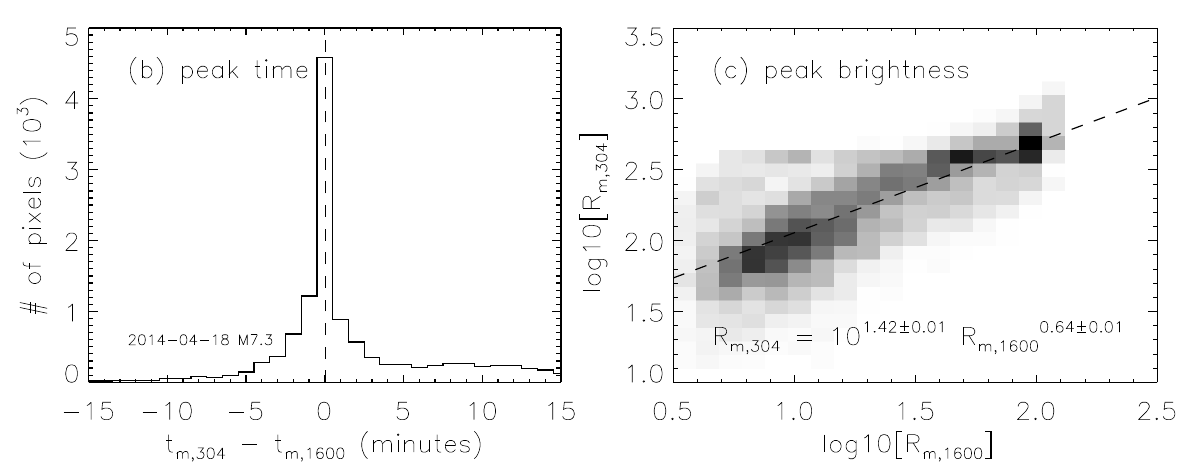}
    \includegraphics[width=0.98\textwidth,clip=]{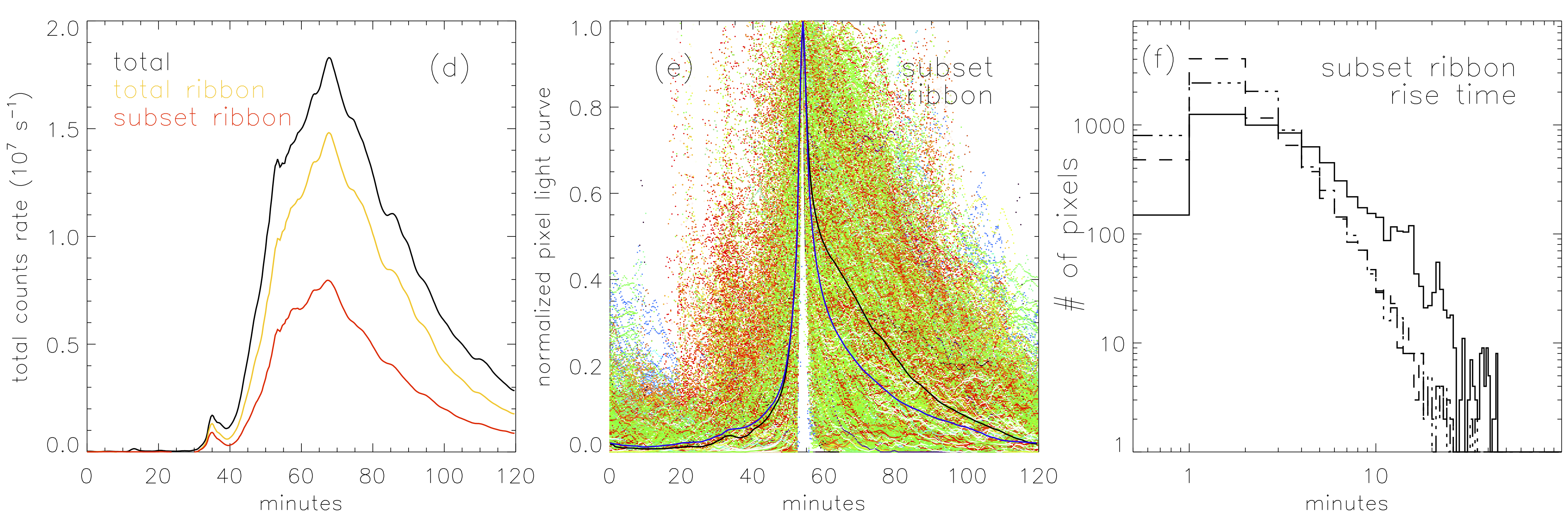}

	\caption{Statistics of pixel brightness in 304~\AA\ passband. (a) Sample pixel light curves of the residual brightness $I'$ in 304~\AA\ (black) in comparison with the brightness $I'$ in 1600~\AA\ (blue) in the same pixels. The two horizontal dotted lines mark the residual brightness at $I' = 9I_q$ and $I' = 19 I_q$ in 304~\AA, respectively.  
    (b) Histogram of the peak time difference between 304~\AA\ and 1600~\AA\ brightness, $\Delta t_m \equiv t_{m,304} - t_{m,1600}$ in flaring pixels identified in both passbands (or ribbon pixels). (c) 2D histogram of the normalized peak brightness $R_{m,304}$ and $R_{m,1600}$ in the two passbands from a subset of ribbon pixels whose $|\Delta t_m| \le 2$~min. The dashed line shows the fit of the brightness pair in the two passbands to a power-law function.
    (d) Total light curves of all flaring pixels found in 304~\AA\ images (black), of flaring pixels (or ribbon pixels) identified in both passbands (orange), and of the subset of ribbon pixels that satisfy $|\Delta t_{m}| \le 2$~min (red). (e) Epoch plots of the pixel light curves of the subset of the ribbon pixels. Each light curve $I'$ is normalized to its own peak, and is shifted in the time domain so that the peak times of all light curves are aligned. Color indicates the time of the peak brightness of the original unshifted light curve: earlier brightened pixels are marked in cold colors, and later brightened pixels in warm colors. The solid black curve is the average of all the displayed epoch light curves in 304~\AA, and the solid blue curve is the average of the epoch light curves of the same pixels in 1600~\AA. (f) Histograms of the rise time $\tau_{r}$ of the 304~\AA\ light curves of the subset of the ribbon pixels. The rise times are estimated in two different ways. Histograms in solid lines show the rise times determined as the time interval during which the pixel brightness rises from $19I_q$ to the peak. Histograms in dashed lines show the half rise time, same as in Figure~\ref{fig:histogram1600}d.
    } 
    \label{fig:pxlcv304}
   \end{figure}

Figure~\ref{fig:rxflx304} shows the flaring pixels identified for the SOL2014-04-18 M7.3 flare from the 304~\AA\ images. The left panels show the cumulative area of the flare, the cumulative magnetic flux $\psi$, and the global reconnection rate $\dot{\psi}$ measured in positive and negative magnetic fields separately. The identified flaring pixels, using threshold $I' \ge 9I_q$ and $\tau_c = 4$~min, are also mapped on a line-of-sight magnetogram, shown in the bottom right panel. Compared with the measurements using the UV 1600~\AA\ images (Figure~\ref{fig:rxflx1600}), many more flaring pixels are identified in the EUV 304~\AA\ images. As a result, the reconnection flux and reconnection rate are greater than measured using the UV 1600~\AA\ images. 

The increased number of flaring pixels identified in the 304~\AA\ images are from two populations. On the one hand, the EUV 304~\AA\ passband is more sensitive to emission from the upper chromosphere and the low corona, and has much less background emission from the lower atmosphere; therefore the residual brightness of flaring pixels is more pronounced, and weaker {\em ribbon} emission signatures can be picked up in the images. As a prominent example, the entire rectangle-shaped remote ribbon is identified in this passband, whereas only parts of the remote ribbon are identified in the 1600~\AA\ images. On the other hand, flaring pixels in flare {\em loops} connecting the two ribbons are also identified in the EUV 304~\AA\ images, and it is hard to separate the flare loop pixels from flare ribbon pixels with the 304~\AA\ images alone. These effects explain the greater reconnection flux measured with the 304~\AA\ images, which should be considered as an over-estimate of $\psi$. Figure~\ref{fig:rxflx304} further shows that inclusion of the loop pixels also generates a larger imbalance between the positive (red) and negative (blue) fluxes in comparison with Figure~\ref{fig:rxflx1600}.

The total light curve in the 304~\AA\ passband evolves along with the UV 1600~\AA\ light curve during the rise of the flare till the peak of the UV 1600~\AA\ light curve, and then continues to grow and peak after the soft X-ray peak (Figure~\ref{fig:overview}a). The additional (and later) emission is produced in flare loops that have cooled down to 1-2~MK. 

\subsection{Comparison of Ribbon Brightness in 1600~\AA\ and 304~\AA}
\label{subsec:1600304}

Flaring pixels identified in the 304~\AA\ images can come from flaring {\em ribbons} or {\em loops}. On the other hand, nearly all ribbon pixels identified in 1600~\AA\ images are also identified in the 304~\AA\ images. To understand the chromosphere response to flare energy release in these two passbands, we compare the brightness of the pixels identified in both passbands, or {\em ribbon} pixels. Figure~\ref{fig:pxlcv304}a shows the 304~\AA\ residual brightness $I'_{304}$ in a few ribbon pixels also identified in the 1600~\AA\ images. It is seen that, given the cadence and image scale, the ribbon brightness often rises and peaks simultaneously in the two passbands, but the decay of the 304~\AA\ brightness is slower, and exhibits multiple peaks due to emission from overlying coronal loops that brighten subsequently. Sometimes the overlying loops become brighter than the underlying ribbon, as shown in the rightmost panel of Figure~\ref{fig:pxlcv304}a. 

For a meaningful comparison of the ribbon properties in the two passbands, we then down-select a subset of pixels whose brightness in 1600~\AA\ passband and in 304~\AA\ passband reach maxima within $\pm 2$~min.
The peak times of the 1600~\AA\ and 304~\AA\ brightness of flaring pixels identified in both passbands are compared in Figure~\ref{fig:pxlcv304}b. The histogram of the time difference $\Delta t_m = t_{m,304} - t_{m, 1600}$ peaks at $0\pm0.5$ min, and in about one half of the pixels, $|\Delta t| \le 2$~min. We consider that in this subset of pixels, peak brightness in 304~\AA\ passband is from the ribbons, rather than from loops.  The normalized peak brightnesses $R_{m, 304} \equiv I'_{m,304}/I_{q,304}$
 and $R_{m,1600} \equiv I'_{m,1600}/I_{q, 1600}$ in the two passbands of this subset of ribbon pixels are compared in Figure~\ref{fig:pxlcv304}c, suggesting that $R_{m, 304}$ and $R_{m,1600}$ can be scaled by a power-law 
$R_{m,304} = 10^{1.42 \pm 0.01} R_{m,1600}^{0.64 \pm 0.01}$ .

Next, we examine the timescales of the 304~\AA\ light curves in this subset of ribbon pixels. Figure~\ref{fig:pxlcv304}d shows the total light curve of this subset of ribbon pixels (red), in comparison with the total light curve from all flaring pixels identified in 304~\AA\ images (black), and the total light curve of all ribbon pixels identified in both passbands (orange). Figure~\ref{fig:pxlcv304}e presents the epoch plots of the 304~\AA\ brightness of this subset of ribbon pixels, which exhibit a similar rise and generally longer decay compared with the 1600~\AA\ light curves, from the same pixels, due to the complicated contributions to the 304~\AA\ passband from overlying flare loops particularly during the decay phase.
Therefore, we only analyze the rise of the $I'_{304}$ in this subset, shown in Figure~\ref{fig:pxlcv304}f. Again, the rise time is measured in three different ways, the half-time, the threshold-time, and the gaussian time. The histograms of the half-time (dashed line) and of the gaussian time (dash-dotted line) are comparable, with the median at 1.5~min and 2.3~min respectively. The histogram of the rise time, derived as the interval from $I' = 19I_q$ to the peak brightness, shows a larger number of pixels with a longer rise time, due to the more structurous time profile of the 304~\AA\ brightness. Still, in all histograms, the peak population of the rise time is at 1-2~min.  Given the cadences of the two passbands (24~s in 1600~\AA\ and 12~s in 304~\AA), we consider that the rise times, however measured, are comparable in the two passbands, and timescales below 1~min cannot be properly determined. 


\begin{figure}    
   \includegraphics[width=0.98\textwidth,clip=]{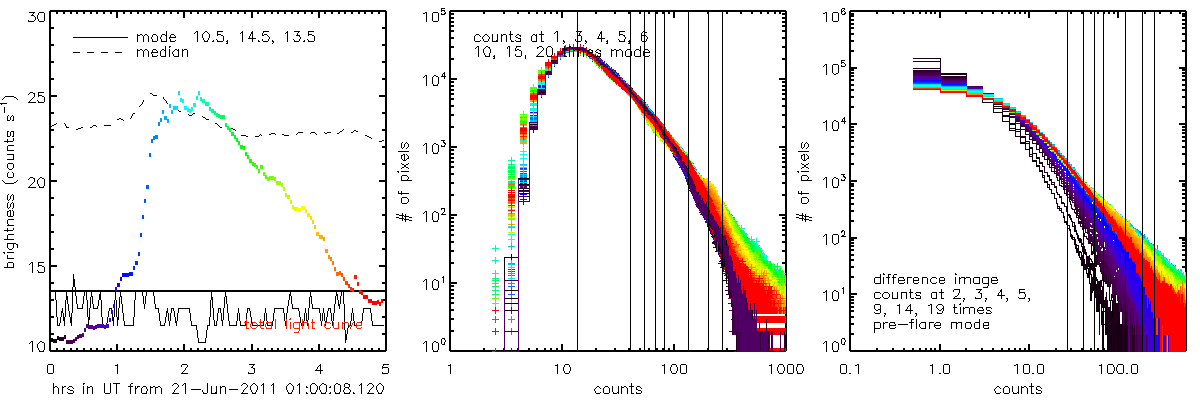}
   \includegraphics[width=0.98\textwidth,clip=]{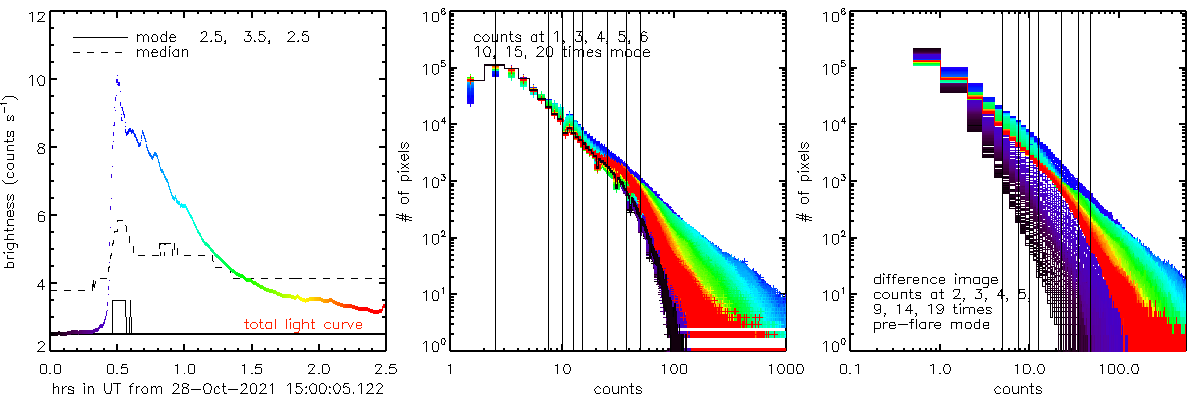}
	\caption{Statistics of EUV 304 photometry in the active regions producing the SOL2011-06-21 C7.8 flare (top) and the SOL2021-10-28 X1.0 flare (bottom), respectively, same as in Figure~\ref{fig:histogram304}.} 
    \label{fig:histogram304more}
   \end{figure}

\begin{figure}    
    \includegraphics[width=0.98\textwidth,clip=]{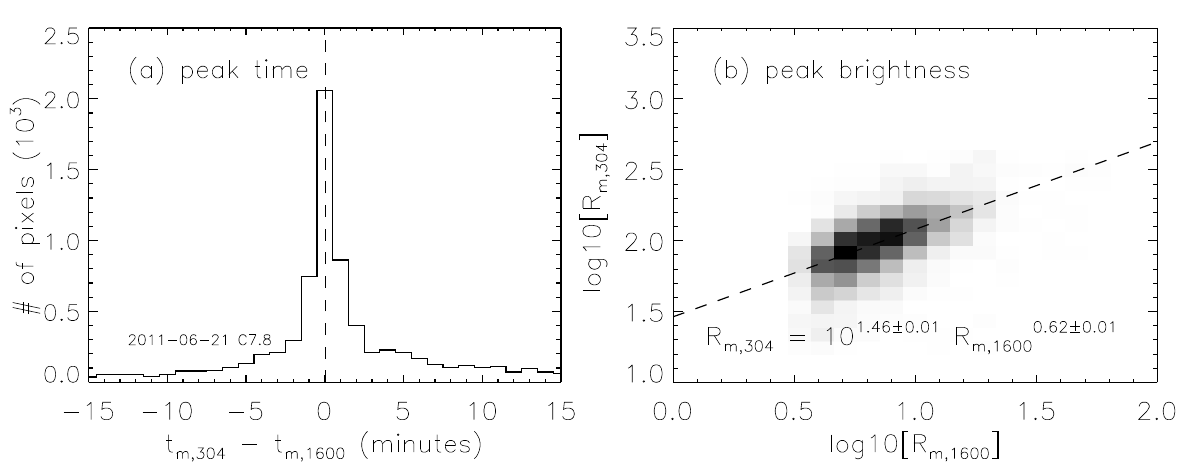}
 \includegraphics[width=0.98\textwidth,clip=]{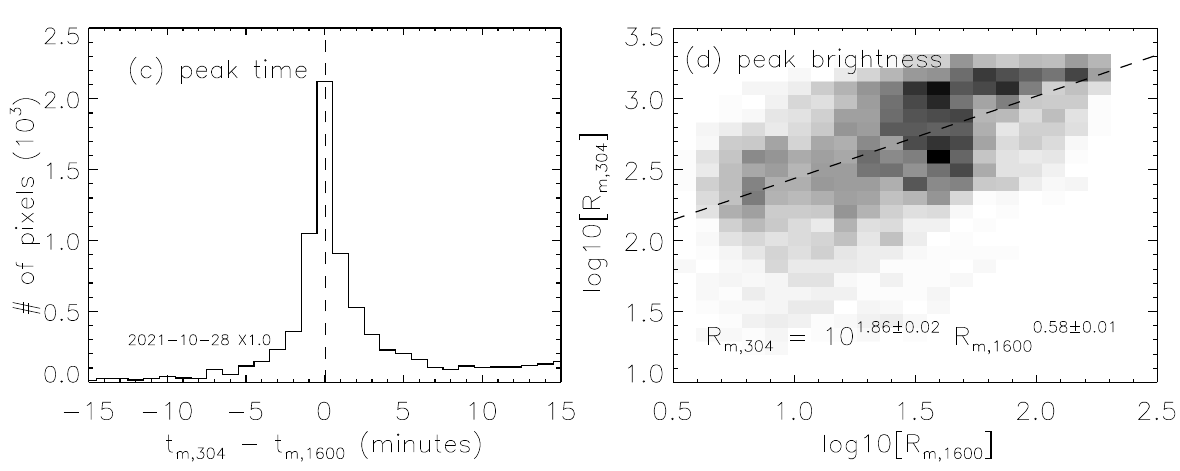}
	\caption{The distribution of the peak time difference $\Delta t_m \equiv t_{m,304} - t_{m,1600} $ of ribbon pixels identified in both passbands (left), and the 2D histogram of the normalized peak brightness $R_{m,304}$ and $R_{m,1600}$ of a subset of pixels with $|\Delta t_m| \le 2$~min (right), for the SOL2011-06-21 C7.8 flare (top), and the SOL2021-10-28 X1.0 flare (bottom). The dashed lines in the right panels show the fit of the pair brightness to a power-law relation.
    } 
    \label{fig:1600_304}
   \end{figure}

To understand whether the pixel brightness comparison is consistent for different flares, we repeat the analysis for the other two flares, the SOL2011-06-21 C7.8 flare and the SOL2021-10-28 X1.0 flare. Figure~\ref{fig:histogram304more} shows the pixel brightness distribution for these two flares. Comparing the three events, the quiescent brightness, $I_q = 13.5, 11.5, 2.5$ DN s$^{-1}$ for the three flares occurred in 2011, 2014, and 2021, varies significantly from flare to flare over a decade. 
In this passband, the center-to-limb variation of the brightness is not evident; therefore, the change of $I_q$ reflects CCD degradation over a decade.  Regardless of the large change of $I_q$, during the flare, the brightness distribution flattens at the similar relative brightness $I \ge 10 I_q$ for all three flares of different magnitudes and occurred over a decade. And again, using residual brightness better distinguishes flaring pixels from non-flaring pixels. Therefore, we use the same relative thresholding $I' \ge 9I_q$ and $\tau_c = 4$~min to identify flaring pixels in 304~\AA. 

Again, we repeat the comparison of pixel brightness in the two passbands, $I'_{304}$ and $I'_{1600}$, for these two flares. Figure~\ref{fig:1600_304} shows the 
comparison of the peak time (left) and peak brightness 
(right) between two passbands, for the two flares respectively. The results are similar to the SOL2014-04-18 M7.3 flare. Again, for flaring pixels (or ribbon pixels) identified in both passbands, $\Delta t_{m}$ peaks at $0\pm0.5$min; and in a subset of about half of these ribbon pixels, $|\Delta t_m| \le 2$~min. For these two flares, the scaling law of the normalized ribbon peak brightness for the subset is found to be $R_{m,304} = 10^{1.46\pm0.01} R_{m,1600}^{0.62\pm0.01}, $ and $R_{m,304} = 10^{1.86\pm0.02} R_{m,1600}^{0.58\pm0.01}$, respectively.

In summary, the comparison between the EUV 304~\AA\ and UV 1600~\AA\ brightness suggests that, despite the significant variations in the absolute brightness (in units of DN s$^{-1}$) from flare to flare over a decade, the relative thresholding brightness from the flare ribbons is much less variant. Flaring {\em ribbon} brightness in these two passbands rise and peak on similar timescales, and their peak brightness is scaled by power laws with very similar power index. The EUV 304~\AA\ brightness decays more slowly, however, largely due to emission from overlying coronal loops.

\section{Discussions and Conclusions}
\label{sec:discussion}

Flare ribbon brightening is the proxy of energy release by magnetic reconnection in the corona. A flare is a collection of multiple reconnection energy release events, i.e., elementary bursts, that may be distinguished in spatially resolved chromosphere observations. In the past decade, AIA/SDO has observed tens of thousands of flares, and have provided a unique database for large-sample statistical studies of solar flares. In this paper, we re-examine the practice of using flare ribbon signatures observed in the AIA UV 1600~\AA\ images to infer global magnetic reconnection rates \citep[e.g.][]{Kazachenko2017}, and further investigate the tempo-spatial properties of flare ribbon brightness to explore the potential for more quantitative studies of flare energetics using AIA imaging observations of the flare chromosphere, such as with the UV Foot-point Calorimetry (UFC) method \citep[][and references therein]{Qiu2021}.

\subsection{Summary of Findings}
\label{subsec:summary}

We have conducted a statistical study of the pixel brightness (in units of DN s$^{-1}$) 
of the UV 1600~\AA\ broadband images for each of 18 active regions observed by AIA from 2010 
through 2021. To understand the nature of the observed brightness variation,
we repeat the experiment by \citet{Simoes2019} that synthesizes the AIA observed brightness using the 
high-resolution UV spectra of an active region from the legacy Skylab database. We find that
the synthetic non-flaring brightness agrees with AIA observations of the active regions studied in this paper.
Therefore, after correcting the time-dependent instrument response and the center-to-limb variation, the brightness of the quiescent 
region is rather stable in AIA observations for over a decade. 

The active regions studied in this paper have produced flares with their magnitude ranging from B4 to X1 class. Our study 
finds that the observed plage brightness is established at about three times the quiescent brightness $I_q$, 
so that flaring pixels can be identified with a stable relative threshold $I' \ge 3I_q$, where $I'$ is the residual brightness with 
the pre-flare brightness subtracted. These results confirm that the AIA observed flare brightness relative to $I_q$ is 
a meaningful measurement of the flare chromosphere photometry, and AIA observations for over a decade thus provide an 
extensive database for systematic and semi-quantitative study of UV emission from the flaring chromosphere, either in 
the context of the Sun as a star or in spatially resolved manner down to the AIA scale.

We have then identified thousands of flaring ribbon pixels in the UV 1600~\AA\ images for each flare,  measured their peak brightness $I'_m$, and the rise and decay timescales in several different ways characterizing the rapid and gradual parts of the pixel light curves respectively. However measured, the decay time is longer than the rise time, and flaring pixels with a longer rise time usually also have a longer decay time. Our analysis indicates that flares with the peak pixel brightness $\langle I'_m \rangle$ larger than ten times $I_q$ are {\em bright} flares that exhibit sharp pixel light curves with an impulsive rise, and the median rise time $\langle \tau_r \rangle$ is within 2~min. The decay of pixel light curves of {\em bright} flares ($I'_m \ge 10I_q$) often exhibit two phases with a rapid decay on timescales comparable with the rise time followed by a more gradual decay on longer timescales. These properties of the pixel brightness light curves give insight into the tempo-spatial structure of flare energy release and help constrain numerical models of flare heating.

We have also examined the pixel brightness of the EUV 304~\AA\ images in three active regions, in comparison with the UV 1600~\AA\ observations. Flare ribbon pixels identified in both UV and EUV images exhibit similar time profiles in the two passbands during the rise, and the peak brightness in the EUV 304~\AA\ and UV 1600~\AA\ appear to be related following a very similar scaling law, despite the large differences in the magnitude of the three flares, which were observed in 2011, 2014, and 2021, respectively. The result from the cross comparison of flare ribbon brightness in two spectral ranges, though limited in scope, provides further observational constraint on modeling flare heating of the lower solar atmosphere, the boundary of the solar corona.

\subsection{Discussions}
\label{subsec:discussion}

This study exploits the large database of the AIA imaging observations, and examines the temporal properties of spatially resolved (down to the AIA pixel scale) UV emission at flare ribbons. A flare is considered to be comprised of many discrete energy release events, or ``elementary bursts", perhaps through the so-called ``patchy" reconnection, or reflecting magnetic islands associated with the tearing mode in the reconnection current sheet \citep{Furth1963, Shibata2001, Wyper2021, Huang2016, Dahlin2025}. It has not been clear what are the tempo-spatial scales of these discrete energy release events. In this study, we have estimated the characteristic timescales of ribbon pixel (of 0.6") light curves to be of order 1-2~min. We note that, limited by the spatial resolution and the cadence of the AIA observations (12-24~s), the timescale below 30~s cannot be properly revealed in this analysis. For example, \citet{Graham2015, Graham2020} have analyzed an X1.0 flare observed by IRIS with a time cadence 5~s and pixel scale 0.3". 
They suggested that IRIS's capability may be close to resolving the spatial scale of ``elementary bursts" \citep[also see][]{Brannon2015}, with the rise time of the pixel light curve $\tau_{r} \approx 10-20$~s, about an order of magnitude smaller than $\tau_{r}$ found in this study. 
A recent study by \citet{Faber2025} using diffraction-limited imaging observations by Swedish 1-m Solar Telescope \citep[SST;][]{Scharmer2003} of pixel scale 0.04" and time cadence 7~s has found bright ``blobs" of radius 140-200~km (0.2 - 0.3") resolved in the $H_{\beta}$ red-wing ($+0.8$\AA) images, with the brightening in the $H_{\beta}$ red-wing lasting for $\le$ 15~s. The spatial scale of these $H_{\beta}$ blobs is similar to the cross-section of post-reconnection flare loops in H$\alpha$ images obtained by GST \citep{Jing2016}. These observations with very high resolution, however, do not have a high cadence to refine the temporal scales of flare energy release at those fine spatial scales.
The advent of DKIST imaging and spectral observations with still higher cadence (around 1~s) and spatial resolution (below 0.1"), will help further explore the fundamental spatio-temporal scales of chromosphere heating, and understand the nature of these scales \citep{Rast2021}.

In the time domain, the total flare emission consists of multiple bursts, loosely referred to as Quasi-Periodic Pulsations \citep[QPPs; see review by][]{Zimovets2021}. QPPs have been identified in flare emissions at nearly all wavelengths, in SXRs, HXRs, and optical and UV emissions, and their ``periods" range from seconds to minutes. To date, it has been unclear what the nature of these bursts is, or whether QPPs are global or local. Globally, QPPs may be associated with modulations by flare excited MHD waves of various kinds \citep[e.g.][]{Ofman2011, Takasao2016, Li2024} or the global MHD process governing the entire flare-CME system \citep[e.g.][]{Takahashi2017}. If governed by the global mode, QPPs should be observed nearly coherently at all or most locations. QPPs may also be excited locally, such as by unsteady tearing mode reconnection in a current sheet \citep{Brannon2015, Parker2017}, and each burst in the total emission light curve may be characteristic of a single energy release event at a different location. \citet{Vievering2023} have analyzed QPPs in nine solar eruptive events, and have found that QPPs identified in the total hard X-ray light curve are correlated with multiple peaks in flare reconnection rate $\dot{\psi}(t)$ derived by tracking newly brightened flare ribbon pixels. Their findings imply that HXR bursts occurring at different times may be produced at different locations, namely the locations of newly brightened flare ribbons which map new reconnection events in the corona. In this study, we analyze the UV light curve from each ribbon pixel and recognize that a fraction of the ribbon pixels (but not all pixels) exhibit multiple significant peaks. These multiple peaks may be co-spatial due to recurring energy release (and deposition) at the same location \citep[e.g.][]{Zhu2025}, or, may be characteristic of different energy release events in different structures within one pixel which are not resolved by AIA. Observations with much enhanced spatial resolution can help clarify the picture. On the other hand, we note that the average of the normalized epoch plots of thousands of flaring ribbon pixels does not exhibit evident structures beyond the single burst, indicating that the multiple peaks are either not global (i.e., not present in the majority of the pixels), or more likely not in phase so that all but the strongest peak are smoothed out in the superposed epoch plot. More dedicated study on this topic will be conducted to probe the nature of multiple peaks in spatially resolved flare ribbon pixel light curves.

Finally, the comparison between UV 1600~\AA\ and EUV 304~\AA\ observations show the advantage of using the UV images to monitor the chromosphere and transition region photometry consistently. The EUV 304~\AA\ broadband images exhibit a mixture of mostly optically-thick emissions from the upper chromosphere to the corona by plasmas of temperatures from a few tens kK to 1-2 MK \citep{Odwyer2010, Antolin2024, Sahin2024}. As such, the 304~\AA\ images show both flare ribbons and overlying loops, which are often difficult to separate. On the other hand, the 304~\AA\ passband is more sensitive to subtle brightening as well as dimming signatures, some of which may occur prior to flare eruption. Whereas brightening in the UV and EUV passbands is often the proxy for reconnection energy release and plasma heating, dimming signatures are produced by reduced density (and often times temperature) of plasmas along magnetic field lines that become longer due to expansion or reconnection \citep{Qiu2024}. Therefore, the pre-eruption signatures identified in EUV 304~\AA\ images can help diagnose evolution, either ideal or non-ideal, of magnetic structures, which leads to catastrophic energy release in flares and CMEs. To take advantage of the diagnostic potential of the EUV 304~\AA\ images, much effort is needed to develop approaches to systematically and quantitatively analyze signatures in these images, and infer physical properties from them.

\begin{acks}
We thank the referee for insightful comments that have helped improve the manuscript.
The completion of the study is motivated by discussions with Dr. M. Rempel. With this study, authors also wish to pay tribute to late Dr. T. Tarbell, who had sustained interest in and support of this effort, and had provided precious guidance in calibrating the UV observations by TRACE and AIA. The work is supported by the NASA Heliophysics Guest Investigator program with the grants 80NSSC22K0519 and 80NSSC23K0414, and by the NSF Astronomical Sciences program and Atmospheric and Geospace Sciences program through the grants AST-2407849 and AGS-2401228. RF's work was also part of the MaxMillennium program at Montana State University supported by the NASA Heliophysics Data and Model Consortium through the grant 80NSSC20K0873.
\end{acks}

\section*{Disclosure of Potential Conflicts of Interest}
The authors declare that they have no conflicts of interest.



\bibliographystyle{spr-mp-sola}
\bibliography{chromo}

\begin{thebibliography}{92}
\ifx\bisbn     \undefined \def\bisbn  #1{ISBN #1}\fi
\ifx\binits    \undefined \def\binits#1{#1}\fi
\ifx\bauthor   \undefined \def\bauthor#1{#1}\fi
\ifx\batitle   \undefined \def\batitle#1{#1}\fi
\ifx\bjtitle   \undefined \def\bjtitle#1{\textit{#1}}\fi
\ifx\bvolume   \undefined \def\bvolume#1{\textbf{#1}}\fi
\ifx\byear     \undefined \def\byear#1{#1}\fi
\ifx\bissue    \undefined \def\bissue#1{#1}\fi
\ifx\bfpage    \undefined \def\bfpage#1{#1}\fi
\ifx\blpage    \undefined \def\blpage #1{#1}\fi
\ifx\burl      \undefined \def\burl#1{#1}\fi
\ifx\href      \undefined \def\href#1#2{#2}\fi
\ifx\betal     \undefined \def\betal{et al.}\fi
\ifx\bctitle   \undefined \def\bctitle#1{#1}\fi
\ifx\beditor   \undefined \def\beditor#1{#1}\fi
\ifx\bbtitle   \undefined \def\bbtitle#1{\textit{#1}}\fi
\ifx\bedition  \undefined \def\bedition#1{#1}\fi
\ifx\bseriesno \undefined \def\bseriesno#1{\textbf{#1}}\fi
\ifx\blocation \undefined \def\blocation#1{#1}\fi
\ifx\bsertitle \undefined \def\bsertitle#1{\textit{#1}}\fi
\ifx\bsnm      \undefined \def\bsnm#1{#1}\fi
\ifx\bsuffix   \undefined \def\bsuffix#1{#1}\fi
\ifx\bparticle \undefined \def\bparticle#1{#1}\fi
\ifx\barticle  \undefined \def\barticle#1{}\fi
\ifx\binstitute  \undefined \def\binstitute#1{#1}\fi
\ifx\bpublisher  \undefined \def\bpublisher#1{#1}\fi
\ifx\doiurl    \undefined \def\doiurl#1{\href{#1}{DOI}}\fi
\makeatletter
\def\safeHref#1#2#3{\in@{http}{#2}\ifin@\href{#2}{#3}\else\href{#1#2}{#3}\fi}
\makeatother
\ifx\adsurl    \undefined
  \def\adsurl#1{\safeHref{https://ui.adsabs.harvard.edu/abs/}{#1}{ADS}}\fi
\ifx\arxivurl  \undefined
  \def\arxivurl#1{\safeHref{http://arxiv.org/abs/}{#1}{arXiv}}\fi
\ifx\botherref \undefined \def\botherref#1{}\fi
\ifx\url       \undefined \def\url#1{#1}\fi
\ifx\bchapter  \undefined \def\bchapter#1{}\fi
\ifx\bbook     \undefined \def\bbook#1{}\fi
\ifx\bcomment  \undefined \def\bcomment#1{#1}\fi
\ifx\oauthor   \undefined \def\oauthor#1{#1}\fi
\ifx\citeauthoryear \undefined\def \citeauthoryear#1{#1}\fi
\def\endbibitem {}
\ifx\bconflocation  \undefined \def\bconflocation#1{#1} \fi

\bibitem[\protect\citeauthoryear{{Alexander} and
  {Coyner}}{2006}]{Alexander2006}
\begin{barticle}
\bauthor{\bsnm{{Alexander}}, \binits{D.}},
\bauthor{\bsnm{{Coyner}}, \binits{A.J.}}:
\byear{2006},
\batitle{{Temporal and Spatial Relationships between Ultraviolet and Hard X-Ray
  Emission in Solar Flares}}.
\bjtitle{\apj}
\bvolume{640},
\bfpage{505}.
\doiurl{https://doi.org/10.1086/500076}.
\adsurl{2006ApJ...640..505A}.
\end{barticle}
\endbibitem

\bibitem[\protect\citeauthoryear{{Allred}, {Kowalski}, and
  {Carlsson}}{2015}]{Allred2015}
\begin{barticle}
\bauthor{\bsnm{{Allred}}, \binits{J.C.}},
\bauthor{\bsnm{{Kowalski}}, \binits{A.F.}},
\bauthor{\bsnm{{Carlsson}}, \binits{M.}}:
\byear{2015},
\batitle{{A Unified Computational Model for Solar and Stellar Flares}}.
\bjtitle{\apj}
\bvolume{809},
\bfpage{104}.
\doiurl{https://doi.org/10.1088/0004-637X/809/1/104}.
\adsurl{2015ApJ...809..104A}.
\end{barticle}
\endbibitem

\bibitem[\protect\citeauthoryear{{Antolin} et~al.}{2024}]{Antolin2024}
\begin{barticle}
\bauthor{\bsnm{{Antolin}}, \binits{P.}},
\bauthor{\bsnm{{Auch{\`e}re}}, \binits{F.}},
\bauthor{\bsnm{{Winch}}, \binits{E.}},
\bauthor{\bsnm{{Soubri{\'e}}}, \binits{E.}},
\bauthor{\bsnm{{Oliver}}, \binits{R.}}:
\byear{2024},
\batitle{{Decomposing the AIA 304 {\r{A}} Channel into Its Cool and Hot
  Components}}.
\bjtitle{\solphys}
\bvolume{299},
\bfpage{94}.
\doiurl{https://doi.org/10.1007/s11207-024-02337-4}.
\adsurl{2024SoPh..299...94A}.
\end{barticle}
\endbibitem

\bibitem[\protect\citeauthoryear{{Asai} et~al.}{2004}]{Asai2004}
\begin{barticle}
\bauthor{\bsnm{{Asai}}, \binits{A.}},
\bauthor{\bsnm{{Yokoyama}}, \binits{T.}},
\bauthor{\bsnm{{Shimojo}}, \binits{M.}},
\bauthor{\bsnm{{Masuda}}, \binits{S.}},
\bauthor{\bsnm{{Kurokawa}}, \binits{H.}},
\bauthor{\bsnm{{Shibata}}, \binits{K.}}:
\byear{2004},
\batitle{{Flare Ribbon Expansion and Energy Release Rate}}.
\bjtitle{\apj}
\bvolume{611},
\bfpage{557}.
\doiurl{https://doi.org/10.1086/422159}.
\adsurl{2004ApJ...611..557A}.
\end{barticle}
\endbibitem

\bibitem[\protect\citeauthoryear{{Aschwanden} and
  {Alexander}}{2001}]{Aschwanden2001}
\begin{barticle}
\bauthor{\bsnm{{Aschwanden}}, \binits{M.J.}},
\bauthor{\bsnm{{Alexander}}, \binits{D.}}:
\byear{2001},
\batitle{{Flare Plasma Cooling from 30 MK down to 1 MK modeled from Yohkoh,
  GOES, and TRACE observations during the Bastille Day Event (14 July 2000)}}.
\bjtitle{\solphys}
\bvolume{204},
\bfpage{91}.
\doiurl{https://doi.org/10.1023/A:1014257826116}.
\adsurl{2001SoPh..204...91A}.
\end{barticle}
\endbibitem

\bibitem[\protect\citeauthoryear{{Bobra} et~al.}{2014}]{Bobra2014}
\begin{barticle}
\bauthor{\bsnm{{Bobra}}, \binits{M.G.}},
\bauthor{\bsnm{{Sun}}, \binits{X.}},
\bauthor{\bsnm{{Hoeksema}}, \binits{J.T.}},
\bauthor{\bsnm{{Turmon}}, \binits{M.}},
\bauthor{\bsnm{{Liu}}, \binits{Y.}},
\bauthor{\bsnm{{Hayashi}}, \binits{K.}},
\bauthor{\bsnm{{Barnes}}, \binits{G.}},
\bauthor{\bsnm{{Leka}}, \binits{K.D.}}:
\byear{2014},
\batitle{{The Helioseismic and Magnetic Imager (HMI) Vector Magnetic Field
  Pipeline: SHARPs - Space-Weather HMI Active Region Patches}}.
\bjtitle{\solphys}
\bvolume{289},
\bfpage{3549}.
\doiurl{https://doi.org/10.1007/s11207-014-0529-3}.
\adsurl{2014SoPh..289.3549B}.
\end{barticle}
\endbibitem

\bibitem[\protect\citeauthoryear{{Brannon}, {Longcope}, and
  {Qiu}}{2015}]{Brannon2015}
\begin{barticle}
\bauthor{\bsnm{{Brannon}}, \binits{S.R.}},
\bauthor{\bsnm{{Longcope}}, \binits{D.W.}},
\bauthor{\bsnm{{Qiu}}, \binits{J.}}:
\byear{2015},
\batitle{{Spectroscopic Observations of an Evolving Flare Ribbon Substructure
  Suggesting Origin in Current Sheet Waves}}.
\bjtitle{\apj}
\bvolume{810},
\bfpage{4}.
\doiurl{https://doi.org/10.1088/0004-637X/810/1/4}.
\adsurl{2015ApJ...810....4B}.
\end{barticle}
\endbibitem

\bibitem[\protect\citeauthoryear{{Brekke} and
  {Kjeldseth-Moe}}{1994}]{Brekke1994}
\begin{barticle}
\bauthor{\bsnm{{Brekke}}, \binits{P.}},
\bauthor{\bsnm{{Kjeldseth-Moe}}, \binits{O.}}:
\byear{1994},
\batitle{{The solar UV continuum 1440-1680 {\r{A}} and its center-to-limb
  variation}}.
\bjtitle{\solphys}
\bvolume{150},
\bfpage{19}.
\doiurl{https://doi.org/10.1007/BF00712874}.
\adsurl{1994SoPh..150...19B}.
\end{barticle}
\endbibitem

\bibitem[\protect\citeauthoryear{{Brosius}, {Daw}, and
  {Inglis}}{2016}]{Brosius2016}
\begin{barticle}
\bauthor{\bsnm{{Brosius}}, \binits{J.W.}},
\bauthor{\bsnm{{Daw}}, \binits{A.N.}},
\bauthor{\bsnm{{Inglis}}, \binits{A.R.}}:
\byear{2016},
\batitle{{Quasi-periodic Fluctuations and Chromospheric Evaporation in a Solar
  Flare Ribbon Observed by Hinode/EIS, IRIS, and RHESSI}}.
\bjtitle{\apj}
\bvolume{830},
\bfpage{101}.
\doiurl{https://doi.org/10.3847/0004-637X/830/2/101}.
\adsurl{2016ApJ...830..101B}.
\end{barticle}
\endbibitem

\bibitem[\protect\citeauthoryear{{Cao} et~al.}{2010}]{Cao2010}
\begin{barticle}
\bauthor{\bsnm{{Cao}}, \binits{W.}},
\bauthor{\bsnm{{Gorceix}}, \binits{N.}},
\bauthor{\bsnm{{Coulter}}, \binits{R.}},
\bauthor{\bsnm{{Ahn}}, \binits{K.}},
\bauthor{\bsnm{{Rimmele}}, \binits{T.R.}},
\bauthor{\bsnm{{Goode}}, \binits{P.R.}}:
\byear{2010},
\batitle{{Scientific instrumentation for the 1.6 m New Solar Telescope in Big
  Bear}}.
\bjtitle{Astronomische Nachrichten}
\bvolume{331},
\bfpage{636}.
\doiurl{https://doi.org/10.1002/asna.201011390}.
\adsurl{2010AN....331..636C}.
\end{barticle}
\endbibitem

\bibitem[\protect\citeauthoryear{{Cheng}}{1990}]{Cheng1990}
\begin{barticle}
\bauthor{\bsnm{{Cheng}}, \binits{C.-C.}}:
\byear{1990},
\batitle{{Ultraviolet Observations of the Preimpulsive Phase in a Solar Flare:
  Enhanced Turbulence and Heating}}.
\bjtitle{\apj}
\bvolume{349},
\bfpage{362}.
\doiurl{https://doi.org/10.1086/168319}.
\adsurl{1990ApJ...349..362C}.
\end{barticle}
\endbibitem

\bibitem[\protect\citeauthoryear{{Cheung} et~al.}{2015}]{Cheung2015}
\begin{barticle}
\bauthor{\bsnm{{Cheung}}, \binits{M.C.M.}},
\bauthor{\bsnm{{Boerner}}, \binits{P.}},
\bauthor{\bsnm{{Schrijver}}, \binits{C.J.}},
\bauthor{\bsnm{{Testa}}, \binits{P.}},
\bauthor{\bsnm{{Chen}}, \binits{F.}},
\bauthor{\bsnm{{Peter}}, \binits{H.}},
\bauthor{\bsnm{{Malanushenko}}, \binits{A.}}:
\byear{2015},
\batitle{{Thermal Diagnostics with the Atmospheric Imaging Assembly on board
  the Solar Dynamics Observatory: A Validated Method for Differential Emission
  Measure Inversions}}.
\bjtitle{\apj}
\bvolume{807},
\bfpage{143}.
\doiurl{https://doi.org/10.1088/0004-637X/807/2/143}.
\adsurl{2015ApJ...807..143C}.
\end{barticle}
\endbibitem

\bibitem[\protect\citeauthoryear{{Cheung} et~al.}{2019}]{Cheung2019}
\begin{barticle}
\bauthor{\bsnm{{Cheung}}, \binits{M.C.M.}},
\bauthor{\bsnm{{Rempel}}, \binits{M.}},
\bauthor{\bsnm{{Chintzoglou}}, \binits{G.}},
\bauthor{\bsnm{{Chen}}, \binits{F.}},
\bauthor{\bsnm{{Testa}}, \binits{P.}},
\bauthor{\bsnm{{Mart{\'{\i}}nez-Sykora}}, \binits{J.}},
\bauthor{\bsnm{{Sainz Dalda}}, \binits{A.}},
\bauthor{\bsnm{{DeRosa}}, \binits{M.L.}},
\bauthor{\bsnm{{Malanushenko}}, \binits{A.}},
\bauthor{\bsnm{{Hansteen}}, \binits{V.}},
\bauthor{\bsnm{{De Pontieu}}, \binits{B.}},
\bauthor{\bsnm{{Carlsson}}, \binits{M.}},
\bauthor{\bsnm{{Gudiksen}}, \binits{B.}},
\bauthor{\bsnm{{McIntosh}}, \binits{S.W.}}:
\byear{2019},
\batitle{{A comprehensive three-dimensional radiative magnetohydrodynamic
  simulation of a solar flare}}.
\bjtitle{Nature Astronomy}
\bvolume{3},
\bfpage{160}.
\doiurl{https://doi.org/10.1038/s41550-018-0629-3}.
\adsurl{2019NatAs...3..160C}.
\end{barticle}
\endbibitem

\bibitem[\protect\citeauthoryear{{Chikunova} et~al.}{2023}]{Chikunova2023}
\begin{barticle}
\bauthor{\bsnm{{Chikunova}}, \binits{G.}},
\bauthor{\bsnm{{Podladchikova}}, \binits{T.}},
\bauthor{\bsnm{{Dissauer}}, \binits{K.}},
\bauthor{\bsnm{{Veronig}}, \binits{A.M.}},
\bauthor{\bsnm{{Dumbovi{\'c}}}, \binits{M.}},
\bauthor{\bsnm{{Temmer}}, \binits{M.}},
\bauthor{\bsnm{{Dickson}}, \binits{E.C.M.}}:
\byear{2023},
\batitle{{Three-dimensional relation between coronal dimming, filament
  eruption, and CME. A case study of the 28 October 2021 X1.0 event}}.
\bjtitle{\aap}
\bvolume{678},
\bfpage{A166}.
\doiurl{https://doi.org/10.1051/0004-6361/202347011}.
\adsurl{2023A&A...678A.166C}.
\end{barticle}
\endbibitem

\bibitem[\protect\citeauthoryear{{Cook} and {Brueckner}}{1979}]{Cook1979}
\begin{barticle}
\bauthor{\bsnm{{Cook}}, \binits{J.W.}},
\bauthor{\bsnm{{Brueckner}}, \binits{G.E.}}:
\byear{1979},
\batitle{{EUV continua of solar flares 1420 - 1960 {\r{A}}.}}
\bjtitle{\apj}
\bvolume{227},
\bfpage{645}.
\doiurl{https://doi.org/10.1086/156775}.
\adsurl{1979ApJ...227..645C}.
\end{barticle}
\endbibitem

\bibitem[\protect\citeauthoryear{{{\c{S}}ahin} and {Antolin}}{2024}]{Sahin2024}
\begin{barticle}
\bauthor{\bsnm{{{\c{S}}ahin}}, \binits{S.}},
\bauthor{\bsnm{{Antolin}}, \binits{P.}}:
\byear{2024},
\batitle{{From Chromospheric Evaporation to Coronal Rain: An Investigation of
  the Mass and Energy Cycle of a Flare}}.
\bjtitle{\apj}
\bvolume{970},
\bfpage{106}.
\doiurl{https://doi.org/10.3847/1538-4357/ad4ed9}.
\adsurl{2024ApJ...970..106S}.
\end{barticle}
\endbibitem

\bibitem[\protect\citeauthoryear{{Dahlin} et~al.}{2025}]{Dahlin2025}
\begin{botherref}
\oauthor{\bsnm{{Dahlin}}, \binits{J.T.}},
\oauthor{\bsnm{{Antiochos}}, \binits{S.K.}},
\oauthor{\bsnm{{DeVore}}, \binits{C.R.}},
\oauthor{\bsnm{{Wyper}}, \binits{P.F.}},
\oauthor{\bsnm{{Qiu}}, \binits{J.}}:
2025,
{Determining the 3D Dynamics of Solar Flare Magnetic Reconnection}.
\textit{arXiv e-prints},
arXiv:2504.00913.
\doiurl{https://doi.org/10.48550/arXiv.2504.00913}.
\adsurl{2025arXiv250400913D}.
\end{botherref}
\endbibitem

\bibitem[\protect\citeauthoryear{{De Pontieu} et~al.}{2014}]{DePontieu2014}
\begin{barticle}
\bauthor{\bsnm{{De Pontieu}}, \binits{B.}},
\bauthor{\bsnm{{Title}}, \binits{A.M.}},
\bauthor{\bsnm{{Lemen}}, \binits{J.R.}},
\bauthor{\bsnm{{Kushner}}, \binits{G.D.}},
\bauthor{\bsnm{{Akin}}, \binits{D.J.}},
\bauthor{\bsnm{{Allard}}, \binits{B.}},
\bauthor{\bsnm{{Berger}}, \binits{T.}},
\bauthor{\bsnm{{Boerner}}, \binits{P.}},
\bauthor{\bsnm{{Cheung}}, \binits{M.}},
\bauthor{\bsnm{{Chou}}, \binits{C.}},
\bauthor{\bsnm{{Drake}}, \binits{J.F.}},
\bauthor{\bsnm{{Duncan}}, \binits{D.W.}},
\bauthor{\bsnm{{Freeland}}, \binits{S.}},
\bauthor{\bsnm{{Heyman}}, \binits{G.F.}},
\bauthor{\bsnm{{Hoffman}}, \binits{C.}},
\bauthor{\bsnm{{Hurlburt}}, \binits{N.E.}},
\bauthor{\bsnm{{Lindgren}}, \binits{R.W.}},
\bauthor{\bsnm{{Mathur}}, \binits{D.}},
\bauthor{\bsnm{{Rehse}}, \binits{R.}},
\bauthor{\bsnm{{Sabolish}}, \binits{D.}},
\bauthor{\bsnm{{Seguin}}, \binits{R.}},
\bauthor{\bsnm{{Schrijver}}, \binits{C.J.}},
\bauthor{\bsnm{{Tarbell}}, \binits{T.D.}},
\bauthor{\bsnm{{W{\"u}lser}}, \binits{J.-P.}},
\bauthor{\bsnm{{Wolfson}}, \binits{C.J.}},
\bauthor{\bsnm{{Yanari}}, \binits{C.}},
\bauthor{\bsnm{{Mudge}}, \binits{J.}},
\bauthor{\bsnm{{Nguyen-Phuc}}, \binits{N.}},
\bauthor{\bsnm{{Timmons}}, \binits{R.}},
\bauthor{\bsnm{{van Bezooijen}}, \binits{R.}},
\bauthor{\bsnm{{Weingrod}}, \binits{I.}},
\bauthor{\bsnm{{Brookner}}, \binits{R.}},
\bauthor{\bsnm{{Butcher}}, \binits{G.}},
\bauthor{\bsnm{{Dougherty}}, \binits{B.}},
\bauthor{\bsnm{{Eder}}, \binits{J.}},
\bauthor{\bsnm{{Knagenhjelm}}, \binits{V.}},
\bauthor{\bsnm{{Larsen}}, \binits{S.}},
\bauthor{\bsnm{{Mansir}}, \binits{D.}},
\bauthor{\bsnm{{Phan}}, \binits{L.}},
\bauthor{\bsnm{{Boyle}}, \binits{P.}},
\bauthor{\bsnm{{Cheimets}}, \binits{P.N.}},
\bauthor{\bsnm{{DeLuca}}, \binits{E.E.}},
\bauthor{\bsnm{{Golub}}, \binits{L.}},
\bauthor{\bsnm{{Gates}}, \binits{R.}},
\bauthor{\bsnm{{Hertz}}, \binits{E.}},
\bauthor{\bsnm{{McKillop}}, \binits{S.}},
\bauthor{\bsnm{{Park}}, \binits{S.}},
\bauthor{\bsnm{{Perry}}, \binits{T.}},
\bauthor{\bsnm{{Podgorski}}, \binits{W.A.}},
\bauthor{\bsnm{{Reeves}}, \binits{K.}},
\bauthor{\bsnm{{Saar}}, \binits{S.}},
\bauthor{\bsnm{{Testa}}, \binits{P.}},
\bauthor{\bsnm{{Tian}}, \binits{H.}},
\bauthor{\bsnm{{Weber}}, \binits{M.}},
\bauthor{\bsnm{{Dunn}}, \binits{C.}},
\bauthor{\bsnm{{Eccles}}, \binits{S.}},
\bauthor{\bsnm{{Jaeggli}}, \binits{S.A.}},
\bauthor{\bsnm{{Kankelborg}}, \binits{C.C.}},
\bauthor{\bsnm{{Mashburn}}, \binits{K.}},
\bauthor{\bsnm{{Pust}}, \binits{N.}},
\bauthor{\bsnm{{Springer}}, \binits{L.}},
\bauthor{\bsnm{{Carvalho}}, \binits{R.}},
\bauthor{\bsnm{{Kleint}}, \binits{L.}},
\bauthor{\bsnm{{Marmie}}, \binits{J.}},
\bauthor{\bsnm{{Mazmanian}}, \binits{E.}},
\bauthor{\bsnm{{Pereira}}, \binits{T.M.D.}},
\bauthor{\bsnm{{Sawyer}}, \binits{S.}},
\bauthor{\bsnm{{Strong}}, \binits{J.}},
\bauthor{\bsnm{{Worden}}, \binits{S.P.}},
\bauthor{\bsnm{{Carlsson}}, \binits{M.}},
\bauthor{\bsnm{{Hansteen}}, \binits{V.H.}},
\bauthor{\bsnm{{Leenaarts}}, \binits{J.}},
\bauthor{\bsnm{{Wiesmann}}, \binits{M.}},
\bauthor{\bsnm{{Aloise}}, \binits{J.}},
\bauthor{\bsnm{{Chu}}, \binits{K.-C.}},
\bauthor{\bsnm{{Bush}}, \binits{R.I.}},
\bauthor{\bsnm{{Scherrer}}, \binits{P.H.}},
\bauthor{\bsnm{{Brekke}}, \binits{P.}},
\bauthor{\bsnm{{Martinez-Sykora}}, \binits{J.}},
\bauthor{\bsnm{{Lites}}, \binits{B.W.}},
\bauthor{\bsnm{{McIntosh}}, \binits{S.W.}},
\bauthor{\bsnm{{Uitenbroek}}, \binits{H.}},
\bauthor{\bsnm{{Okamoto}}, \binits{T.J.}},
\bauthor{\bsnm{{Gummin}}, \binits{M.A.}},
\bauthor{\bsnm{{Auker}}, \binits{G.}},
\bauthor{\bsnm{{Jerram}}, \binits{P.}},
\bauthor{\bsnm{{Pool}}, \binits{P.}},
\bauthor{\bsnm{{Waltham}}, \binits{N.}}:
\byear{2014},
\batitle{{The Interface Region Imaging Spectrograph (IRIS)}}.
\bjtitle{\solphys}
\bvolume{289},
\bfpage{2733}.
\doiurl{https://doi.org/10.1007/s11207-014-0485-y}.
\adsurl{2014SoPh..289.2733D}.
\end{barticle}
\endbibitem

\bibitem[\protect\citeauthoryear{{Doyle} and {Cook}}{1992}]{Doyle1992}
\begin{barticle}
\bauthor{\bsnm{{Doyle}}, \binits{J.G.}},
\bauthor{\bsnm{{Cook}}, \binits{J.W.}}:
\byear{1992},
\batitle{{The Sun as a Star: High Spectral Resolution Solar Data Degraded to
  Low-Dispersion IUE Resolution}}.
\bjtitle{\apj}
\bvolume{391},
\bfpage{393}.
\doiurl{https://doi.org/10.1086/171354}.
\adsurl{1992ApJ...391..393D}.
\end{barticle}
\endbibitem

\bibitem[\protect\citeauthoryear{{Fisher}, {Canfield}, and
  {McClymont}}{1985a}]{Fisher1985c}
\begin{barticle}
\bauthor{\bsnm{{Fisher}}, \binits{G.H.}},
\bauthor{\bsnm{{Canfield}}, \binits{R.C.}},
\bauthor{\bsnm{{McClymont}}, \binits{A.N.}}:
\byear{1985}a,
\batitle{{Flare Loop Radiative Hydrodynamics - Part Seven - Dynamics of the
  Thick Target Heated Chromosphere}}.
\bjtitle{\apj}
\bvolume{289},
\bfpage{434}.
\doiurl{https://doi.org/10.1086/162903}.
\adsurl{1985ApJ...289..434F}.
\end{barticle}
\endbibitem

\bibitem[\protect\citeauthoryear{{Fisher}, {Canfield}, and
  {McClymont}}{1985b}]{Fisher1985b}
\begin{barticle}
\bauthor{\bsnm{{Fisher}}, \binits{G.H.}},
\bauthor{\bsnm{{Canfield}}, \binits{R.C.}},
\bauthor{\bsnm{{McClymont}}, \binits{A.N.}}:
\byear{1985}b,
\batitle{{Flare Loop Radiative Hydrodynamics - Part Six - Chromospheric
  Evaporation due to Heating by Nonthermal Electrons}}.
\bjtitle{\apj}
\bvolume{289},
\bfpage{425}.
\doiurl{https://doi.org/10.1086/162902}.
\adsurl{1985ApJ...289..425F}.
\end{barticle}
\endbibitem

\bibitem[\protect\citeauthoryear{{Fisher}, {Canfield}, and
  {McClymont}}{1985c}]{Fisher1985a}
\begin{barticle}
\bauthor{\bsnm{{Fisher}}, \binits{G.H.}},
\bauthor{\bsnm{{Canfield}}, \binits{R.C.}},
\bauthor{\bsnm{{McClymont}}, \binits{A.N.}}:
\byear{1985}c,
\batitle{{Flare loop radiative hydrodynamics. V - Response to thick-target
  heating. VI - Chromospheric evaporation due to heating by nonthermal
  electrons. VII - Dynamics of the thick-target heated chromosphere}}.
\bjtitle{\apj}
\bvolume{289},
\bfpage{414}.
\doiurl{https://doi.org/10.1086/162901}.
\adsurl{1985ApJ...289..414F}.
\end{barticle}
\endbibitem

\bibitem[\protect\citeauthoryear{{Fletcher}}{2024}]{Fletcher2024}
\begin{barticle}
\bauthor{\bsnm{{Fletcher}}, \binits{L.}}:
\byear{2024},
\batitle{{Solar Flare Spectroscopy}}.
\bjtitle{Annual Review of Astronomy and Astrophysics}
\bvolume{62},
\bfpage{437}.
\doiurl{https://doi.org/10.1146/annurev-astro-052920-010547}.
\adsurl{2024ARA&A..62..437F}.
\end{barticle}
\endbibitem

\bibitem[\protect\citeauthoryear{{Fletcher} and {Hudson}}{2001}]{Fletcher2001}
\begin{barticle}
\bauthor{\bsnm{{Fletcher}}, \binits{L.}},
\bauthor{\bsnm{{Hudson}}, \binits{H.}}:
\byear{2001},
\batitle{{The Magnetic Structure and Generation of EUV Flare Ribbons}}.
\bjtitle{\solphys}
\bvolume{204},
\bfpage{69}.
\doiurl{https://doi.org/10.1023/A:1014275821318}.
\adsurl{2001SoPh..204...69F}.
\end{barticle}
\endbibitem

\bibitem[\protect\citeauthoryear{{Fletcher} and {Hudson}}{2008}]{Fletcher2008}
\begin{barticle}
\bauthor{\bsnm{{Fletcher}}, \binits{L.}},
\bauthor{\bsnm{{Hudson}}, \binits{H.S.}}:
\byear{2008},
\batitle{{Impulsive Phase Flare Energy Transport by Large-Scale Alfv{\'e}n
  Waves and the Electron Acceleration Problem}}.
\bjtitle{\apj}
\bvolume{675},
\bfpage{1645}.
\doiurl{https://doi.org/10.1086/527044}.
\adsurl{2008ApJ...675.1645F}.
\end{barticle}
\endbibitem

\bibitem[\protect\citeauthoryear{{Furth}, {Killeen}, and
  {Rosenbluth}}{1963}]{Furth1963}
\begin{barticle}
\bauthor{\bsnm{{Furth}}, \binits{H.P.}},
\bauthor{\bsnm{{Killeen}}, \binits{J.}},
\bauthor{\bsnm{{Rosenbluth}}, \binits{M.N.}}:
\byear{1963},
\batitle{{Finite-Resistivity Instabilities of a Sheet Pinch}}.
\bjtitle{Physics of Fluids}
\bvolume{6},
\bfpage{459}.
\doiurl{https://doi.org/10.1063/1.1706761}.
\adsurl{1963PhFl....6..459F}.
\end{barticle}
\endbibitem

\bibitem[\protect\citeauthoryear{{Gan}, {Zhang}, and {Fang}}{1991}]{Gan1991}
\begin{barticle}
\bauthor{\bsnm{{Gan}}, \binits{W.Q.}},
\bauthor{\bsnm{{Zhang}}, \binits{H.Q.}},
\bauthor{\bsnm{{Fang}}, \binits{C.}}:
\byear{1991},
\batitle{{A hydrodynamic model of the impulsive phase of a solar flare loop}}.
\bjtitle{\aap}
\bvolume{241},
\bfpage{618}.
\adsurl{1991A\%26A...241..618G}.
\end{barticle}
\endbibitem

\bibitem[\protect\citeauthoryear{{Graham} and {Cauzzi}}{2015}]{Graham2015}
\begin{barticle}
\bauthor{\bsnm{{Graham}}, \binits{D.R.}},
\bauthor{\bsnm{{Cauzzi}}, \binits{G.}}:
\byear{2015},
\batitle{{Temporal Evolution of Multiple Evaporating Ribbon Sources in a Solar
  Flare}}.
\bjtitle{\apjl}
\bvolume{807},
\bfpage{L22}.
\doiurl{https://doi.org/10.1088/2041-8205/807/2/L22}.
\adsurl{2015ApJ...807L..22G}.
\end{barticle}
\endbibitem

\bibitem[\protect\citeauthoryear{{Graham} et~al.}{2020}]{Graham2020}
\begin{barticle}
\bauthor{\bsnm{{Graham}}, \binits{D.R.}},
\bauthor{\bsnm{{Cauzzi}}, \binits{G.}},
\bauthor{\bsnm{{Zangrilli}}, \binits{L.}},
\bauthor{\bsnm{{Kowalski}}, \binits{A.}},
\bauthor{\bsnm{{Sim{\~o}es}}, \binits{P.}},
\bauthor{\bsnm{{Allred}}, \binits{J.}}:
\byear{2020},
\batitle{{Spectral Signatures of Chromospheric Condensation in a Major Solar
  Flare}}.
\bjtitle{\apj}
\bvolume{895},
\bfpage{6}.
\doiurl{https://doi.org/10.3847/1538-4357/ab88ad}.
\adsurl{2020ApJ...895....6G}.
\end{barticle}
\endbibitem

\bibitem[\protect\citeauthoryear{{Guo} et~al.}{2023}]{Guo2023}
\begin{barticle}
\bauthor{\bsnm{{Guo}}, \binits{J.H.}},
\bauthor{\bsnm{{Ni}}, \binits{Y.W.}},
\bauthor{\bsnm{{Zhong}}, \binits{Z.}},
\bauthor{\bsnm{{Guo}}, \binits{Y.}},
\bauthor{\bsnm{{Xia}}, \binits{C.}},
\bauthor{\bsnm{{Li}}, \binits{H.T.}},
\bauthor{\bsnm{{Poedts}}, \binits{S.}},
\bauthor{\bsnm{{Schmieder}}, \binits{B.}},
\bauthor{\bsnm{{Chen}}, \binits{P.F.}}:
\byear{2023},
\batitle{{Thermodynamic and Magnetic Topology Evolution of the X1.0 Flare on
  2021 October 28 Simulated by a Data-driven Radiative Magnetohydrodynamic
  Model}}.
\bjtitle{\apjs}
\bvolume{266},
\bfpage{3}.
\doiurl{https://doi.org/10.3847/1538-4365/acc797}.
\adsurl{2023ApJS..266....3G}.
\end{barticle}
\endbibitem

\bibitem[\protect\citeauthoryear{{Handy} et~al.}{1999}]{Handy1999}
\begin{barticle}
\bauthor{\bsnm{{Handy}}, \binits{B.N.}},
\bauthor{\bsnm{{Acton}}, \binits{L.W.}},
\bauthor{\bsnm{{Kankelborg}}, \binits{C.C.}},
\bauthor{\bsnm{{Wolfson}}, \binits{C.J.}},
\bauthor{\bsnm{{Akin}}, \binits{D.J.}},
\bauthor{\bsnm{{Bruner}}, \binits{M.E.}},
\bauthor{\bsnm{{Caravalho}}, \binits{R.}},
\bauthor{\bsnm{{Catura}}, \binits{R.C.}},
\bauthor{\bsnm{{Chevalier}}, \binits{R.}},
\bauthor{\bsnm{{Duncan}}, \binits{D.W.}},
\bauthor{\bsnm{{Edwards}}, \binits{C.G.}},
\bauthor{\bsnm{{Feinstein}}, \binits{C.N.}},
\bauthor{\bsnm{{Freeland}}, \binits{S.L.}},
\bauthor{\bsnm{{Friedlaender}}, \binits{F.M.}},
\bauthor{\bsnm{{Hoffmann}}, \binits{C.H.}},
\bauthor{\bsnm{{Hurlburt}}, \binits{N.E.}},
\bauthor{\bsnm{{Jurcevich}}, \binits{B.K.}},
\bauthor{\bsnm{{Katz}}, \binits{N.L.}},
\bauthor{\bsnm{{Kelly}}, \binits{G.A.}},
\bauthor{\bsnm{{Lemen}}, \binits{J.R.}},
\bauthor{\bsnm{{Levay}}, \binits{M.}},
\bauthor{\bsnm{{Lindgren}}, \binits{R.W.}},
\bauthor{\bsnm{{Mathur}}, \binits{D.P.}},
\bauthor{\bsnm{{Meyer}}, \binits{S.B.}},
\bauthor{\bsnm{{Morrison}}, \binits{S.J.}},
\bauthor{\bsnm{{Morrison}}, \binits{M.D.}},
\bauthor{\bsnm{{Nightingale}}, \binits{R.W.}},
\bauthor{\bsnm{{Pope}}, \binits{T.P.}},
\bauthor{\bsnm{{Rehse}}, \binits{R.A.}},
\bauthor{\bsnm{{Schrijver}}, \binits{C.J.}},
\bauthor{\bsnm{{Shine}}, \binits{R.A.}},
\bauthor{\bsnm{{Shing}}, \binits{L.}},
\bauthor{\bsnm{{Strong}}, \binits{K.T.}},
\bauthor{\bsnm{{Tarbell}}, \binits{T.D.}},
\bauthor{\bsnm{{Title}}, \binits{A.M.}},
\bauthor{\bsnm{{Torgerson}}, \binits{D.D.}},
\bauthor{\bsnm{{Golub}}, \binits{L.}},
\bauthor{\bsnm{{Bookbinder}}, \binits{J.A.}},
\bauthor{\bsnm{{Caldwell}}, \binits{D.}},
\bauthor{\bsnm{{Cheimets}}, \binits{P.N.}},
\bauthor{\bsnm{{Davis}}, \binits{W.N.}},
\bauthor{\bsnm{{Deluca}}, \binits{E.E.}},
\bauthor{\bsnm{{McMullen}}, \binits{R.A.}},
\bauthor{\bsnm{{Warren}}, \binits{H.P.}},
\bauthor{\bsnm{{Amato}}, \binits{D.}},
\bauthor{\bsnm{{Fisher}}, \binits{R.}},
\bauthor{\bsnm{{Maldonado}}, \binits{H.}},
\bauthor{\bsnm{{Parkinson}}, \binits{C.}}:
\byear{1999},
\batitle{{The transition region and coronal explorer}}.
\bjtitle{\solphys}
\bvolume{187},
\bfpage{229}.
\doiurl{https://doi.org/10.1023/A:1005166902804}.
\adsurl{1999SoPh..187..229H}.
\end{barticle}
\endbibitem

\bibitem[\protect\citeauthoryear{{Hannah} and {Kontar}}{2012}]{Hannah2012}
\begin{barticle}
\bauthor{\bsnm{{Hannah}}, \binits{I.G.}},
\bauthor{\bsnm{{Kontar}}, \binits{E.P.}}:
\byear{2012},
\batitle{{Differential emission measures from the regularized inversion of
  Hinode and SDO data}}.
\bjtitle{\aap}
\bvolume{539},
\bfpage{A146}.
\doiurl{https://doi.org/10.1051/0004-6361/201117576}.
\adsurl{2012A&A...539A.146H}.
\end{barticle}
\endbibitem

\bibitem[\protect\citeauthoryear{{Hinterreiter}
  et~al.}{2018}]{Hinterreiter2018}
\begin{barticle}
\bauthor{\bsnm{{Hinterreiter}}, \binits{J.}},
\bauthor{\bsnm{{Veronig}}, \binits{A.M.}},
\bauthor{\bsnm{{Thalmann}}, \binits{J.K.}},
\bauthor{\bsnm{{Tschernitz}}, \binits{J.}},
\bauthor{\bsnm{{P{\"o}tzi}}, \binits{W.}}:
\byear{2018},
\batitle{{Statistical Properties of Ribbon Evolution and Reconnection Electric
  Fields in Eruptive and Confined Flares}}.
\bjtitle{\solphys}
\bvolume{293},
\bfpage{38}.
\doiurl{https://doi.org/10.1007/s11207-018-1253-1}.
\adsurl{2018SoPh..293...38H}.
\end{barticle}
\endbibitem

\bibitem[\protect\citeauthoryear{{Hofmeister}, {Savin}, and
  {Hahn}}{2024}]{Hofmeister2024}
\begin{botherref}
\oauthor{\bsnm{{Hofmeister}}, \binits{S.}},
\oauthor{\bsnm{{Savin}}, \binits{D.W.}},
\oauthor{\bsnm{{Hahn}}, \binits{M.}}:
2024,
{Revised Point-Spread Functions for the Atmospheric Imaging Assembly onboard
  the Solar Dynamics Observatory}.
\textit{arXiv e-prints},
arXiv:2410.08967.
\doiurl{https://doi.org/10.48550/arXiv.2410.08967}.
\adsurl{2024arXiv241008967H}.
\end{botherref}
\endbibitem

\bibitem[\protect\citeauthoryear{{Hori} et~al.}{1997}]{Hori1997}
\begin{barticle}
\bauthor{\bsnm{{Hori}}, \binits{K.}},
\bauthor{\bsnm{{Yokoyama}}, \binits{T.}},
\bauthor{\bsnm{{Kosugi}}, \binits{T.}},
\bauthor{\bsnm{{Shibata}}, \binits{K.}}:
\byear{1997},
\batitle{{Pseudo-Two-dimensional Hydrodynamic Modeling of Solar Flare Loops}}.
\bjtitle{\apj}
\bvolume{489},
\bfpage{426}.
\doiurl{https://doi.org/10.1086/304754}.
\adsurl{1997ApJ...489..426H}.
\end{barticle}
\endbibitem

\bibitem[\protect\citeauthoryear{{Huang} and {Bhattacharjee}}{2016}]{Huang2016}
\begin{barticle}
\bauthor{\bsnm{{Huang}}, \binits{Y.-M.}},
\bauthor{\bsnm{{Bhattacharjee}}, \binits{A.}}:
\byear{2016},
\batitle{{Turbulent Magnetohydrodynamic Reconnection Mediated by the Plasmoid
  Instability}}.
\bjtitle{\apj}
\bvolume{818},
\bfpage{20}.
\doiurl{https://doi.org/10.3847/0004-637X/818/1/20}.
\adsurl{2016ApJ...818...20H}.
\end{barticle}
\endbibitem

\bibitem[\protect\citeauthoryear{{Jing} et~al.}{2016}]{Jing2016}
\begin{barticle}
\bauthor{\bsnm{{Jing}}, \binits{J.}},
\bauthor{\bsnm{{Xu}}, \binits{Y.}},
\bauthor{\bsnm{{Cao}}, \binits{W.}},
\bauthor{\bsnm{{Liu}}, \binits{C.}},
\bauthor{\bsnm{{Gary}}, \binits{D.}},
\bauthor{\bsnm{{Wang}}, \binits{H.}}:
\byear{2016},
\batitle{{Unprecedented Fine Structure of a Solar Flare Revealed by the
  1.6{\,}m New Solar Telescope}}.
\bjtitle{Scientific Reports}
\bvolume{6},
\bfpage{24319}.
\doiurl{https://doi.org/10.1038/srep24319}.
\adsurl{2016NatSR...624319J}.
\end{barticle}
\endbibitem

\bibitem[\protect\citeauthoryear{{Kazachenko} et~al.}{2017}]{Kazachenko2017}
\begin{barticle}
\bauthor{\bsnm{{Kazachenko}}, \binits{M.D.}},
\bauthor{\bsnm{{Lynch}}, \binits{B.J.}},
\bauthor{\bsnm{{Welsch}}, \binits{B.T.}},
\bauthor{\bsnm{{Sun}}, \binits{X.}}:
\byear{2017},
\batitle{{A Database of Flare Ribbon Properties from the Solar Dynamics
  Observatory. I. Reconnection Flux}}.
\bjtitle{\apj}
\bvolume{845},
\bfpage{49}.
\doiurl{https://doi.org/10.3847/1538-4357/aa7ed6}.
\adsurl{2017ApJ...845...49K}.
\end{barticle}
\endbibitem

\bibitem[\protect\citeauthoryear{{Kerr}, {Allred}, and
  {Polito}}{2020}]{Kerr2020}
\begin{botherref}
\oauthor{\bsnm{{Kerr}}, \binits{G.S.}},
\oauthor{\bsnm{{Allred}}, \binits{J.C.}},
\oauthor{\bsnm{{Polito}}, \binits{V.}}:
2020,
{Solar Flare Arcade Modelling: Bridging the gap from 1D to 3D Simulations of
  Optically Thin Radiation}.
\textit{arXiv e-prints},
arXiv:2007.13856.
\adsurl{2020arXiv200713856K}.
\end{botherref}
\endbibitem

\bibitem[\protect\citeauthoryear{{Kerr} et~al.}{2016}]{Kerr2016}
\begin{barticle}
\bauthor{\bsnm{{Kerr}}, \binits{G.S.}},
\bauthor{\bsnm{{Fletcher}}, \binits{L.}},
\bauthor{\bsnm{{Russell}}, \binits{A.e.J.B.}},
\bauthor{\bsnm{{Allred}}, \binits{J.C.}}:
\byear{2016},
\batitle{{Simulations of the Mg II k and Ca II 8542 lines from an Alfv{\'E}n
  Wave-heated Flare Chromosphere}}.
\bjtitle{\apj}
\bvolume{827},
\bfpage{101}.
\doiurl{https://doi.org/10.3847/0004-637X/827/2/101}.
\adsurl{2016ApJ...827..101K}.
\end{barticle}
\endbibitem

\bibitem[\protect\citeauthoryear{{Kowalski}}{2024}]{Kowalski2024}
\begin{barticle}
\bauthor{\bsnm{{Kowalski}}, \binits{A.F.}}:
\byear{2024},
\batitle{{Stellar flares}}.
\bjtitle{Living Reviews in Solar Physics}
\bvolume{21},
\bfpage{1}.
\doiurl{https://doi.org/10.1007/s41116-024-00039-4}.
\adsurl{2024LRSP...21....1K}.
\end{barticle}
\endbibitem

\bibitem[\protect\citeauthoryear{{Kowalski} et~al.}{2013}]{Kowalski2013}
\begin{barticle}
\bauthor{\bsnm{{Kowalski}}, \binits{A.F.}},
\bauthor{\bsnm{{Hawley}}, \binits{S.L.}},
\bauthor{\bsnm{{Wisniewski}}, \binits{J.P.}},
\bauthor{\bsnm{{Osten}}, \binits{R.A.}},
\bauthor{\bsnm{{Hilton}}, \binits{E.J.}},
\bauthor{\bsnm{{Holtzman}}, \binits{J.A.}},
\bauthor{\bsnm{{Schmidt}}, \binits{S.J.}},
\bauthor{\bsnm{{Davenport}}, \binits{J.R.A.}}:
\byear{2013},
\batitle{{Time-resolved Properties and Global Trends in dMe Flares from
  Simultaneous Photometry and Spectra}}.
\bjtitle{\apjs}
\bvolume{207},
\bfpage{15}.
\doiurl{https://doi.org/10.1088/0067-0049/207/1/15}.
\adsurl{2013ApJS..207...15K}.
\end{barticle}
\endbibitem

\bibitem[\protect\citeauthoryear{{Kowalski} et~al.}{2017}]{Kowalski2017}
\begin{barticle}
\bauthor{\bsnm{{Kowalski}}, \binits{A.F.}},
\bauthor{\bsnm{{Allred}}, \binits{J.C.}},
\bauthor{\bsnm{{Daw}}, \binits{A.}},
\bauthor{\bsnm{{Cauzzi}}, \binits{G.}},
\bauthor{\bsnm{{Carlsson}}, \binits{M.}}:
\byear{2017},
\batitle{{The Atmospheric Response to High Nonthermal Electron Beam Fluxes in
  Solar Flares. I. Modeling the Brightest NUV Footpoints in the X1 Solar Flare
  of 2014 March 29}}.
\bjtitle{\apj}
\bvolume{836},
\bfpage{12}.
\doiurl{https://doi.org/10.3847/1538-4357/836/1/12}.
\adsurl{2017ApJ...836...12K}.
\end{barticle}
\endbibitem

\bibitem[\protect\citeauthoryear{{Lemen} et~al.}{2012}]{Lemen2012}
\begin{barticle}
\bauthor{\bsnm{{Lemen}}, \binits{J.R.}},
\bauthor{\bsnm{{Title}}, \binits{A.M.}},
\bauthor{\bsnm{{Akin}}, \binits{D.J.}},
\bauthor{\bsnm{{Boerner}}, \binits{P.F.}},
\bauthor{\bsnm{{Chou}}, \binits{C.}},
\bauthor{\bsnm{{Drake}}, \binits{J.F.}},
\bauthor{\bsnm{{Duncan}}, \binits{D.W.}},
\bauthor{\bsnm{{Edwards}}, \binits{C.G.}},
\bauthor{\bsnm{{Friedlaender}}, \binits{F.M.}},
\bauthor{\bsnm{{Heyman}}, \binits{G.F.}}:
\byear{2012},
\batitle{{The Atmospheric Imaging Assembly (AIA) on the Solar Dynamics
  Observatory (SDO)}}.
\bjtitle{\solphys}
\bvolume{275},
\bfpage{17}.
\doiurl{https://doi.org/10.1007/s11207-011-9776-8}.
\adsurl{2012SoPh..275...17L}.
\end{barticle}
\endbibitem

\bibitem[\protect\citeauthoryear{{Li} et~al.}{2024}]{Li2024}
\begin{barticle}
\bauthor{\bsnm{{Li}}, \binits{D.}},
\bauthor{\bsnm{{Hong}}, \binits{Z.}},
\bauthor{\bsnm{{Hou}}, \binits{Z.}},
\bauthor{\bsnm{{Su}}, \binits{Y.}}:
\byear{2024},
\batitle{{Localizing Quasiperiodic Pulsations in Hard X-Ray, Microwave, and
  Ly{\ensuremath{\alpha}} Emissions of an X6.4 Flare}}.
\bjtitle{\apj}
\bvolume{970},
\bfpage{77}.
\doiurl{https://doi.org/10.3847/1538-4357/ad566c}.
\adsurl{2024ApJ...970...77L}.
\end{barticle}
\endbibitem

\bibitem[\protect\citeauthoryear{{Liu} et~al.}{2013}]{Liu2013}
\begin{barticle}
\bauthor{\bsnm{{Liu}}, \binits{W.-J.}},
\bauthor{\bsnm{{Qiu}}, \binits{J.}},
\bauthor{\bsnm{{Longcope}}, \binits{D.W.}},
\bauthor{\bsnm{{Caspi}}, \binits{A.}}:
\byear{2013},
\batitle{{Determining Heating Rates in Reconnection Formed Flare Loops of the
  M8.0 Flare on 2005 May 13}}.
\bjtitle{\apj}
\bvolume{770},
\bfpage{111}.
\doiurl{https://doi.org/10.1088/0004-637X/770/2/111}.
\adsurl{2013ApJ...770..111L}.
\end{barticle}
\endbibitem

\bibitem[\protect\citeauthoryear{{Longcope}}{2014}]{Longcope2014}
\begin{barticle}
\bauthor{\bsnm{{Longcope}}, \binits{D.W.}}:
\byear{2014},
\batitle{{A Simple Model of Chromospheric Evaporation and Condensation Driven
  Conductively in a Solar Flare}}.
\bjtitle{\apj}
\bvolume{795},
\bfpage{10}.
\doiurl{https://doi.org/10.1088/0004-637X/795/1/10}.
\adsurl{2014ApJ...795...10L}.
\end{barticle}
\endbibitem

\bibitem[\protect\citeauthoryear{{Massa} et~al.}{2023}]{Massa2023}
\begin{barticle}
\bauthor{\bsnm{{Massa}}, \binits{P.}},
\bauthor{\bsnm{{Emslie}}, \binits{A.G.}},
\bauthor{\bsnm{{Hannah}}, \binits{I.G.}},
\bauthor{\bsnm{{Kontar}}, \binits{E.P.}}:
\byear{2023},
\batitle{{Robust construction of differential emission measure profiles using a
  regularized maximum likelihood method}}.
\bjtitle{\aap}
\bvolume{672},
\bfpage{A120}.
\doiurl{https://doi.org/10.1051/0004-6361/202345883}.
\adsurl{2023A&A...672A.120M}.
\end{barticle}
\endbibitem

\bibitem[\protect\citeauthoryear{{Meegan} et~al.}{2009}]{Meegan2009}
\begin{barticle}
\bauthor{\bsnm{{Meegan}}, \binits{C.}},
\bauthor{\bsnm{{Lichti}}, \binits{G.}},
\bauthor{\bsnm{{Bhat}}, \binits{P.N.}},
\bauthor{\bsnm{{Bissaldi}}, \binits{E.}},
\bauthor{\bsnm{{Briggs}}, \binits{M.S.}},
\bauthor{\bsnm{{Connaughton}}, \binits{V.}},
\bauthor{\bsnm{{Diehl}}, \binits{R.}},
\bauthor{\bsnm{{Fishman}}, \binits{G.}},
\bauthor{\bsnm{{Greiner}}, \binits{J.}},
\bauthor{\bsnm{{Hoover}}, \binits{A.S.}},
\bauthor{\bsnm{{van der Horst}}, \binits{A.J.}},
\bauthor{\bsnm{{von Kienlin}}, \binits{A.}},
\bauthor{\bsnm{{Kippen}}, \binits{R.M.}},
\bauthor{\bsnm{{Kouveliotou}}, \binits{C.}},
\bauthor{\bsnm{{McBreen}}, \binits{S.}},
\bauthor{\bsnm{{Paciesas}}, \binits{W.S.}},
\bauthor{\bsnm{{Preece}}, \binits{R.}},
\bauthor{\bsnm{{Steinle}}, \binits{H.}},
\bauthor{\bsnm{{Wallace}}, \binits{M.S.}},
\bauthor{\bsnm{{Wilson}}, \binits{R.B.}},
\bauthor{\bsnm{{Wilson-Hodge}}, \binits{C.}}:
\byear{2009},
\batitle{{The Fermi Gamma-ray Burst Monitor}}.
\bjtitle{\apj}
\bvolume{702},
\bfpage{791}.
\doiurl{https://doi.org/10.1088/0004-637X/702/1/791}.
\adsurl{2009ApJ...702..791M}.
\end{barticle}
\endbibitem

\bibitem[\protect\citeauthoryear{{Milligan}}{2015}]{Milligan2015}
\begin{barticle}
\bauthor{\bsnm{{Milligan}}, \binits{R.O.}}:
\byear{2015},
\batitle{{Extreme Ultra-Violet Spectroscopy of the Lower Solar Atmosphere
  During Solar Flares (Invited Review)}}.
\bjtitle{\solphys}
\bvolume{290},
\bfpage{3399}.
\doiurl{https://doi.org/10.1007/s11207-015-0748-2}.
\adsurl{2015SoPh..290.3399M}.
\end{barticle}
\endbibitem

\bibitem[\protect\citeauthoryear{{Naus} et~al.}{2022}]{Naus2022}
\begin{barticle}
\bauthor{\bsnm{{Naus}}, \binits{S.J.}},
\bauthor{\bsnm{{Qiu}}, \binits{J.}},
\bauthor{\bsnm{{DeVore}}, \binits{C.R.}},
\bauthor{\bsnm{{Antiochos}}, \binits{S.K.}},
\bauthor{\bsnm{{Dahlin}}, \binits{J.T.}},
\bauthor{\bsnm{{Drake}}, \binits{J.F.}},
\bauthor{\bsnm{{Swisdak}}, \binits{M.}}:
\byear{2022},
\batitle{{Correlated Spatio-temporal Evolution of Extreme-Ultraviolet Ribbons
  and Hard X-Rays in a Solar Flare}}.
\bjtitle{\apj}
\bvolume{926},
\bfpage{218}.
\doiurl{https://doi.org/10.3847/1538-4357/ac4028}.
\adsurl{2022ApJ...926..218N}.
\end{barticle}
\endbibitem

\bibitem[\protect\citeauthoryear{{O'Dwyer} et~al.}{2010}]{Odwyer2010}
\begin{barticle}
\bauthor{\bsnm{{O'Dwyer}}, \binits{B.}},
\bauthor{\bsnm{{Del Zanna}}, \binits{G.}},
\bauthor{\bsnm{{Mason}}, \binits{H.E.}},
\bauthor{\bsnm{{Weber}}, \binits{M.A.}},
\bauthor{\bsnm{{Tripathi}}, \binits{D.}}:
\byear{2010},
\batitle{{SDO/AIA response to coronal hole, quiet Sun, active region, and flare
  plasma}}.
\bjtitle{\aap}
\bvolume{521},
\bfpage{A21}.
\doiurl{https://doi.org/10.1051/0004-6361/201014872}.
\adsurl{2010A\%26A...521A..21O}.
\end{barticle}
\endbibitem

\bibitem[\protect\citeauthoryear{{Ofman} and {Thompson}}{2011}]{Ofman2011}
\begin{barticle}
\bauthor{\bsnm{{Ofman}}, \binits{L.}},
\bauthor{\bsnm{{Thompson}}, \binits{B.J.}}:
\byear{2011},
\batitle{{SDO/AIA Observation of Kelvin-Helmholtz Instability in the Solar
  Corona}}.
\bjtitle{\apjl}
\bvolume{734},
\bfpage{L11}.
\doiurl{https://doi.org/10.1088/2041-8205/734/1/L11}.
\adsurl{2011ApJ...734L..11O}.
\end{barticle}
\endbibitem

\bibitem[\protect\citeauthoryear{{Parker} and {Longcope}}{2017}]{Parker2017}
\begin{barticle}
\bauthor{\bsnm{{Parker}}, \binits{J.}},
\bauthor{\bsnm{{Longcope}}, \binits{D.}}:
\byear{2017},
\batitle{{Modeling a Propagating Sawtooth Flare Ribbon Structure as a Tearing
  Mode in the Presence of Velocity Shear}}.
\bjtitle{\apj}
\bvolume{847},
\bfpage{30}.
\doiurl{https://doi.org/10.3847/1538-4357/aa8908}.
\adsurl{2017ApJ...847...30P}.
\end{barticle}
\endbibitem

\bibitem[\protect\citeauthoryear{{Pesnell}, {Thompson}, and
  {Chamberlin}}{2012}]{Pesnell2012}
\begin{barticle}
\bauthor{\bsnm{{Pesnell}}, \binits{W.D.}},
\bauthor{\bsnm{{Thompson}}, \binits{B.J.}},
\bauthor{\bsnm{{Chamberlin}}, \binits{P.C.}}:
\byear{2012},
\batitle{{The Solar Dynamics Observatory (SDO)}}.
\bjtitle{\solphys}
\bvolume{275},
\bfpage{3}.
\doiurl{https://doi.org/10.1007/s11207-011-9841-3}.
\adsurl{2012SoPh..275....3P}.
\end{barticle}
\endbibitem

\bibitem[\protect\citeauthoryear{{Plowman}, {Kankelborg}, and
  {Martens}}{2013}]{Plowman2013}
\begin{barticle}
\bauthor{\bsnm{{Plowman}}, \binits{J.}},
\bauthor{\bsnm{{Kankelborg}}, \binits{C.}},
\bauthor{\bsnm{{Martens}}, \binits{P.}}:
\byear{2013},
\batitle{{Fast Differential Emission Measure Inversion of Solar Coronal Data}}.
\bjtitle{\apj}
\bvolume{771},
\bfpage{2}.
\doiurl{https://doi.org/10.1088/0004-637X/771/1/2}.
\adsurl{2013ApJ...771....2P}.
\end{barticle}
\endbibitem

\bibitem[\protect\citeauthoryear{{Poduval} et~al.}{2013}]{Poduval2013}
\begin{barticle}
\bauthor{\bsnm{{Poduval}}, \binits{B.}},
\bauthor{\bsnm{{DeForest}}, \binits{C.E.}},
\bauthor{\bsnm{{Schmelz}}, \binits{J.T.}},
\bauthor{\bsnm{{Pathak}}, \binits{S.}}:
\byear{2013},
\batitle{{Point-spread Functions for the Extreme-ultraviolet Channels of
  SDO/AIA Telescopes}}.
\bjtitle{\apj}
\bvolume{765},
\bfpage{144}.
\doiurl{https://doi.org/10.1088/0004-637X/765/2/144}.
\adsurl{2013ApJ...765..144P}.
\end{barticle}
\endbibitem

\bibitem[\protect\citeauthoryear{{Qiu}}{2021}]{Qiu2021}
\begin{barticle}
\bauthor{\bsnm{{Qiu}}, \binits{J.}}:
\byear{2021},
\batitle{{The Neupert Effect of Flare Ultraviolet and Soft X-Ray Emissions}}.
\bjtitle{\apj}
\bvolume{909},
\bfpage{99}.
\doiurl{https://doi.org/10.3847/1538-4357/abe0b3}.
\adsurl{2021ApJ...909...99Q}.
\end{barticle}
\endbibitem

\bibitem[\protect\citeauthoryear{{Qiu}}{2024}]{Qiu2024}
\begin{barticle}
\bauthor{\bsnm{{Qiu}}, \binits{J.}}:
\byear{2024},
\batitle{{Tracing field lines that are reconnecting, or expanding, or both}}.
\bjtitle{Frontiers in Astronomy and Space Sciences}
\bvolume{11},
\bfpage{1401846}.
\doiurl{https://doi.org/10.3389/fspas.2024.1401846}.
\adsurl{2024FrASS..1101846Q}.
\end{barticle}
\endbibitem

\bibitem[\protect\citeauthoryear{{Qiu} and {Cheng}}{2022}]{Qiu2022}
\begin{barticle}
\bauthor{\bsnm{{Qiu}}, \binits{J.}},
\bauthor{\bsnm{{Cheng}}, \binits{J.}}:
\byear{2022},
\batitle{{Properties and Energetics of Magnetic Reconnection: I. Evolution of
  Flare Ribbons}}.
\bjtitle{\solphys}
\bvolume{297},
\bfpage{80}.
\doiurl{https://doi.org/10.1007/s11207-022-02003-7}.
\adsurl{2022SoPh..297...80Q}.
\end{barticle}
\endbibitem

\bibitem[\protect\citeauthoryear{{Qiu} and {Longcope}}{2016}]{Qiu2016}
\begin{barticle}
\bauthor{\bsnm{{Qiu}}, \binits{J.}},
\bauthor{\bsnm{{Longcope}}, \binits{D.W.}}:
\byear{2016},
\batitle{{Long Duration Flare Emission: Impulsive Heating or Gradual Heating?}}
\bjtitle{\apj}
\bvolume{820},
\bfpage{14}.
\doiurl{https://doi.org/10.3847/0004-637X/820/1/14}.
\adsurl{2016ApJ...820...14Q}.
\end{barticle}
\endbibitem

\bibitem[\protect\citeauthoryear{{Qiu}, {Liu}, and {Longcope}}{2012}]{Qiu2012}
\begin{barticle}
\bauthor{\bsnm{{Qiu}}, \binits{J.}},
\bauthor{\bsnm{{Liu}}, \binits{W.-J.}},
\bauthor{\bsnm{{Longcope}}, \binits{D.W.}}:
\byear{2012},
\batitle{{Heating of Flare Loops with Observationally Constrained Heating
  Functions}}.
\bjtitle{\apj}
\bvolume{752},
\bfpage{124}.
\doiurl{https://doi.org/10.1088/0004-637X/752/2/124}.
\adsurl{2012ApJ...752..124Q}.
\end{barticle}
\endbibitem

\bibitem[\protect\citeauthoryear{{Qiu} et~al.}{2002}]{Qiu2002}
\begin{barticle}
\bauthor{\bsnm{{Qiu}}, \binits{J.}},
\bauthor{\bsnm{{Lee}}, \binits{J.}},
\bauthor{\bsnm{{Gary}}, \binits{D.E.}},
\bauthor{\bsnm{{Wang}}, \binits{H.}}:
\byear{2002},
\batitle{{Motion of Flare Footpoint Emission and Inferred Electric Field in
  Reconnecting Current Sheets}}.
\bjtitle{\apj}
\bvolume{565},
\bfpage{1335}.
\doiurl{https://doi.org/10.1086/324706}.
\adsurl{2002ApJ...565.1335Q}.
\end{barticle}
\endbibitem

\bibitem[\protect\citeauthoryear{{Qiu} et~al.}{2010}]{Qiu2010}
\begin{barticle}
\bauthor{\bsnm{{Qiu}}, \binits{J.}},
\bauthor{\bsnm{{Liu}}, \binits{W.}},
\bauthor{\bsnm{{Hill}}, \binits{N.}},
\bauthor{\bsnm{{Kazachenko}}, \binits{M.}}:
\byear{2010},
\batitle{{Reconnection and Energetics in Two-ribbon Flares: A Revisit of the
  Bastille-day Flare}}.
\bjtitle{\apj}
\bvolume{725},
\bfpage{319}.
\doiurl{https://doi.org/10.1088/0004-637X/725/1/319}.
\adsurl{2010ApJ...725..319Q}.
\end{barticle}
\endbibitem

\bibitem[\protect\citeauthoryear{{Qiu} et~al.}{2013}]{Qiu2013}
\begin{barticle}
\bauthor{\bsnm{{Qiu}}, \binits{J.}},
\bauthor{\bsnm{{Sturrock}}, \binits{Z.}},
\bauthor{\bsnm{{Longcope}}, \binits{D.W.}},
\bauthor{\bsnm{{Klimchuk}}, \binits{J.A.}},
\bauthor{\bsnm{{Liu}}, \binits{W.-J.}}:
\byear{2013},
\batitle{{Ultraviolet and Extreme-ultraviolet Emissions at the Flare Footpoints
  Observed by Atmosphere Imaging Assembly}}.
\bjtitle{\apj}
\bvolume{774},
\bfpage{14}.
\doiurl{https://doi.org/10.1088/0004-637X/774/1/14}.
\adsurl{2013ApJ...774...14Q}.
\end{barticle}
\endbibitem

\bibitem[\protect\citeauthoryear{{Rast} et~al.}{2021}]{Rast2021}
\begin{barticle}
\bauthor{\bsnm{{Rast}}, \binits{M.P.}},
\bauthor{\bsnm{{Bello Gonz{\'a}lez}}, \binits{N.}},
\bauthor{\bsnm{{Bellot Rubio}}, \binits{L.}},
\bauthor{\bsnm{{Cao}}, \binits{W.}},
\bauthor{\bsnm{{Cauzzi}}, \binits{G.}},
\bauthor{\bsnm{{Deluca}}, \binits{E.}},
\bauthor{\bsnm{{de Pontieu}}, \binits{B.}},
\bauthor{\bsnm{{Fletcher}}, \binits{L.}},
\bauthor{\bsnm{{Gibson}}, \binits{S.E.}},
\bauthor{\bsnm{{Judge}}, \binits{P.G.}},
\bauthor{\bsnm{{Katsukawa}}, \binits{Y.}},
\bauthor{\bsnm{{Kazachenko}}, \binits{M.D.}},
\bauthor{\bsnm{{Khomenko}}, \binits{E.}},
\bauthor{\bsnm{{Landi}}, \binits{E.}},
\bauthor{\bsnm{{Mart{\'\i}nez Pillet}}, \binits{V.}},
\bauthor{\bsnm{{Petrie}}, \binits{G.J.D.}},
\bauthor{\bsnm{{Qiu}}, \binits{J.}},
\bauthor{\bsnm{{Rachmeler}}, \binits{L.A.}},
\bauthor{\bsnm{{Rempel}}, \binits{M.}},
\bauthor{\bsnm{{Schmidt}}, \binits{W.}},
\bauthor{\bsnm{{Scullion}}, \binits{E.}},
\bauthor{\bsnm{{Sun}}, \binits{X.}},
\bauthor{\bsnm{{Welsch}}, \binits{B.T.}},
\bauthor{\bsnm{{Andretta}}, \binits{V.}},
\bauthor{\bsnm{{Antolin}}, \binits{P.}},
\bauthor{\bsnm{{Ayres}}, \binits{T.R.}},
\bauthor{\bsnm{{Balasubramaniam}}, \binits{K.S.}},
\bauthor{\bsnm{{Ballai}}, \binits{I.}},
\bauthor{\bsnm{{Berger}}, \binits{T.E.}},
\bauthor{\bsnm{{Bradshaw}}, \binits{S.J.}},
\bauthor{\bsnm{{Campbell}}, \binits{R.J.}},
\bauthor{\bsnm{{Carlsson}}, \binits{M.}},
\bauthor{\bsnm{{Casini}}, \binits{R.}},
\bauthor{\bsnm{{Centeno}}, \binits{R.}},
\bauthor{\bsnm{{Cranmer}}, \binits{S.R.}},
\bauthor{\bsnm{{Criscuoli}}, \binits{S.}},
\bauthor{\bsnm{{Deforest}}, \binits{C.}},
\bauthor{\bsnm{{Deng}}, \binits{Y.}},
\bauthor{\bsnm{{Erd{\'e}lyi}}, \binits{R.}},
\bauthor{\bsnm{{Fedun}}, \binits{V.}},
\bauthor{\bsnm{{Fischer}}, \binits{C.E.}},
\bauthor{\bsnm{{Gonz{\'a}lez Manrique}}, \binits{S.J.}},
\bauthor{\bsnm{{Hahn}}, \binits{M.}},
\bauthor{\bsnm{{Harra}}, \binits{L.}},
\bauthor{\bsnm{{Henriques}}, \binits{V.M.J.}},
\bauthor{\bsnm{{Hurlburt}}, \binits{N.E.}},
\bauthor{\bsnm{{Jaeggli}}, \binits{S.}},
\bauthor{\bsnm{{Jafarzadeh}}, \binits{S.}},
\bauthor{\bsnm{{Jain}}, \binits{R.}},
\bauthor{\bsnm{{Jefferies}}, \binits{S.M.}},
\bauthor{\bsnm{{Keys}}, \binits{P.H.}},
\bauthor{\bsnm{{Kowalski}}, \binits{A.F.}},
\bauthor{\bsnm{{Kuckein}}, \binits{C.}},
\bauthor{\bsnm{{Kuhn}}, \binits{J.R.}},
\bauthor{\bsnm{{Kuridze}}, \binits{D.}},
\bauthor{\bsnm{{Liu}}, \binits{J.}},
\bauthor{\bsnm{{Liu}}, \binits{W.}},
\bauthor{\bsnm{{Longcope}}, \binits{D.}},
\bauthor{\bsnm{{Mathioudakis}}, \binits{M.}},
\bauthor{\bsnm{{McAteer}}, \binits{R.T.J.}},
\bauthor{\bsnm{{McIntosh}}, \binits{S.W.}},
\bauthor{\bsnm{{McKenzie}}, \binits{D.E.}},
\bauthor{\bsnm{{Miralles}}, \binits{M.P.}},
\bauthor{\bsnm{{Morton}}, \binits{R.J.}},
\bauthor{\bsnm{{Muglach}}, \binits{K.}},
\bauthor{\bsnm{{Nelson}}, \binits{C.J.}},
\bauthor{\bsnm{{Panesar}}, \binits{N.K.}},
\bauthor{\bsnm{{Parenti}}, \binits{S.}},
\bauthor{\bsnm{{Parnell}}, \binits{C.E.}},
\bauthor{\bsnm{{Poduval}}, \binits{B.}},
\bauthor{\bsnm{{Reardon}}, \binits{K.P.}},
\bauthor{\bsnm{{Reep}}, \binits{J.W.}},
\bauthor{\bsnm{{Schad}}, \binits{T.A.}},
\bauthor{\bsnm{{Schmit}}, \binits{D.}},
\bauthor{\bsnm{{Sharma}}, \binits{R.}},
\bauthor{\bsnm{{Socas-Navarro}}, \binits{H.}},
\bauthor{\bsnm{{Srivastava}}, \binits{A.K.}},
\bauthor{\bsnm{{Sterling}}, \binits{A.C.}},
\bauthor{\bsnm{{Suematsu}}, \binits{Y.}},
\bauthor{\bsnm{{Tarr}}, \binits{L.A.}},
\bauthor{\bsnm{{Tiwari}}, \binits{S.}},
\bauthor{\bsnm{{Tritschler}}, \binits{A.}},
\bauthor{\bsnm{{Verth}}, \binits{G.}},
\bauthor{\bsnm{{Vourlidas}}, \binits{A.}},
\bauthor{\bsnm{{Wang}}, \binits{H.}},
\bauthor{\bsnm{{Wang}}, \binits{Y.-M.}},
\bauthor{\bsnm{{NSO and DKIST Project}}},
\bauthor{\bsnm{{DKIST Instrument Scientists}}},
\bauthor{\bsnm{{DKIST Science Working Group}}},
\bauthor{\bsnm{{DKIST Critical Science Plan Community}}}:
\byear{2021},
\batitle{{Critical Science Plan for the Daniel K. Inouye Solar Telescope
  (DKIST)}}.
\bjtitle{\solphys}
\bvolume{296},
\bfpage{70}.
\doiurl{https://doi.org/10.1007/s11207-021-01789-2}.
\adsurl{2021SoPh..296...70R}.
\end{barticle}
\endbibitem

\bibitem[\protect\citeauthoryear{{Reale}}{2014}]{Reale2014}
\begin{barticle}
\bauthor{\bsnm{{Reale}}, \binits{F.}}:
\byear{2014},
\batitle{{Coronal Loops: Observations and Modeling of Confined Plasma}}.
\bjtitle{Living Reviews in Solar Physics}
\bvolume{11},
\bfpage{4}.
\doiurl{https://doi.org/10.12942/lrsp-2014-4}.
\adsurl{2014LRSP...11....4R}.
\end{barticle}
\endbibitem

\bibitem[\protect\citeauthoryear{{Reep} and {Knizhnik}}{2019}]{Reep2019}
\begin{barticle}
\bauthor{\bsnm{{Reep}}, \binits{J.W.}},
\bauthor{\bsnm{{Knizhnik}}, \binits{K.J.}}:
\byear{2019},
\batitle{{What Determines the X-Ray Intensity and Duration of a Solar Flare?}}
\bjtitle{\apj}
\bvolume{874},
\bfpage{157}.
\doiurl{https://doi.org/10.3847/1538-4357/ab0ae7}.
\adsurl{2019ApJ...874..157R}.
\end{barticle}
\endbibitem

\bibitem[\protect\citeauthoryear{{Reep} and {Russell}}{2016}]{Reep2016}
\begin{barticle}
\bauthor{\bsnm{{Reep}}, \binits{J.W.}},
\bauthor{\bsnm{{Russell}}, \binits{A.J.B.}}:
\byear{2016},
\batitle{{Alfv{\'e}nic Wave Heating of the Upper Chromosphere in Flares}}.
\bjtitle{\apjl}
\bvolume{818},
\bfpage{L20}.
\doiurl{https://doi.org/10.3847/2041-8205/818/1/L20}.
\adsurl{2016ApJ...818L..20R}.
\end{barticle}
\endbibitem

\bibitem[\protect\citeauthoryear{{Rempel}}{2017}]{Rempel2017}
\begin{barticle}
\bauthor{\bsnm{{Rempel}}, \binits{M.}}:
\byear{2017},
\batitle{{Extension of the MURaM Radiative MHD Code for Coronal Simulations}}.
\bjtitle{\apj}
\bvolume{834},
\bfpage{10}.
\doiurl{https://doi.org/10.3847/1538-4357/834/1/10}.
\adsurl{2017ApJ...834...10R}.
\end{barticle}
\endbibitem

\bibitem[\protect\citeauthoryear{{Rempel} et~al.}{2023}]{Rempel2023}
\begin{barticle}
\bauthor{\bsnm{{Rempel}}, \binits{M.}},
\bauthor{\bsnm{{Chintzoglou}}, \binits{G.}},
\bauthor{\bsnm{{Cheung}}, \binits{M.C.M.}},
\bauthor{\bsnm{{Fan}}, \binits{Y.}},
\bauthor{\bsnm{{Kleint}}, \binits{L.}}:
\byear{2023},
\batitle{{Comprehensive Radiative MHD Simulations of Eruptive Flares above
  Collisional Polarity Inversion Lines}}.
\bjtitle{\apj}
\bvolume{955},
\bfpage{105}.
\doiurl{https://doi.org/10.3847/1538-4357/aced4d}.
\adsurl{2023ApJ...955..105R}.
\end{barticle}
\endbibitem

\bibitem[\protect\citeauthoryear{{Rimmele} et~al.}{2020}]{Rimmele2020}
\begin{barticle}
\bauthor{\bsnm{{Rimmele}}, \binits{T.R.}},
\bauthor{\bsnm{{Warner}}, \binits{M.}},
\bauthor{\bsnm{{Keil}}, \binits{S.L.}},
\bauthor{\bsnm{{Goode}}, \binits{P.R.}},
\bauthor{\bsnm{{Kn{\"o}lker}}, \binits{M.}},
\bauthor{\bsnm{{Kuhn}}, \binits{J.R.}},
\bauthor{\bsnm{{Rosner}}, \binits{R.R.}},
\bauthor{\bsnm{{McMullin}}, \binits{J.P.}},
\bauthor{\bsnm{{Casini}}, \binits{R.}},
\bauthor{\bsnm{{Lin}}, \binits{H.}},
\bauthor{\bsnm{{W{\"o}ger}}, \binits{F.}},
\bauthor{\bsnm{{von der L{\"u}he}}, \binits{O.}},
\bauthor{\bsnm{{Tritschler}}, \binits{A.}},
\bauthor{\bsnm{{Davey}}, \binits{A.}},
\bauthor{\bsnm{{de Wijn}}, \binits{A.}},
\bauthor{\bsnm{{Elmore}}, \binits{D.F.}},
\bauthor{\bsnm{{Fehlmann}}, \binits{A.}},
\bauthor{\bsnm{{Harrington}}, \binits{D.M.}},
\bauthor{\bsnm{{Jaeggli}}, \binits{S.A.}},
\bauthor{\bsnm{{Rast}}, \binits{M.P.}},
\bauthor{\bsnm{{Schad}}, \binits{T.A.}},
\bauthor{\bsnm{{Schmidt}}, \binits{W.}},
\bauthor{\bsnm{{Mathioudakis}}, \binits{M.}},
\bauthor{\bsnm{{Mickey}}, \binits{D.L.}},
\bauthor{\bsnm{{Anan}}, \binits{T.}},
\bauthor{\bsnm{{Beck}}, \binits{C.}},
\bauthor{\bsnm{{Marshall}}, \binits{H.K.}},
\bauthor{\bsnm{{Jeffers}}, \binits{P.F.}},
\bauthor{\bsnm{{Oschmann}}, \binits{J.M.}},
\bauthor{\bsnm{{Beard}}, \binits{A.}},
\bauthor{\bsnm{{Berst}}, \binits{D.C.}},
\bauthor{\bsnm{{Cowan}}, \binits{B.A.}},
\bauthor{\bsnm{{Craig}}, \binits{S.C.}},
\bauthor{\bsnm{{Cross}}, \binits{E.}},
\bauthor{\bsnm{{Cummings}}, \binits{B.K.}},
\bauthor{\bsnm{{Donnelly}}, \binits{C.}},
\bauthor{\bsnm{{de Vanssay}}, \binits{J.-B.}},
\bauthor{\bsnm{{Eigenbrot}}, \binits{A.D.}},
\bauthor{\bsnm{{Ferayorni}}, \binits{A.}},
\bauthor{\bsnm{{Foster}}, \binits{C.}},
\bauthor{\bsnm{{Galapon}}, \binits{C.A.}},
\bauthor{\bsnm{{Gedrites}}, \binits{C.}},
\bauthor{\bsnm{{Gonzales}}, \binits{K.}},
\bauthor{\bsnm{{Goodrich}}, \binits{B.D.}},
\bauthor{\bsnm{{Gregory}}, \binits{B.S.}},
\bauthor{\bsnm{{Guzman}}, \binits{S.S.}},
\bauthor{\bsnm{{Guzzo}}, \binits{S.}},
\bauthor{\bsnm{{Hegwer}}, \binits{S.}},
\bauthor{\bsnm{{Hubbard}}, \binits{R.P.}},
\bauthor{\bsnm{{Hubbard}}, \binits{J.R.}},
\bauthor{\bsnm{{Johansson}}, \binits{E.M.}},
\bauthor{\bsnm{{Johnson}}, \binits{L.C.}},
\bauthor{\bsnm{{Liang}}, \binits{C.}},
\bauthor{\bsnm{{Liang}}, \binits{M.}},
\bauthor{\bsnm{{McQuillen}}, \binits{I.}},
\bauthor{\bsnm{{Mayer}}, \binits{C.}},
\bauthor{\bsnm{{Newman}}, \binits{K.}},
\bauthor{\bsnm{{Onodera}}, \binits{B.}},
\bauthor{\bsnm{{Phelps}}, \binits{L.}},
\bauthor{\bsnm{{Puentes}}, \binits{M.M.}},
\bauthor{\bsnm{{Richards}}, \binits{C.}},
\bauthor{\bsnm{{Rimmele}}, \binits{L.M.}},
\bauthor{\bsnm{{Sekulic}}, \binits{P.}},
\bauthor{\bsnm{{Shimko}}, \binits{S.R.}},
\bauthor{\bsnm{{Simison}}, \binits{B.E.}},
\bauthor{\bsnm{{Smith}}, \binits{B.}},
\bauthor{\bsnm{{Starman}}, \binits{E.}},
\bauthor{\bsnm{{Sueoka}}, \binits{S.R.}},
\bauthor{\bsnm{{Summers}}, \binits{R.T.}},
\bauthor{\bsnm{{Szabo}}, \binits{A.}},
\bauthor{\bsnm{{Szabo}}, \binits{L.}},
\bauthor{\bsnm{{Wampler}}, \binits{S.B.}},
\bauthor{\bsnm{{Williams}}, \binits{T.R.}},
\bauthor{\bsnm{{White}}, \binits{C.}}:
\byear{2020},
\batitle{{The Daniel K. Inouye Solar Telescope - Observatory Overview}}.
\bjtitle{\solphys}
\bvolume{295},
\bfpage{172}.
\doiurl{https://doi.org/10.1007/s11207-020-01736-7}.
\adsurl{2020SoPh..295..172R}.
\end{barticle}
\endbibitem

\bibitem[\protect\citeauthoryear{{Saba}, {Gaeng}, and
  {Tarbell}}{2006}]{Saba2006}
\begin{barticle}
\bauthor{\bsnm{{Saba}}, \binits{J.L.R.}},
\bauthor{\bsnm{{Gaeng}}, \binits{T.}},
\bauthor{\bsnm{{Tarbell}}, \binits{T.D.}}:
\byear{2006},
\batitle{{Analysis of Solar Flare Ribbon Evolution: A Semiautomated Approach}}.
\bjtitle{\apj}
\bvolume{641},
\bfpage{1197}.
\doiurl{https://doi.org/10.1086/500631}.
\adsurl{2006ApJ...641.1197S}.
\end{barticle}
\endbibitem

\bibitem[\protect\citeauthoryear{{Scharmer} et~al.}{2003}]{Scharmer2003}
\begin{bchapter}
\bauthor{\bsnm{{Scharmer}}, \binits{G.B.}},
\bauthor{\bsnm{{Bjelksjo}}, \binits{K.}},
\bauthor{\bsnm{{Korhonen}}, \binits{T.K.}},
\bauthor{\bsnm{{Lindberg}}, \binits{B.}},
\bauthor{\bsnm{{Petterson}}, \binits{B.}}:
\byear{2003},
\bctitle{{The 1-meter Swedish solar telescope}}.
In: \beditor{\bsnm{{Keil}}, \binits{S.L.}},
\beditor{\bsnm{{Avakyan}}, \binits{S.V.}} (eds.)
\bbtitle{Innovative Telescopes and Instrumentation for Solar Astrophysics},
\bsertitle{Society of Photo-Optical Instrumentation Engineers (SPIE) Conference
  Series}
\bseriesno{4853},
\bfpage{341}.
\doiurl{https://doi.org/10.1117/12.460377}.
\adsurl{2003SPIE.4853..341S}.
\end{bchapter}
\endbibitem

\bibitem[\protect\citeauthoryear{{Scherrer} et~al.}{2012}]{Scherrer2012}
\begin{barticle}
\bauthor{\bsnm{{Scherrer}}, \binits{P.H.}},
\bauthor{\bsnm{{Schou}}, \binits{J.}},
\bauthor{\bsnm{{Bush}}, \binits{R.I.}},
\bauthor{\bsnm{{Kosovichev}}, \binits{A.G.}},
\bauthor{\bsnm{{Bogart}}, \binits{R.S.}},
\bauthor{\bsnm{{Hoeksema}}, \binits{J.T.}},
\bauthor{\bsnm{{Liu}}, \binits{Y.}},
\bauthor{\bsnm{{Duvall}}, \binits{T.L.}},
\bauthor{\bsnm{{Zhao}}, \binits{J.}},
\bauthor{\bsnm{{Title}}, \binits{A.M.}},
\bauthor{\bsnm{{Schrijver}}, \binits{C.J.}},
\bauthor{\bsnm{{Tarbell}}, \binits{T.D.}},
\bauthor{\bsnm{{Tomczyk}}, \binits{S.}}:
\byear{2012},
\batitle{{The Helioseismic and Magnetic Imager (HMI) Investigation for the
  Solar Dynamics Observatory (SDO)}}.
\bjtitle{\solphys}
\bvolume{275},
\bfpage{207}.
\doiurl{https://doi.org/10.1007/s11207-011-9834-2}.
\adsurl{2012SoPh..275..207S}.
\end{barticle}
\endbibitem

\bibitem[\protect\citeauthoryear{{Shibata} and {Tanuma}}{2001}]{Shibata2001}
\begin{barticle}
\bauthor{\bsnm{{Shibata}}, \binits{K.}},
\bauthor{\bsnm{{Tanuma}}, \binits{S.}}:
\byear{2001},
\batitle{{Plasmoid-induced-reconnection and fractal reconnection}}.
\bjtitle{Earth, Planets, and Space}
\bvolume{53},
\bfpage{473}.
\doiurl{https://doi.org/10.1186/BF03353258}.
\adsurl{2001EP\%26S...53..473S}.
\end{barticle}
\endbibitem

\bibitem[\protect\citeauthoryear{{Sim{\~o}es} et~al.}{2019}]{Simoes2019}
\begin{barticle}
\bauthor{\bsnm{{Sim{\~o}es}}, \binits{P.J.A.}},
\bauthor{\bsnm{{Reid}}, \binits{H.A.S.}},
\bauthor{\bsnm{{Milligan}}, \binits{R.O.}},
\bauthor{\bsnm{{Fletcher}}, \binits{L.}}:
\byear{2019},
\batitle{{The Spectral Content of SDO/AIA 1600 and 1700 {\r{A}} Filters from
  Flare and Plage Observations}}.
\bjtitle{\apj}
\bvolume{870},
\bfpage{114}.
\doiurl{https://doi.org/10.3847/1538-4357/aaf28d}.
\adsurl{2019ApJ...870..114S}.
\end{barticle}
\endbibitem

\bibitem[\protect\citeauthoryear{{Takahashi}, {Qiu}, and
  {Shibata}}{2017}]{Takahashi2017}
\begin{barticle}
\bauthor{\bsnm{{Takahashi}}, \binits{T.}},
\bauthor{\bsnm{{Qiu}}, \binits{J.}},
\bauthor{\bsnm{{Shibata}}, \binits{K.}}:
\byear{2017},
\batitle{{Quasi-periodic Oscillations in Flares and Coronal Mass Ejections
  Associated with Magnetic Reconnection}}.
\bjtitle{\apj}
\bvolume{848},
\bfpage{102}.
\doiurl{https://doi.org/10.3847/1538-4357/aa8f97}.
\adsurl{2017ApJ...848..102T}.
\end{barticle}
\endbibitem

\bibitem[\protect\citeauthoryear{{Takasao} and {Shibata}}{2016}]{Takasao2016}
\begin{barticle}
\bauthor{\bsnm{{Takasao}}, \binits{S.}},
\bauthor{\bsnm{{Shibata}}, \binits{K.}}:
\byear{2016},
\batitle{{Above-the-loop-top Oscillation and Quasi-periodic Coronal Wave
  Generation in Solar Flares}}.
\bjtitle{\apj}
\bvolume{823},
\bfpage{150}.
\doiurl{https://doi.org/10.3847/0004-637X/823/2/150}.
\adsurl{2016ApJ...823..150T}.
\end{barticle}
\endbibitem

\bibitem[\protect\citeauthoryear{{Tamburri}, {Kazachenko}, and
  {Kowalski}}{2024}]{Tamburri2024}
\begin{barticle}
\bauthor{\bsnm{{Tamburri}}, \binits{C.A.}},
\bauthor{\bsnm{{Kazachenko}}, \binits{M.D.}},
\bauthor{\bsnm{{Kowalski}}, \binits{A.F.}}:
\byear{2024},
\batitle{{The Relationships among Solar Flare Impulsiveness, Energy Release,
  and Ribbon Development}}.
\bjtitle{\apj}
\bvolume{966},
\bfpage{94}.
\doiurl{https://doi.org/10.3847/1538-4357/ad3047}.
\adsurl{2024ApJ...966...94T}.
\end{barticle}
\endbibitem

\bibitem[\protect\citeauthoryear{{Temmer} et~al.}{2007}]{Temmer2007}
\begin{barticle}
\bauthor{\bsnm{{Temmer}}, \binits{M.}},
\bauthor{\bsnm{{Veronig}}, \binits{A.M.}},
\bauthor{\bsnm{{Vr{\v s}nak}}, \binits{B.}},
\bauthor{\bsnm{{Miklenic}}, \binits{C.}}:
\byear{2007},
\batitle{{Energy Release Rates along H{$\alpha$} Flare Ribbons and the Location
  of Hard X-Ray Sources}}.
\bjtitle{\apj}
\bvolume{654},
\bfpage{665}.
\doiurl{https://doi.org/10.1086/509634}.
\adsurl{2007ApJ...654..665T}.
\end{barticle}
\endbibitem

\bibitem[\protect\citeauthoryear{{Thoen Faber} et~al.}{2025}]{Faber2025}
\begin{barticle}
\bauthor{\bsnm{{Thoen Faber}}, \binits{J.}},
\bauthor{\bsnm{{Joshi}}, \binits{R.}},
\bauthor{\bsnm{{Rouppe van der Voort}}, \binits{L.}},
\bauthor{\bsnm{{Wedemeyer}}, \binits{S.}},
\bauthor{\bsnm{{Fletcher}}, \binits{L.}},
\bauthor{\bsnm{{Aulanier}}, \binits{G.}},
\bauthor{\bsnm{{N{\'o}brega-Siverio}}, \binits{D.}}:
\byear{2025},
\batitle{{High-resolution observational analysis of flare ribbon fine
  structures}}.
\bjtitle{\aap}
\bvolume{693},
\bfpage{A8}.
\doiurl{https://doi.org/10.1051/0004-6361/202452370}.
\adsurl{2025A&A...693A...8T}.
\end{barticle}
\endbibitem

\bibitem[\protect\citeauthoryear{{Vernazza}, {Avrett}, and
  {Loeser}}{1981}]{Vernazza1981}
\begin{barticle}
\bauthor{\bsnm{{Vernazza}}, \binits{J.E.}},
\bauthor{\bsnm{{Avrett}}, \binits{E.H.}},
\bauthor{\bsnm{{Loeser}}, \binits{R.}}:
\byear{1981},
\batitle{{Structure of the solar chromosphere. III. Models of the EUV
  brightness components of the quiet sun.}}
\bjtitle{\apjs}
\bvolume{45},
\bfpage{635}.
\doiurl{https://doi.org/10.1086/190731}.
\adsurl{1981ApJS...45..635V}.
\end{barticle}
\endbibitem

\bibitem[\protect\citeauthoryear{{Vievering} et~al.}{2023}]{Vievering2023}
\begin{barticle}
\bauthor{\bsnm{{Vievering}}, \binits{J.T.}},
\bauthor{\bsnm{{Vourlidas}}, \binits{A.}},
\bauthor{\bsnm{{Zhu}}, \binits{C.}},
\bauthor{\bsnm{{Qiu}}, \binits{J.}},
\bauthor{\bsnm{{Glesener}}, \binits{L.}}:
\byear{2023},
\batitle{{Evolution of Solar Eruptive Events: Investigating the Relationships
  among Magnetic Reconnection, Flare Energy Release, and Coronal Mass
  Ejections}}.
\bjtitle{\apj}
\bvolume{946},
\bfpage{81}.
\doiurl{https://doi.org/10.3847/1538-4357/acbe3d}.
\adsurl{2023ApJ...946...81V}.
\end{barticle}
\endbibitem

\bibitem[\protect\citeauthoryear{{Warren}}{2006}]{Warren2006}
\begin{barticle}
\bauthor{\bsnm{{Warren}}, \binits{H.P.}}:
\byear{2006},
\batitle{{Multithread Hydrodynamic Modeling of a Solar Flare}}.
\bjtitle{\apj}
\bvolume{637},
\bfpage{522}.
\doiurl{https://doi.org/10.1086/497904}.
\adsurl{2006ApJ...637..522W}.
\end{barticle}
\endbibitem

\bibitem[\protect\citeauthoryear{{Warren} and {Warshall}}{2001}]{Warren2001}
\begin{barticle}
\bauthor{\bsnm{{Warren}}, \binits{H.P.}},
\bauthor{\bsnm{{Warshall}}, \binits{A.D.}}:
\byear{2001},
\batitle{{Ultraviolet Flare Ribbon Brightenings and the Onset of Hard X-Ray
  Emission}}.
\bjtitle{\apjl}
\bvolume{560},
\bfpage{L87}.
\doiurl{https://doi.org/10.1086/324060}.
\adsurl{2001ApJ...560L..87W}.
\end{barticle}
\endbibitem

\bibitem[\protect\citeauthoryear{{Weber} et~al.}{2004}]{Weber2004}
\begin{bchapter}
\bauthor{\bsnm{{Weber}}, \binits{M.A.}},
\bauthor{\bsnm{{Deluca}}, \binits{E.E.}},
\bauthor{\bsnm{{Golub}}, \binits{L.}},
\bauthor{\bsnm{{Sette}}, \binits{A.L.}}:
\byear{2004},
\bctitle{{Temperature diagnostics with multichannel imaging telescopes}}.
In: \beditor{\bsnm{{Stepanov}}, \binits{A.V.}},
\beditor{\bsnm{{Benevolenskaya}}, \binits{E.E.}},
\beditor{\bsnm{{Kosovichev}}, \binits{A.G.}} (eds.)
\bbtitle{Multi-Wavelength Investigations of Solar Activity},
\bsertitle{IAU Symposium}
\bseriesno{223},
\bfpage{321}.
\doiurl{https://doi.org/10.1017/S1743921304006088}.
\adsurl{2004IAUS..223..321W}.
\end{bchapter}
\endbibitem

\bibitem[\protect\citeauthoryear{{Woods} et~al.}{2011}]{Woods2011}
\begin{barticle}
\bauthor{\bsnm{{Woods}}, \binits{T.N.}},
\bauthor{\bsnm{{Hock}}, \binits{R.}},
\bauthor{\bsnm{{Eparvier}}, \binits{F.}},
\bauthor{\bsnm{{Jones}}, \binits{A.R.}},
\bauthor{\bsnm{{Chamberlin}}, \binits{P.C.}},
\bauthor{\bsnm{{Klimchuk}}, \binits{J.A.}},
\bauthor{\bsnm{{Didkovsky}}, \binits{L.}},
\bauthor{\bsnm{{Judge}}, \binits{D.}},
\bauthor{\bsnm{{Mariska}}, \binits{J.}},
\bauthor{\bsnm{{Warren}}, \binits{H.}},
\bauthor{\bsnm{{Schrijver}}, \binits{C.J.}},
\bauthor{\bsnm{{Webb}}, \binits{D.F.}},
\bauthor{\bsnm{{Bailey}}, \binits{S.}},
\bauthor{\bsnm{{Tobiska}}, \binits{W.K.}}:
\byear{2011},
\batitle{{New Solar Extreme-ultraviolet Irradiance Observations during
  Flares}}.
\bjtitle{\apj}
\bvolume{739},
\bfpage{59}.
\doiurl{https://doi.org/10.1088/0004-637X/739/2/59}.
\adsurl{2011ApJ...739...59W}.
\end{barticle}
\endbibitem

\bibitem[\protect\citeauthoryear{{Wyper} and {Pontin}}{2021}]{Wyper2021}
\begin{barticle}
\bauthor{\bsnm{{Wyper}}, \binits{P.F.}},
\bauthor{\bsnm{{Pontin}}, \binits{D.I.}}:
\byear{2021},
\batitle{{Is Flare Ribbon Fine Structure Related to Tearing in the Flare
  Current Sheet?}}
\bjtitle{\apj}
\bvolume{920},
\bfpage{102}.
\doiurl{https://doi.org/10.3847/1538-4357/ac1943}.
\adsurl{2021ApJ...920..102W}.
\end{barticle}
\endbibitem

\bibitem[\protect\citeauthoryear{{Zhu} et~al.}{2020}]{Zhu2020}
\begin{barticle}
\bauthor{\bsnm{{Zhu}}, \binits{C.}},
\bauthor{\bsnm{{Qiu}}, \binits{J.}},
\bauthor{\bsnm{{Liewer}}, \binits{P.}},
\bauthor{\bsnm{{Vourlidas}}, \binits{A.}},
\bauthor{\bsnm{{Spiegel}}, \binits{M.}},
\bauthor{\bsnm{{Hu}}, \binits{Q.}}:
\byear{2020},
\batitle{{How Does Magnetic Reconnection Drive the Early-stage Evolution of
  Coronal Mass Ejections?}}
\bjtitle{\apj}
\bvolume{893},
\bfpage{141}.
\doiurl{https://doi.org/10.3847/1538-4357/ab838a}.
\adsurl{2020ApJ...893..141Z}.
\end{barticle}
\endbibitem

\bibitem[\protect\citeauthoryear{{Zhu} et~al.}{2025}]{Zhu2025}
\begin{botherref}
\oauthor{\bsnm{{Zhu}}, \binits{C.}},
\oauthor{\bsnm{{DeVore}}, \binits{C.R.}},
\oauthor{\bsnm{{Dahlin}}, \binits{J.T.}},
\oauthor{\bsnm{{Qiu}}, \binits{J.}},
\oauthor{\bsnm{{Kazachenko}}, \binits{M.D.}},
\oauthor{\bsnm{{Uritsky}}, \binits{V.M.}}:
2025,
{Genesis of a Confined X6.4 Flare in a Quadrupolar Magnetic Configuration}.
\textit{\apj}
\textbf{in preparation}.
\end{botherref}
\endbibitem

\bibitem[\protect\citeauthoryear{{Zimovets} et~al.}{2021}]{Zimovets2021}
\begin{barticle}
\bauthor{\bsnm{{Zimovets}}, \binits{I.V.}},
\bauthor{\bsnm{{McLaughlin}}, \binits{J.A.}},
\bauthor{\bsnm{{Srivastava}}, \binits{A.K.}},
\bauthor{\bsnm{{Kolotkov}}, \binits{D.Y.}},
\bauthor{\bsnm{{Kuznetsov}}, \binits{A.A.}},
\bauthor{\bsnm{{Kupriyanova}}, \binits{E.G.}},
\bauthor{\bsnm{{Cho}}, \binits{I.-H.}},
\bauthor{\bsnm{{Inglis}}, \binits{A.R.}},
\bauthor{\bsnm{{Reale}}, \binits{F.}},
\bauthor{\bsnm{{Pascoe}}, \binits{D.J.}},
\bauthor{\bsnm{{Tian}}, \binits{H.}},
\bauthor{\bsnm{{Yuan}}, \binits{D.}},
\bauthor{\bsnm{{Li}}, \binits{D.}},
\bauthor{\bsnm{{Zhang}}, \binits{Q.M.}}:
\byear{2021},
\batitle{{Quasi-Periodic Pulsations in Solar and Stellar Flares: A Review of
  Underpinning Physical Mechanisms and Their Predicted Observational
  Signatures}}.
\bjtitle{\ssr}
\bvolume{217},
\bfpage{66}.
\doiurl{https://doi.org/10.1007/s11214-021-00840-9}.
\adsurl{2021SSRv..217...66Z}.
\end{barticle}
\endbibitem

\end{thebibliography}


\newpage

\end{article}

\end{document}